\documentclass[reqno]{amsart}

\usepackage{lineno}
\usepackage[toc]{appendix}
\modulolinenumbers[5]
\usepackage{comment}

\usepackage{graphicx}
\usepackage{amsmath,amssymb}
\usepackage{amsthm}
\usepackage{xcolor}
\usepackage{mathrsfs}
\usepackage{paralist}
\usepackage{multirow}
\usepackage[T1]{fontenc}
\usepackage[utf8]{inputenc}
\usepackage[english]{babel}
\DeclareMathAlphabet\mathbfcal{OMS}{cmsy}{b}{n}
\usepackage[figurename=Figure]{caption}
\usepackage{lineno}
\usepackage{bbm}
\usepackage[normalem]{ulem}
\usepackage[percent]{overpic}
\usepackage{color}
\usepackage{tikz}
\usepackage{yhmath}

\usepackage{wrapfig}
\usepackage{float}
\usepackage{booktabs}
\usepackage{makecell}
\usepackage{subcaption}
\usepackage{comment}

\usepackage{hyperref}
\hypersetup{
    colorlinks=true,     
    linkcolor=red,       
    citecolor=green,     
    filecolor=magenta,   
    urlcolor=blue        
}

\newcommand{\ud}{d}

\newcommand{\ie}{\emph{i.e.}} 
\newcommand{\eg}{\emph{e.g.}} 

\newcommand{\f}[1]{\overline{#1}} 
\newcommand{\ff}[1]{\widetilde{#1}} 

\newcommand{\pd}[2]{\frac{\partial{#1}}{\partial{#2}}} 
\newcommand{\der}[2]{\frac{\ud{#1}}{\ud{#2}}} 

\newcommand{\del}[1]{}

\newcommand{\CSMAGO}{C_s}
\newcommand{\TWINDOW}{\Delta T}
\newcommand{\NDOF}{N_{\mathrm{dof}}}
\newcommand{\QOI}{\Psi}

\usepackage{authblk}

\usepackage{amsaddr}
\usepackage{siunitx}
        
\usepackage[a4paper,bindingoffset=0.0in,%
            left=1.0in,right=1.0in,top=1.in,bottom=1.in,%
            footskip=.4in]{geometry}

\title[Optimization of SGS models for DSE schemes based on the discrete adjoint method]{End-to-end optimization of subgrid scale models for discontinuous spectral element schemes based on the discrete adjoint method}

\author{Nicola Clinco$^{1}$, Niccol{\`o} Tonicello$^{1}$, Paola Cinnella$^{2}$ and
Gianluigi Rozza$^{1}$}
\address{$^1$ mathLab, Mathematics Area, SISSA, via Bonomea 265, I-34136 Trieste, Italy}
\address{$^2$ Institut Jean Le Rond D'Alembert, Sorbonne Universit\'e, CNRS, 4 Place Jussieu, Paris, France}

\begin{document}
\begin{abstract}
In computational fluid dynamics, among turbulence modeling strategies, the Large Eddy Simulation (LES) formalism offers a compelling balance between accuracy and computational cost by resolving large-scale flow structures while modeling the effects of unresolved subgrid scales. However, the predictive capacity of LES is critically dependent on the choice and calibration of subgrid-scale (SGS) models, which often involve problem-dependent parameters and exhibit intricate interactions with the underlying numerical discretization.
In this work, we propose a discrete-adjoint framework to optimize the control parameters of SGS models \emph{in the loop}, leveraging automatic differentiation within a high-order Spectral Difference (SD) solver. Coarse-grained simulations of Forced Homogeneous Isotropic Turbulence (FHIT), together with reference filtered Direct Numerical Simulation (DNS) data, are employed to optimize a limited set of parameters for classical SGS models, including the Smagorinsky model and non-linear tensor-basis formulations. When applied to chaotic systems, such as LES, the choice of the objective function plays a crucial role in the stability and overall accuracy of the proposed approach. In this work, we consider the spatio-temporally averaged decay of the Legendre modal coefficients as the relevant quantity of interest for the SD scheme. The optimization is performed across different grid resolutions and polynomial orders, highlighting the impact of the numerical discretization on the resulting model performance. The same methodology has been applied to one-dimensional Burgers and to fully three-dimensional turbulence.
The trained models are subsequently assessed on a range of out-of-sample configurations, including Decaying Homogeneous Isotropic turbulence (DHIT) and the Taylor-Green vortex. Variations in polynomial order, grid resolution, and Reynolds number are considered to evaluate the robustness and generalization capabilities of the approach. In all test cases, the optimized models consistently demonstrate significant improvements over baseline SGS closures.
\end{abstract}

\maketitle

\section{Introduction}\label{sec:introduction}

The continuous advancement of computational resources has considerably increased the capabilities in performing high-fidelity simulations of turbulent flows~\cite{slotnick2014cfd}. In particular, Direct Numerical Simulation (DNS) and Large-Eddy Simulation (LES) have become two fundamental approaches for the simulation of turbulent flows at different levels of detail. DNS resolves all turbulent structures down to the \emph{Kolmogorov scale}, providing the most accurate representation of the flow dynamics. However, its computational cost is still prohibitive for many practical applications. LES, on the other hand, offers a more affordable alternative by explicitly resolving the large-scale structures while modeling the effects of the smaller, sub-grid scales~\cite{Pierre_Sagaut_2001_INC,eric_garnier_2009}. The modelization of the unresolved sub-grid scales is commonly referred to as Sub-Grid Scale (SGS) modeling.

Approaches such as DNS and LES can significantly benefit from highly resolving spatial discretizations, such as the high-order Spectral Element Methods (SEMs). These include Discontinuous Spectral Element Methods (DSEMs) such as the Discontinuous Galerkin (DG) \cite{hesthaven:book,cockburn:98,cockburn:98b}, the Flux Reconstruction (FR) \cite{huynh2007flux} and the Spectral Difference (SD) \cite{kopriva1996conservative,liu:06a,wang:07} schemes.

In the context of LES paradigm, two main approaches are commonly distinguished in the literature. The first is referred as Explicit Large-Eddy Simulation (ELES), which consists in modeling the unresolved scales by an analytical model, function of the large-scale, resolved variables and their gradients (\eg $\,$Smagorisnky ~\cite{smagorinsky1963}, WALE~\cite{nicoud1999subgrid}, Vreman~\cite{vreman2004eddy}, Sigma~\cite{nicoud1999subgrid} and many others). The second approach, known as Implicit Large-Eddy Simulation approach (ILES), instead, aims at modeling the unresolved scales through the numerical dispersion and diffusion of the numerical scheme itself~\cite{Grinstein_Margolin_Rider_2007,gloerfelt2019large}.
In this work, we focus on ELES. SGS closure models generally involve free parameters that require calibration in order to achieve satisfactory performance. The optimal choice of these parameters tends to be problem-dependent, as it can vary significantly with both the numerical discretization and the characteristics of the flow under consideration. In particular, the interaction between the SGS model and the numerical scheme may strongly influence the resulting flow dynamics. Consequently, parameter values that provide accurate results for a given numerical method or flow configuration may not necessarily remain optimal when different discretizations or flow regimes are considered~\cite{MeyersSagaut06,SagautSensivityAnalisys08,LUCOR_MEYERS_SAGAUT_2007}.

In recent years, several methodologies have been proposed for the development of SGS models. In particular, the rapid growth of machine learning techniques has generated significant interest within the CFD community~\cite{DuraisamyIaccarino,sanderse2024scientific}.
In this context, data-driven SGS modeling strategies can be broadly divided into two categories.
The first one, commonly referred to as \emph{a-priori} modeling, constructs SGS closures without direct interaction with the CFD solver. In this setting, the training procedure relies entirely on precomputed high-fidelity data (\ie filtered DNS) and, upon a proper definition of the model architecture, a usable function between resolved and under-resolved scales can be learned from data and utilized as SGS model. Although this methodology has demonstrated promising results in various applications~\cite{XieZelongWang20,Prakash2022115457,ClincoToniRo2025114302,beck2019deep,huang2026consistency,srinivasan2019predictions}, it does not explicitly account for the discretization errors introduced by the numerical solver. This limitation is particularly relevant in LES, where discretization errors can be of the same order of magnitude as the modeled SGS contributions~\cite{GHOSAL1996187,KRAVCHENKO1997310,CHOW2003366}. This is closely related to the well known differences between \emph{a-priori} and \emph{a-posteriori} analyses of LES: models which correlate well with filtered DNS data generally perform poorly when used in \emph{a-posteriori} simulations. At the same time, models which are not particularly excellent in \emph{a-priori} analyses have shown to be robust and perform well in a-posteriori simulations ~\cite{MeneveauKatz2000,piomelli1988model}.

As a consequence, developing SGS models that remain consistent with the underlying numerical discretization is of paramount importance.
Training SGS models directly \emph{in the loop} (\ie, by employing the LES solver within the training process) overcomes the main limitations of \emph{a-priori} training. This methodology allows the development of models that are not only physics-informed by the specific PDE at hand, but are also aware of their discretization and thus incorporate the numerical error of the LES solver directly during training. In this paradigm, two methodologies that have been proven effective can be found in the use of Reinforcement Learning (RL) and Ensemble Kalman Filter (EnKF). Within the RL framework, the model parameters, or control policies, are iteratively adapted based on feedback from the environment (\ie, the LES solver). Alternatively, EnKF-based methodologies provide a gradient-free statistical assimilation framework in which model states and parameters are continuously updated using observational data. Both approaches have been successfully applied in adaptive and closed-loop learning settings, as demonstrated in previous studies on reinforcement learning–based control \cite{KURZ2023109094,beck2023toward} and data assimilation using EnKF methods \cite{WangZelongJianchunEnsemble,VILLANUEVA2024109597}. However, such approaches require a large number of costly LES, which tends to increase with the search space dimensionality.

Another strategy enabling training-in-the-loop models is represented by the broad category of differentiable solvers~\cite{ThuereyCFDloop,Maulik25}. In this setting, the optimization is performed within the PDE solver itself, thereby naturally incorporating discretization effects during the training process. Several differentiable CFD solvers have recently been developed. Bezgin \emph{et al.}~\cite{BEZGIN2025109433} proposed a fully differentiable finite-volume solver implemented in \texttt{JAX}, while Wang \emph{et al.}~\cite{WANGFERRER2025} developed a differentiable discontinuous Galerkin solver in \texttt{JAX} for two-dimensional applications. Similarly, Agdestein \& Sanderse~\cite{ADGESTEINSANDERSE26} implemented a differentiable solver for the incompressible Navier–Stokes equations in \texttt{Julia}. In all these approaches, the optimization relies on the construction of the complete computational graph of the solver, enabling gradient propagation through the entire time-integration procedure. While this strategy provides exact derivatives of the discrete objective function, building and storing the full computational graph may become prohibitively expensive for LES applications, especially for large computational grids and long integration horizons. One way to keep the computational cost due to the training affordable is to compute the sensitivities of the objective functional using adjoint-based methods~\cite{SDadjoint,Giles2000,SIRIGNANO2020109811,KENWAY2019100542}.

The first application of the adjoint method in CFD was pioneered by Jameson in 1988~\cite{Jameson1988}. The continuous adjoint method was employed to optimize aircraft wing configurations using both a potential flow solver and three-dimensional inviscid Euler equations, marking a foundational step toward modern adjoint-based shape optimization frameworks. Among these, applications to optimal boundary control of the compressible Navier-Stokes equations~\cite{CollinsOptimalControl02} and to aero-acoustic simulations for noise reduction~\cite{WeiFreund2006,KimBodonyFreund2014} are worth mentioning.

Within this framework, the continuous adjoint formulation introduces an additional equation for the adjoint variable, derived from the continuous primal equations. This equation is first obtained analytically and then discretized and solved numerically. The discrete-adjoint, instead, generates an additional equation directly from the discretized primal equations. As a consequence, the discrete-adjoint formulation is consistent with the discretization and provides exact gradients of the objective function~\cite{HinzePinnauUlbrichUlbrich2009,NADARAJAH2001,NADARAJAH2000}. Although the continuous adjoint approach has proven effective in previous works~\cite{SDadjoint,KimBodonyFreund2014,CollinsOptimalControl02}, the accuracy of the resulting sensitivities can in some cases depend on the underlying discretization. This is particularly relevant for applications in which the numerical scheme has a strong influence on the overall accuracy of the prediction~\cite{NADARAJAH2001}, such as in LES. Moreover, the determination of appropriate boundary conditions for the adjoint variables introduces additional difficulties for complex flow scenarios~\cite{GunzBurger,HinzePinnauUlbrichUlbrich2009,PETER-NSANALYSIS}. These conditions do not always follow directly from the primal problem and require careful derivation to ensure consistency and well-posedness.
In contrast, the discrete-adjoint method differentiates the primal equations only after they have been discretized as a set of residual equations. On one hand, this approach produces an adjoint system that is consistent with the numerical scheme at the discrete level and yields machine-accurate gradients with respect to the discrete objectives, while also avoiding the ambiguity in adjoint boundary conditions. On the other hand, the discrete approach is considerably more complicated to implement with respect to the continuous setting. Its construction requires differentiating every component of the numerical solver, including flux evaluations, turbulence models, and linear or nonlinear solution procedures, which can lead to a large and highly coupled adjoint system.

Although the adjoint approach is well established in the context of design optimization in CFD, only a limited number of studies have focused on its application to SGS modeling. In particular, the use of adjoint-based methods for the analysis, calibration, or optimization of SGS models remains relatively unexplored compared with their widespread adoption in aerodynamic shape optimization and flow control problems, even more so for discrete-adjoint frameworks. A recent application of the continuous adjoint method for SGS modeling was proposed by Sirignano \emph{et al.}~\cite{SIRIGNANO2020109811}. They trained a data-driven SGS model with a continuous adjoint method using three-dimensional decaying homogeneous isotropic turbulence. The resulting model was shown to be effective in \emph{a-posteriori} simulations and to outperform models built entirely in \emph{a-priori} fashion. Generalization of the methodology to unstructured cases as well as to more challenging flow scenarios were given in \cite{SIRIGNANO23DEEP,MACART2021EMBEDDED}. In a related work, Yuan Zelong et al.~\cite{VOMMadjoint} employed a continuous adjoint framework for the construction of a mixed SGS model, obtaining improved predictions of turbulence statistics compared with classical closures such as the Smagorinsky~\cite{smagorinsky1963} and dynamic Smagorinsky~\cite{germano1991dynamic} models.


To the best of the authors’ knowledge, this is the first work to propose a discrete-adjoint framework for the optimization of subgrid-scale models for fully three-dimensional turbulence. Furthermore, it represents the first application of such an approach within the context of the Discontinuous Spectral Element Method. Even though the specific framework herein proposed is based on the spectral difference scheme, the same concepts are easily extendable to other DSEMs such as DG or FR schemes. Although previous works have developed discrete adjoint-based optimizations of SGS models, they mostly considered two-dimensional problems and a small amount of training data such as a few snapshots. We emphasize that one of the main objectives of LES is the development of models capable of reproducing the long-time statistical behavior of turbulent flows. Consequently, the optimization procedure should rely on a loss function constructed on a sufficiently large number of snapshots to ensure statistical significance. In this work, we then construct a loss function that is based on the time- and space-averaged modal energy in the space of Legendre polynomials rather than individual flow realizations. However, satisfying these requirements requires an efficient computational framework capable of addressing the substantial performance and memory constraints associated with the optimization procedure of large-scale systems. Furthermore, even if the applications herein considered are mostly focused on optimizing a limited number of control parameters in well-established training benchmarks, the real contribution of this work lies in the methodological developments of such a framework. The possibilities in terms of model optimization are significantly wide: from targeted turbulence models, to optimal shock-capturing, de-aliasing techniques and  near wall modeling.

This paper is organized as follows. Section \ref{sec:problemdef} introduces the filtered equations and the concept of Large Eddy Simulations for both one-dimensional forced Burgers turbulence and for the fully compressible three-dimensional Navier–Stokes equations. In Section \ref{sec:sd} we introduce the spectral difference scheme, highlighting the main algorithmic operations to be differentiated and their impact on the overall accuracy of the methodology. In Section \ref{sec:theadjoint}, the discrete-adjoint approach applied to the spectral difference framework is introduced, with emphasis on the implementation details and on the role of the loss function in the optimization process. Section \ref{sec:forcedhit} includes the first part of results of the proposed framework applied to the forced Burgers test case through the optimization of the Smagorinsky coefficient using a loss function based on modal energy content. Finally, Section \ref{sec:ns3d} extends the methodology to the fully compressible Navier–Stokes equations, where a more advanced mixed SGS model, incorporating both eddy-viscosity and structural contributions, is considered and optimized using the proposed approach. The resulting optimized models are shown to generalize successfully to flow configurations outside the training conditions.
\section{Problem Definition: filtered equations}\label{sec:problemdef}
Although the concept of LES is commonly used in CFD, the same rationale can be applied to a large variety of non-linear equations. One classical setting used to study LES and SGS modelling can be found in the one-dimensional forced Burgers turbulence case~\cite{SOLANFUSTERO2021110246,MOURA2015695}:
\begin{equation}
    \frac{\partial u}{\partial t} +\frac{1}{2}\frac{\partial u^2}{\partial x}  = \frac{A}{\Delta t}\sum_{n=1}^{N_c}\frac{\sigma_{n}(t)}{\sqrt{\pi n}}\cos\left( \frac{2\pi n}{L}x\right)
    \label{eq:burgers0}
\end{equation}
where $\sigma_{n}(t)$ is a standard Gaussian random function, $A=0.04$. This source term results from the sum of $N_c$ Fourier modes with amplitude $\sigma_{n}(t)/\sqrt{\pi n}$.

In LES, the governing equations are spatially filtered to separate the large scales, which are resolved, from the small-scale motions, which are modeled. The filtering operator $\overline{(\cdot)}$ denotes the spatial filtering operation which is commonly assumed to be a convolution operation with a kernel function $G$:
\begin{equation}
    \f{\phi}(\mathbf{x},t) = G(\mathbf{x};\Delta) \star \phi(\mathbf{x},t) = \int G(\mathbf{x}-\mathbf{x'};\Delta) \phi(\mathbf{x'}) d \mathbf{x'} \label{eq:spatial_filtering}.
\end{equation}
The kernel function $G$ is influenced by the cutoff length scale $\Delta$, which defines the threshold separating resolved and under-resolved scales.

Applying the above filter to equation \ref{eq:burgers0} leads to:
\begin{equation}
    \frac{\partial \overline{u}}{\partial t} +\frac{1}{2}\frac{\partial \overline{u}^2}{\partial x} = \frac{A}{\Delta t}\sum_{n=1}^{N_c}\frac{\sigma_{n}(t)}{\sqrt{\pi n}}\cos\left( \frac{2\pi n}{L}x\right) - \frac{\partial \tau^{\mathrm{SGS}}}{\partial x}
    \label{eq:burg_filt}
\end{equation}
with
\begin{equation}
    \tau^{\mathrm{SGS}} = \frac{1}{2}\big(\overline{u^{2}}-\overline{u}^{2}\big).
    \label{eq:sgs_burgers}
\end{equation}
The forcing term is typically applied at large scales, so filtering does not modify its functional form.

Rewriting the above equation in terms of a single unknown (namely, $\overline{u}$) leads to what is commonly referred to as the SGS closure problem. The objective is then to determine a suitable model for $\tau^{\mathrm{SGS}}$. The expression in equation \eqref{eq:sgs_burgers}, in fact, depends on $\overline{u^{2}}$ which is, in general, unknown. The classical approach is to relate $\tau^{\mathrm{SGS}}$ to large scale quantities such as $\overline{u}$ and its gradients. Alternatively, it is also possible to fine-tune the parameters of the numerical discretization so that its intrinsic numerical dissipation acts as an implicit SGS model. Both approaches have been successfully applied to the Burgers equations in the past (see for example~\cite{MAULIK201812}). Such concepts will be further expanded below when applied to the compressible Navier-Stokes equations.

In fact, as mentioned above, the use of LES is most often applied to three-dimensional equations of fluid dynamics. In the particular setting of this work, we will consider the compressible Navier-Stokes equations written in conservative form:
\begin{align}
    & \frac{\partial \rho}{\partial t} + \frac{\partial}{\partial x_j} \left( \rho u_j \right)= 0,  \label{eq:mass} \\
    &\frac{\partial}{\partial t} \left( \rho u_i \right) + \frac{\partial}{\partial x_j} \left( \mathbf{\rho} u_i  u_j + p\delta_{ij} \right) = \frac{\partial}{\partial x_j} \left( \Sigma_{ij} \right),\label{eq:momentum} \\
    & \frac{\partial}{\partial t}\left( \rho E \right ) + \frac{\partial}{\partial x_j} \left [ \left( \mathbf{\rho}E +p  \right) u_j \right] = -\frac{\partial q_j}{\partial x_j} + \frac{\partial}{\partial x_j} \left( \Sigma_{ij} u_i\right ),\label{eq:TotalEnergy}
\end{align}
where $\rho$ is the fluid density, $u_i$ is the $i$-th velocity component, $p$ is the pressure, $E$ is the total energy per unit mass, $\Sigma_{ij}$ is the viscous stress tensor and $q_j$ represents the heat flux. Furthermore, $\delta_{ij}$ indicates the Kronecker-delta.
The system \eqref{eq:mass}-\eqref{eq:TotalEnergy} is closed by a thermodynamic equation of state for the pressure which, based on the assumption that air behaves like an ideal gas, is given by:
\begin{equation}
    p  = \rho \left(E-\frac{u_i u_i}{2} \right ) \left( \gamma -1\right), \label{eq:state}
\end{equation}
where $\gamma = c_p/c_v$ is the specific heat ratio and by suitable constitutive relations for the viscous stresses $\Sigma_{ij}$. Hereafter, we consider a Newtonian fluid, with a constant viscosity and a constant Prandtl number.

For compressible turbulent flows, the Favre-filtering technique is applied to equations \eqref{eq:mass}-\eqref{eq:TotalEnergy}. For a general function $\phi$, the Favre-filtered quantity is defined as $\ff{(\phi)} = \f{\rho \phi}/{\f{\rho}}$. 
\begin{align}
    & \frac{\partial \f{\rho}}{\partial t} + \frac{\partial}{\partial x_j} \left( \f{\rho} \ff{u_j} \right)= 0,  \label{eq:fmass} \\
    &\frac{\partial}{\partial t} \left( \f{\rho} \ff{u_i} \right) + \frac{\partial}{\partial x_j} \left( \f{\rho} \ff{u_i}  \ff{u_j} + \f{p} \delta_{ij} \right) = \frac{\partial}{\partial x_j} \left( \f{\Sigma_{ij}} \right) -\frac{\partial \tau^{\mathrm{SGS}}_{ij}}{\partial x_j}, \label{eq:fmomentum} \\
    & \frac{\partial}{\partial t}\left( \f{\rho} \ff{E} \right ) + \frac{\partial}{\partial x_j} \left [ \left( \f{\rho}\ff{E} + \f{p} \right) \ff{u_j} \right] = -\frac{\partial \f{q_j}}{\partial x_j} + \frac{\partial}{\partial x_j} \left( \f{\Sigma_{ij} u_i}\right ) -\frac{\partial q_j^{\mathrm{SGS}}}{\partial x_j}, \label{eq:fTotalEnergy}
\end{align}
where $\tau^{\mathrm{SGS}}_{ij} = \f{\rho u_i u_j}-\f{\rho} \ff{u_i} \ff{u_j}$ represents the subgrid scale tensor and $q^{\mathrm{SGS}}_{j} = \f{\rho E u_j+p u_j} - (\f{\rho} \ff{E} \ff{u_j} +\f{p}\ff{u_j})$ accounts for the subgrid scale energy flux.
Further terms arise from the non-linearity of the viscous term when non-linear dependence of the viscosity with respect to the temperature (such as Sutherland law) is used. However, these terms are often negligible and $\tau^{\mathrm{SGS}}_{ij}$ and $q^{\mathrm{SGS}}_{j}$ are usually the main focus of modeling efforts in compressible LES~\cite{Vreman95Priori}.

It is important to point out that the numerical discretization used to resolve the filtered equations plays a crucial role. In this work, we will focus on the spectral difference scheme, which is a specific variant of the Spectral Element Method~\cite{kopriva1996conservative,liu:06a,wang:07}.

In the context of Spectral Element Methods, the treatment of non-linear terms and numerical flux functions significantly influences the accuracy and stability of $\mathrm{ILES}$. The  Riemann solver, for example, plays a critical role in regulating small-scale features by introducing controlled dissipation, which acts as an implicit turbulence model in ILES~\cite{Fernndez2018OnTA,Manzanero20SVV}. 
Gassner \& Kopriva~\cite{GassnerKopriva} demonstrated that in DG Methods, upwind schemes primarily introduce dissipation errors at high wave numbers, while lower and mid-range wave numbers remain largely unaffected. Moreover, as the accuracy order increases, the dissipation peak shifts toward higher wave numbers. As a result, leveraging polynomial order to regulate numerical dissipation makes DSEM methods well-suited for ILES. Along similar lines, De Wiart \emph{et.al}~\cite{DeWiart2015}  showed that the Riemann solver has a significant impact on the spectral distribution of the dissipation in the case of homogeneous turbulence. They compared the Lax-Friedrichs flux and the Roe flux, and they found that the former was better suited for fully turbulent flows.
Moreover, for high order polynomials, de-aliasing techniques (\eg, over-integration, split forms, explicit filtering) applied to non linear terms can be of significant importance in providing stable results for under-resolved flows within DSEMs~\cite{Beck2013,Beck2014,winters2018comparative}.
\section{The Spectral Difference scheme}\label{sec:sd}
For simplicity, let us consider the general one-dimensional conservation law of the form
\begin{equation}
\pd{u(x,t)}{t}+\pd{f(u)}{x} + \pd{}{x}\bigg( g\bigg( u,\pd{u}{x}\bigg)\bigg)=s(u,x,t),
\label{eq:advection}
\end{equation}
where $f(u)$ denotes the flux depending solely on the conserved variable $u$, $g(u,\partial u/\partial x)$ denotes the first-order flux depending on the variable $u$ and its first derivative and $s(u,x,t)$ represents a source term for simplicity depending only on the conserved variable $u$, spatial coordinate $x$ and time variable $t$. Even though equation \eqref{eq:advection} represents a general conservation law we will often refer to the argument of the first order term as inviscid flux and the argument of the second-order term as viscous flux. Notice that the Burgers equation introduced before can be easily written like equation \eqref{eq:advection}. Also the compressible Navier-Stokes can be seen as a multivariate generalization of the above equation. For more details regarding the SD discretization of the compressible Navier-Stokes equations the reader is referred to~\cite{kopriva1996conservative,liu:06a,wang:07}.

In the SD scheme, the solution is approximated with a polynomial of degree $N$
\begin{equation}
\hat{u}(\hat{x})=\sum_{i=0}^{N}u_{i}l^{s}_{i}(\hat{x}).
\label{eq:solSD}
\end{equation}
within the reference element $\Omega_{n} = \{ \hat{x}| -1 \leq \hat{x} \leq 1\}$. In the one-dimensional case, the map linking the reference element to the physical element, and consequently $\hat{x}$ with $x$, is a simple linear transformation, acting on the length of the element. Let us first focus on the first-order term involving the inviscid flux $f(u)$. 

The values of the conserved variables can be extrapolated at the flux points as
\begin{equation}
\hat{u}(\hat{x}^{f}_{j})=\sum_{i=0}^{N}u_{i}l^{s}_{i}(\hat{x}^{f}_{j}), \qquad j=0,...,N+1,
\end{equation}
and then used to compute fluxes on the same collocation basis:
\begin{equation}
f_{j}= \hat{f}(\hat{x}^f_{j})=\hat{f}(\hat{u}(\hat{x}^{f}_{j})).
\end{equation}
Then, a continuous flux polynomial of degree $N+1$ is constructed, by Lagrange interpolation, using the fluxes evaluated from the interpolated solution at the interior flux points and the numerical fluxes at the element interfaces:
\begin{equation}
\hat{f}(\hat{x})= \hat{f}^{I}_{L}l^f_{0}(\hat{x}) +\sum_{j=1}^{N}f_{j}l^f_{j}(\hat{x}) +\hat{f}^{I}_{R}l^f_{N+1}(\hat{x}).
\end{equation}
In other words, the interpolated values of the flux at elements extrema are substituted by the interface numerical fluxes $\hat{f}^{I}_{L}$ and $\hat{f}^{I}_{R}$.
Finally, the flux divergence is evaluated at the solution points,
\begin{equation}
\der{\hat{f}}{\hat{x}}(\hat{x}^s_{i})= \hat{f}^{I}_{L}\der{l^f_{0}}{\hat{x}}(\hat{x}^{s}_{i}) +\sum_{j=1}^{N}f_{j}\der{l^f_{j}}{\hat{x}}(\hat{x}^{s}_{i}) +\hat{f}^{I}_{R}\der{l^f_{N+1}}{\hat{x}}(\hat{x}^{s}_{i}).
\label{eq:finalSD}
\end{equation}
Second-order fluxes can be computed in a similar fashion. We begin by extrapolating the solution to the flux points as
\begin{equation}
\hat{u}(\hat{x}^{f}_{j})=\sum_{i=0}^{N}u_{i}l^{s}_{i}(\hat{x}^{f}_{j}), \qquad j=0,\dots,N+1.
\end{equation}
At this stage, we can follow the same procedure used for the computation of inviscid fluxes, simply considering the identity flux $f(u)=u$. This leads to the interpolated solution
\begin{equation}
\hat{u}(\hat{x})= \hat{u}^{I}_{L}l^f_{0}(\hat{x}) +\sum_{j=1}^{N}u_{j}l^f_{j}(\hat{x}) +\hat{u}^{I}_{R}l^f_{N+1}(\hat{x}).
\label{eq:dudx}
\end{equation}
By differentiating this expression and evaluating it at the solution points, we obtain the first derivative of the conserved variable at $\hat{x}^{s}_{i}$:
\begin{equation}
\der{\hat{u}}{\hat{x}}(\hat{x}^s_{i})= \hat{u}^{I}_{L}\der{l^f_{0}}{\hat{x}}(\hat{x}^{s}_{i}) +\sum_{j=1}^{N}u_{j}\der{l^f_{j}}{\hat{x}}(\hat{x}^{s}_{i}) +\hat{u}^{I}_{R}\der{l^f_{N+1}}{\hat{x}}(\hat{x}^{s}_{i}).
\label{eq:finalSD_du}
\end{equation}
Unlike the inviscid case, we now need to define interfacial values for the conserved variable in order to compute its first derivative within each element. This choice is flexible, similarly to numerical fluxes, and is often taken as a combination of left and right states. In this work, we adopt a centered evaluation.

Once the first derivative is available at the solution points, the same procedure used for inviscid fluxes can be applied. Defining $\hat{v}:=\partial \hat{u}/\partial\hat{x}$, we extrapolate it to the flux points as
\begin{equation}
\hat{v}(\hat{x}^{f}_{j})=\sum_{i=0}^{N}v_{i}l^{s}_{i}(\hat{x}^{f}_{j}), \qquad j=0,\dots,N+1.
\end{equation}
The second-order fluxes can then be evaluated at the flux points:
\begin{equation}
g_{j}= \hat{g}(\hat{x}^f_{j})=\hat{g}(\hat{u}(\hat{x}^{f}_{j}),\hat{v}(\hat{x}^{f}_{j})).
\end{equation}
By introducing appropriate numerical fluxes at element interfaces, we construct the following interpolant:
\begin{equation}
\hat{g}(\hat{x})= \hat{g}^{I}_{L}l^f_{0}(\hat{x}) +\sum_{j=1}^{N}g_{j}l^f_{j}(\hat{x}) +\hat{g}^{I}_{R}l^f_{N+1}(\hat{x}).
\end{equation}
Finally, differentiating this expression and evaluating it at the solution points yields the derivative of the second-order flux at $\hat{x}^{s}_{i}$:
\begin{equation}
\der{\hat{g}}{\hat{x}}(\hat{x}^s_{i})= \hat{g}^{I}_{L}\der{l^f_{0}}{\hat{x}}(\hat{x}^{s}_{i}) +\sum_{j=1}^{N}g_{j}\der{l^f_{j}}{\hat{x}}(\hat{x}^{s}_{i}) +\hat{g}^{I}_{R}\der{l^f_{N+1}}{\hat{x}}(\hat{x}^{s}_{i}).
\label{eq:finalSD_dg}
\end{equation}
Similarly to the inviscid case, the numerical fluxes $\hat{g}^{I}_{L/R}$ generally depend on the left and right states of the first derivative of the solution and implicitly on the choice of the intermediate interfacial values $\hat{u}^{I}_{L/R}$ introduced in equation \eqref{eq:dudx}. These common fluxes can be computed as a function of the average interface solution and the solution gradients~\cite{sun:07}. Alternatively, an interior penalty (IP) flux can be adopted~\cite{hesthaven2007nodal}. In this case, the numerical viscous flux at each element interface is obtained by combining the arithmetic mean of the (discontinuous) viscous fluxes evaluated to the left and to the right of the interface with an added penalty term proportional to the jump in the solution across the interface.
For any of these choices, while inviscid fluxes in the $n$-th element depend only on the reconstructed solutions of the neighboring elements (\ie, $n-1$ and $n+1$), the computation of viscous fluxes requires information from a wider stencil, involving also the elements $n-2$ and $n+2$. This concept will be crucially important when introducing the adjoint formulation for the SD scheme.

Finally, once first- and second-order terms are evaluated at the solution points, the numerical solution can be advanced in time using a suitable time integration scheme discretizing the following equation:
\begin{equation}
\der{\hat{u}}{t}=-\der{\hat{f}}{\hat{x}}(\hat{x}^s_{i}) -\der{\hat{g}}{\hat{x}}(\hat{x}^s_{i}) +  s(\hat{u}(\hat{x}_{i}^{s}),\hat{x}_{i}^{s},t) = R(\hat{x}^{s}_{i}).
\label{eq:fullSD}
\end{equation}

A specific SD method is then completely defined by:
\begin{enumerate}
\item the location of both solution and flux points;
\item the expression of inviscid interface fluxes $\hat{f}_{L/R}^{I}$. 
\item the expression of viscous interface fluxes $\hat{g}_{L/R}^{I}$.
\item the expression for the intermediate interfacial conserved variables $\hat{u}_{L/R}^{I}$.
\end{enumerate}

As will be discussed in detail in the following sections, each of these choices will be properly defined, as they generally depend on the specific equation being discretized using the SD method.

\begin{figure}[h!]
 \centering  
 \begin{subfigure}{0.48\textwidth}
     \includegraphics[width=\textwidth]{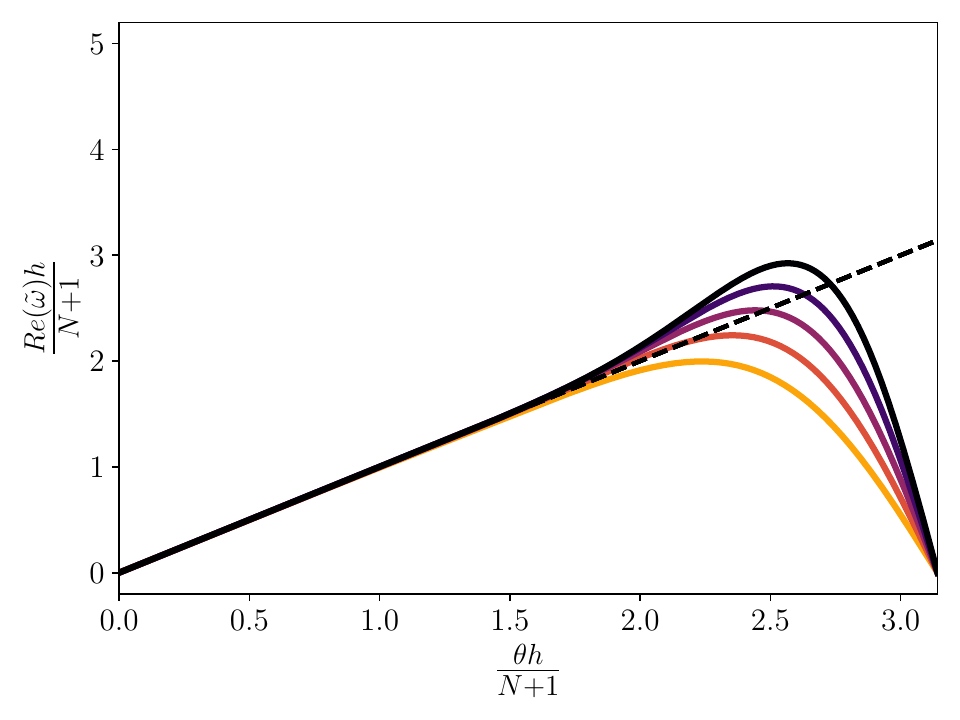}
 \end{subfigure}
 \begin{subfigure}{0.48\textwidth}
     \includegraphics[width=\textwidth]{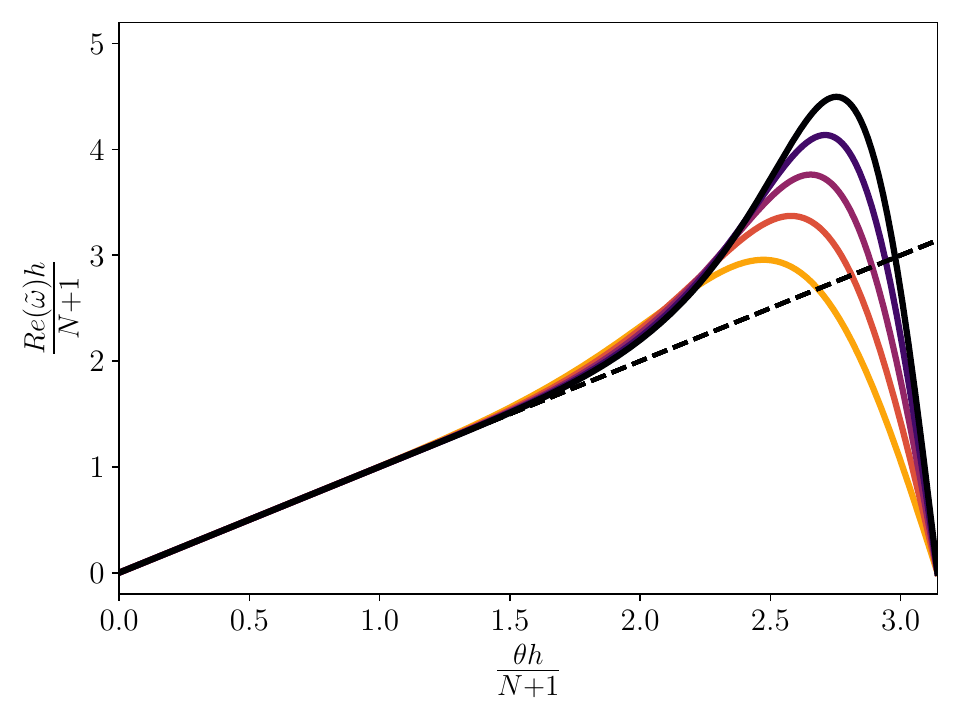}
 \end{subfigure}
\caption{Dispersion curves for Spectral Difference (left) and Discontinuous Galerkin (right) methods for order from $\mathrm{p}=3$ to $\mathrm{p}=7$ (color gradient from ocher to black). Dashed line indicates the analytical solution.}
   \label{fig:eigen_disp}
\end{figure}

\begin{figure}[h!]
 \centering  
 \begin{subfigure}{0.48\textwidth}
     \includegraphics[width=\textwidth]{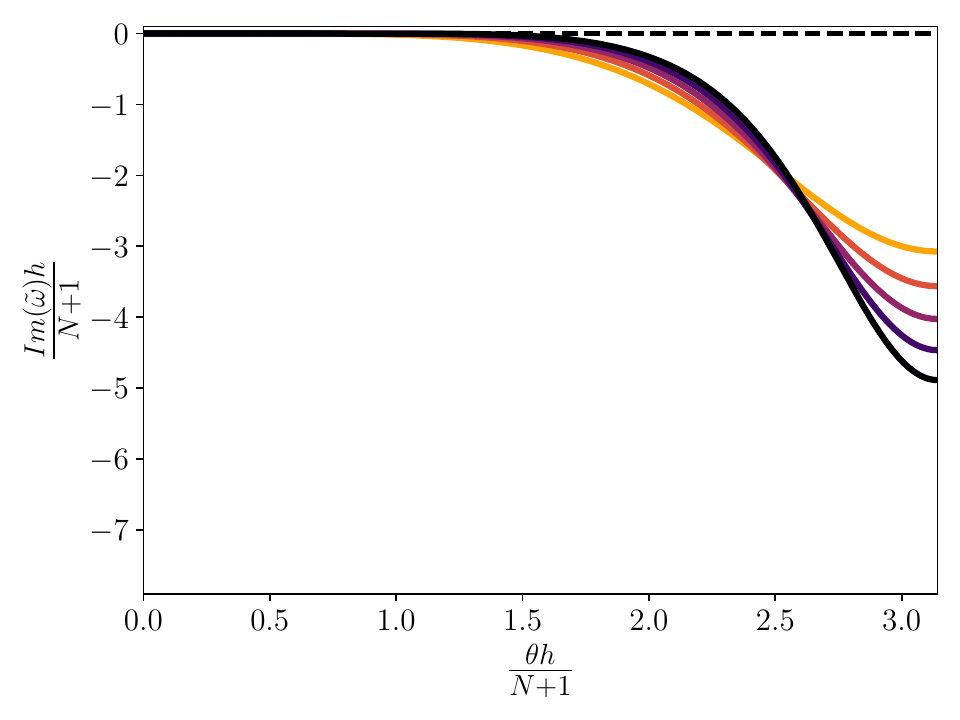}
 \end{subfigure}
 \begin{subfigure}{0.48\textwidth}
     \includegraphics[width=\textwidth]{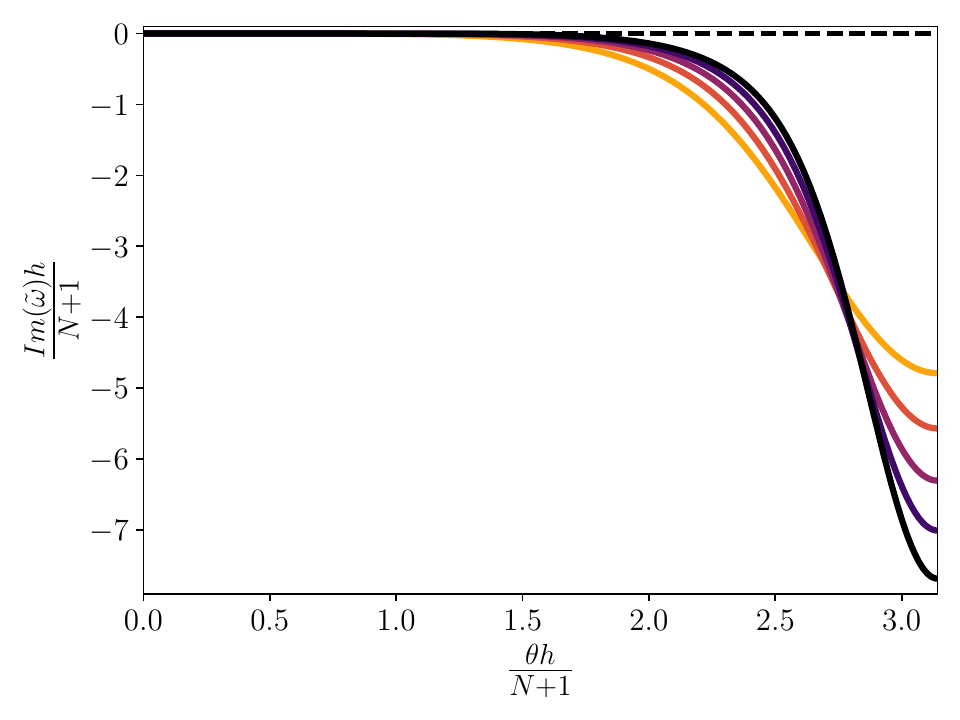}
 \end{subfigure}
\caption{Disssipation curves for Spectral Difference (left) and Discontinuous Galerkin (right) methods for order from $\mathrm{p}=3$ to $\mathrm{p}=7$ (color gradient from ocher to black). Dashed line indicates the analytical solution.}
   \label{fig:eigen_diff}
\end{figure}

As a simple example, it is nevertheless useful to consider the case of homogeneous linear advection, for which a specific SD method is completely defined by the first two components. In this simplified setting, the numerical flux $\hat{f}_{L/R}$ is typically chosen from a family of fluxes of the form:
\begin{equation}
\hat{f}_{L}^{I}=(1-\alpha)\hat{u}_{n-1}(+1)+\alpha \hat{u}_{n}(-1) \quad \mathrm{and} \quad \hat{f}_{R}^{I}=(1-\alpha)\hat{u}_{n}(+1)+\alpha \hat{u}_{n+1}(-1),
\end{equation}
where, for $\alpha=0.5$ we recover a centered flux and for $\alpha=0$ an upwind scheme. It is well known that given a proper discretization of the linear advection equation it is possible to introduce concepts such as numerical dissipation and dispersion of a given scheme. The dispersion and dissipation curves for the SD scheme using a fully upwind numerical flux are respectively reported in figures \ref{fig:eigen_disp} and \ref{fig:eigen_diff} along with the same curves for a classical nodal DG for reference.
In this section we introduced the SD scheme for a general one-dimensional conservation law including both first- and second-order terms. We also briefly mentioned the more simplified setting of linear advection, where specific choices of numerical fluxes and polynomial order are directly linked to the classical numerical dispersion/dissipation eigenanalysis.

There exists a large body of literature discussing dispersion/diffusion analyses of DSEMs, ranging from classical temporal eigenanalysis applied to various formulations~\cite{vincent2011insights,moura2015linear,vermeire2016properties,chapelier2017optimal}, to spatial eigenanalysis~\cite{moura2019spatial,mengaldo2018spatial,mengaldo2018spatial2,tonicello2023fully} and non-modal analysis~\cite{fernandez2019non}, including both constant and non-constant advection velocities~\cite{manzanero2018dispersion,tonicello2021comparative}. All these approaches aim to investigate the role played by the numerical scheme in coarse-grained simulations such as Large-Eddy Simulations.

The specific choices within the numerical scheme, such as polynomial order and numerical fluxes (both inviscid and viscous), can have a significant impact on the numerical description of physical phenomena~\cite{DeWiart2015,moura2020eigenanalysis,kou2023jump}. A large part of the community working on DSEMs for turbulent flows often relies on so-called Implicit Large-Eddy Simulations~\cite{Grinstein_Margolin_Rider_2007,moura2017eddy,fernandez2017subgrid,fernandez2018ability,Manzanero20SVV}, where the numerical scheme is designed to act as an implicit SGS model, providing an appropriate level of numerical dissipation at small scales. Another part of the turbulence modeling community instead relies on numerical schemes that intrinsically preserve kinetic energy (and entropy, for compressible solvers), so that the only mechanism for kinetic energy dissipation is through explicit SGS modeling.

This brief digression, rather than advocating for a universally optimal approach, is meant to highlight the tight connection between the numerical scheme and SGS modeling. We therefore believe that, once the numerical scheme is chosen, any decision regarding turbulence modeling, whether based on implicit or explicit LES approaches, requires tailored optimization for that specific scheme.

We then propose the use of PDE-constrained optimization as a possible way to tailor SGS models to the specific numerical characteristics of the SD method. In a similar way, we expect that applying the same procedure to different schemes, or even to the same scheme with different discretization choices (such as numerical fluxes and polynomial orders), would lead to different optimal models. This concepts will be the focus of following sections and related discussions.

Within the framework of discontinuous spectral element methods, in addition to Fourier modes, the Legendre modes within each element are often considered for modeling purposes~\cite{chapelier2016spectral,lodato2019characteristic,GhiasiKomperda19,MohammadmahiKomperda,Mohammadmahdi23}. The solution within each element, in fact, can be represented as an expansion of orthogonal polynomials in the reference element $[-1,1]$ as:
\begin{equation}
    u(x) = \sum_{i=1}^{N} \beta_i P_i(x), \label{eqn:umodal}
\end{equation}
where $ \{P_i(x)\}_{i=1,...,N}$ is the $i$-th Legendre polynomial and $\beta_{i}$ its corresponding modal coefficient. Being the Legendre polynomials ordered here for increasing polynomial degree, it is often useful, in DSEMs, to investigate how the modal coefficients decay for increasing orders of accuracy in order to assess the effective resolution of the numerical method.

In the next section, the Legendre modal coefficients will be used in the definition of the objective functional used for the discrete adjoint method.
\section{PDE-constrained formulation: the discrete-adjoint method}\label{sec:theadjoint}
As discussed in the introduction, the accuracy of the sensitivities computed using the continuous adjoint approach may depend on the discretization, particularly in applications where the numerical discretization significantly influences the prediction of the continuous system.

Having in mind the specific application of LES, we expect the numerical footprint of the scheme to be relevant on the overall prediction of the problem of interest. In this work, we consequently employ the discrete adjoint approach in order to be consistent with the discrete dynamical system. The numerical solver is differentiated by source-code transformation using the \texttt{Tapenade} tool~\cite{Tapenade}. To assemble the necessary Jacobian matrices for the adjoint formulation in parallel, the method employs ghost cells. The implementation of the discrete adjoint method relies on an interface to the adjoint module library implemented in \texttt{PETSc} \cite{Petsc,TSadjoint}. 

The discrete adjoint method is used to compute sensitivities of an objective function with respect to parameters in systems governed by PDEs in a computationally efficient way.
Let the discrete form of our governing equation be represented as:
\begin{equation}
    R(\mathbf{u};\theta) = 0, \label{eq:DA_residual}
\end{equation}
where $\mathbf u \in \mathbb{R}^{\NDOF\times N_{t}}$ is the vector of discrete state variables with $N_{t}$ the total number of time steps, and $\theta \in \mathbb{R}^{N_p}$ represents design or control parameters. Notice that for time-dependent problems, such as the ones herein considered, the $\mathbf{u}$ variable has to be interpreted as the full spatio-temporal solution of the discrete system of equations.

The objective functional can be written as an integral computed along the trajectory:
\begin{equation}
    \mathcal{J}(\mathbf \Psi;\mathbf \Psi^{\mathrm {ref}}) = \int_{t_0}^{t_f} r(\mathbf{\Psi}(t,\theta),\mathbf \Psi^{\mathrm {ref}}(t);\theta) dt \label{eq:DA_residual}\, ,
\end{equation}
where with $\mathbf{\Psi}$ we denote a quantity of interest computed with the flow-state variables and $\mathbf{\Psi}^{\mathrm{ref}}$ is the reference quantity we would like to match.

A change in the design parameters results in:
\begin{equation}
    {d\mathcal J} = \frac{\partial \mathcal J}{\partial \QOI} \pd{\QOI}{\mathbf u}{d\mathbf u}+\pd{\mathcal J}{\theta}{d\theta} \label{eq:dJdtheta}. 
\end{equation}
Similarly, the variation of the residual is zero: 
\begin{equation}
    dR=\pd{R}{\mathbf u}{d\mathbf u}+\pd{R}{\theta}{d\theta}=0\, .\label{eq:dRdtheta}
\end{equation}
Since the variation of the residual is zero, we can enforce the PDE-constraint in the loss function by multiplying equation \eqref{eq:dRdtheta} by a Lagrangian multiplier $\lambda$ and subtract it to equation \eqref{eq:dJdtheta}:
\begin{equation}
  \begin{aligned}
    {d\mathcal J} &= \frac{\partial \mathcal J}{\partial \QOI} \pd{\QOI}{\mathbf u}{d\mathbf u}+\pd{\mathcal J}{\theta}{d\theta} - \lambda^{T}\left( \pd{R}{\mathbf u}{d\mathbf u}+\pd{R}{\theta}{d\theta} \right) \\
    &=\left( \frac{\partial \mathcal J}{\partial \QOI} \pd{\QOI}{\mathbf u} -\lambda^T \pd{R}{\mathbf u} \right){d\mathbf u}+\left( \pd{\mathcal J}{\theta}{d\theta} -\lambda^{T} \pd{R}{\theta} \right) {d\theta} = 0.
    \label{eq:dJ0}
\end{aligned}
\end{equation}
Choosing $\lambda$ to satisfy the adjoint equation:
\begin{equation}
    \left( \pd{R}{\mathbf u}\right)^{T}\lambda = \frac{\partial \mathcal J}{\partial \QOI} \pd{\QOI}{\mathbf u}, \label{eq:lambdaAdjoint}
\end{equation}
the first term in equation \eqref{eq:dJ0} is eliminated. Consequently, we can compute the derivative of the objective functional with respect to the parameters:
\begin{equation}
    \frac{ d\mathcal J}{d\theta} =\pd{\mathcal J}{\theta} -\lambda^{T}\pd{R}{\theta}. \label{eq:LossGrad}
\end{equation}
Note that equation \eqref{eq:LossGrad} does not depend on the flow-state variables. Consequently, we can compute the gradient of the objective functional with respect to the parameters without further flow-field evaluations. This makes the adjoint-based approach extremely efficient for the computation of the sensitivities.

\begin{figure}[h!]
\centering
\includegraphics[width=.99\textwidth]{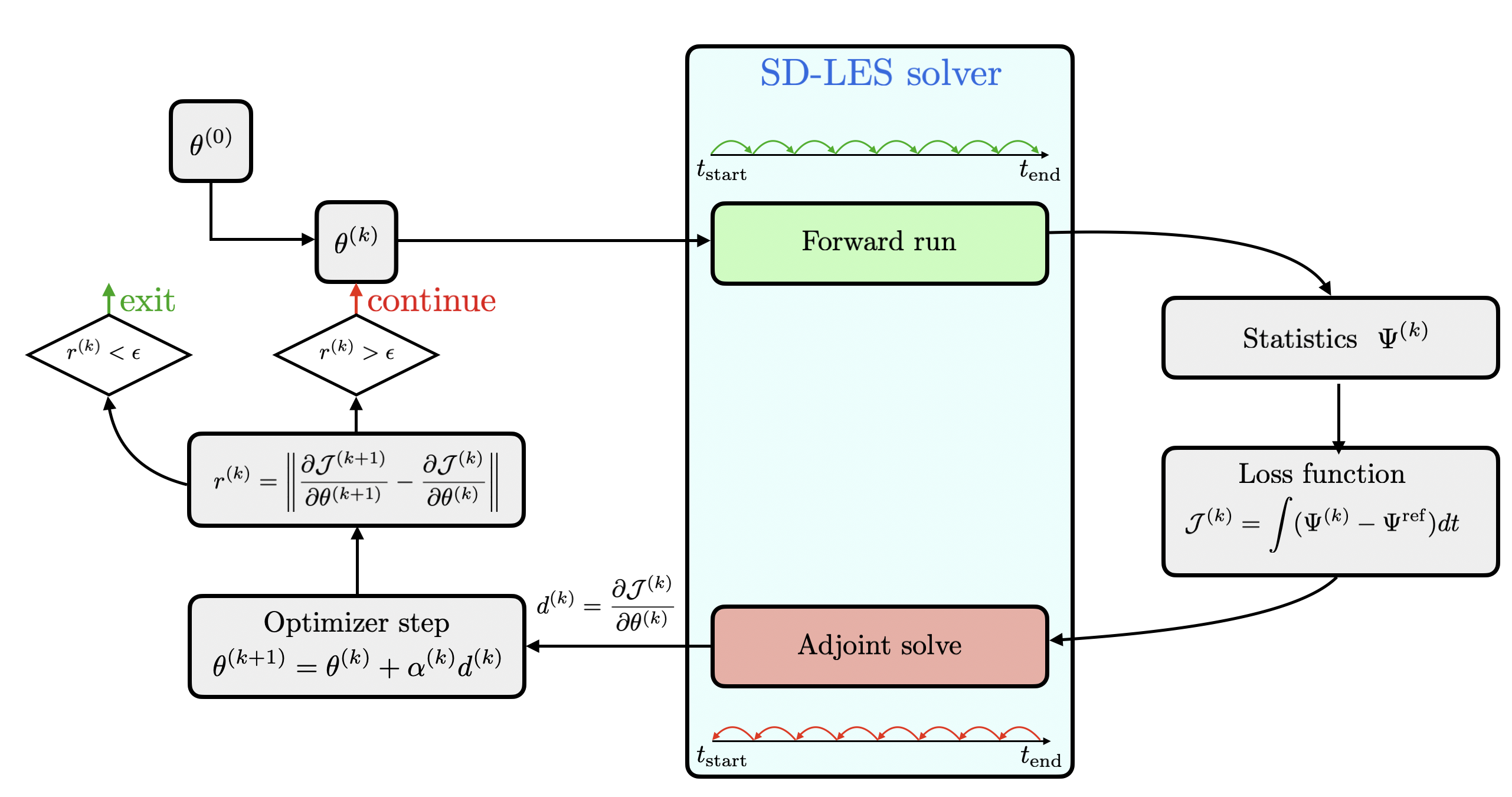}
\caption{A schematic diagram of the optimization algorithm.}
\label{fig:Opt-general-framework0}
\end{figure}

The proposed methodology combines the full-order model solver with a discrete adjoint framework to efficiently propagate the derivatives through the solver. The derivatives are used for optimizing end-to-end turbulence model coefficients of SGS models (\eg, Smagorinsky model). The overall workflow involves:
\begin{enumerate}
\item Starting with an initial estimate of the control parameters.
\item Solving a forward run of a given trajectory.
\item Computing the statistics needed for the definition of the loss function.
\item Computing adjoint sensitivities with respect to model parameters.
\item Iteratively updating the parameters with a gradient-descent method (\eg, Adam optimizer~\cite{Adam14}).
\end{enumerate}
A schematic overview of the methodology is shown in figure \ref{fig:Opt-general-framework0}.
When applying the adjoint-based approach to strongly non-linear and chaotic systems many difficulties arise. Chaotic systems are extremely sensitive to initial conditions. Moreover, the adjoint approach requires a backward in time integration. As a result, adjoint variables can lead to numerical instabilities and meaningless sensitivities. In chaotic systems, small changes in the control parameters of the model may lead to large variations of the trajectories during the forward run. As a result, the practical optimization horizon requires proper temporal windowing and a proper choice of the loss function to avoid spurious effects in the computation of the sensitivities.

In the context of turbulence modeling, the choice of the loss function is of paramount importance. As discussed in~\cite{BEWLEY_MOIN_TEMAM_2001}, considering longer trajectories leads, in general, to better solutions (\ie, solutions which provide more accurate target statistics). However, as the time horizon increases, the possibility of introducing spurious effects due to the chaotic nature of the system increases. As a result, a tradeoff between these limit cases needs to be considered. At the same time, constructing a loss function based only on quantities sampled at the final time is not effective because of the chaotic nature of the solution.

Identifying the mechanisms underlying a given effect is essential for designing an effective loss function. As shown in~\cite{BEWLEY_MOIN_TEMAM_2001}, minimizing a quantity indirectly related to the target objective may yield better results than optimizing the target itself. Therefore, the chosen loss function should capture both the relevant flow physics and numerical features of the problem.
In the present work, we build a loss function which takes into account both numerics and flow physics. 

Due to the crucial relevance of Legendre modes for DSEMs~\cite{chapelier2016spectral,GhiasiKomperda19,MohammadmahiKomperda,Mohammadmahdi23}, we consider an objective functional which relies on the energy associated to Legendre modes. Since modal energies may span several orders of magnitude, the loss function is formulated in logarithmic space in order to balance the contribution of all modes.
For simplicity, we present the loss function definition in the one-dimensional case for a single scalar variable.
Consequently, we consider the following loss function:
\begin{equation}
\mathcal{J} = \sum_{i}\left(\log(\langle \widehat{E}_i \rangle)-\log(\langle \widehat{E}^{\mathrm{ref}}_i \rangle)\right)^2 ,\label{eq:ObjModes}
\end{equation}
where $\langle \widehat{E}_i \rangle$ is computed as:
\begin{equation}
    \langle \widehat{E}_i \rangle = \frac{1}{\TWINDOW \,L}\int_{t_{0}}^{t_{f}} \left(\sum_{e}^{N_e}\beta^2_{e,i}\right)\,dt\, , \label{eq:AvgLegendre}
\end{equation}
where $\TWINDOW=t_{f}-t_{0}$ is the trajectory length, $L$ is the dimension of the domain and $\beta^2_{e,i}$ is the modal energy of the $i$-th mode in the $e$-th element. Furthermore, $\langle\widehat{E}^{\mathrm{ref}}_i \rangle$ denotes the reference modal decay, obtained from time-averaged filtered data across all the elements.
In the Navier–Stokes case, the modal energy is computed by averaging the modal energies associated to the velocities in each element. We refer to the work by Chapelier \& Lodato~\cite{CHAPELIER2016279} for further details.
This quantity is representative of the numerical footprint for the DSEM methods \cite{CHAPELIER2016279,NaddeiPlata21}.

\subsection{Implementation details}

The discrete adjoint method \eqref{eq:lambdaAdjoint} requires the product of the adjoint variable with the Jacobian of the right-hand side of the discretized system.
To assemble this product in parallel, we employ ghost cells to manage the communication of the degrees of freedom (DoFs) needed for local assembly. Each processor is responsible for computing its own local contribution.

\begin{figure}[h!]
 \centering
 \begin{subfigure}[b]{0.38\textwidth}
     \includegraphics[width=\textwidth, trim=3cm 6.5cm 3cm 5cm, clip]{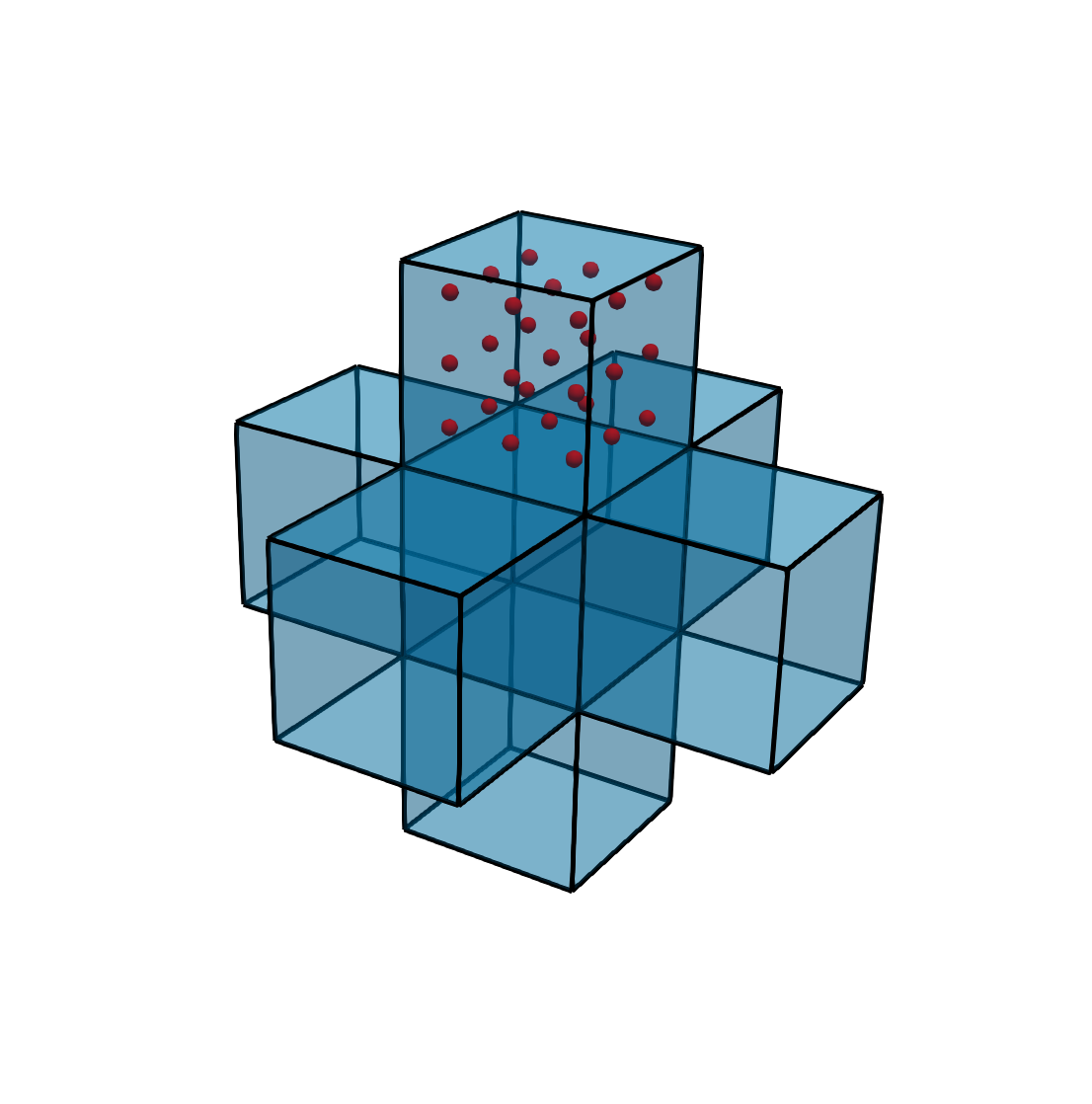}
     \caption{Inviscid operator.}
     \label{fig:ComputationalStencil:a}
 \end{subfigure}
 \begin{subfigure}[b]{0.55\textwidth}
     \includegraphics[width=\textwidth, trim=3cm 6.5cm 3cm 5cm, clip]{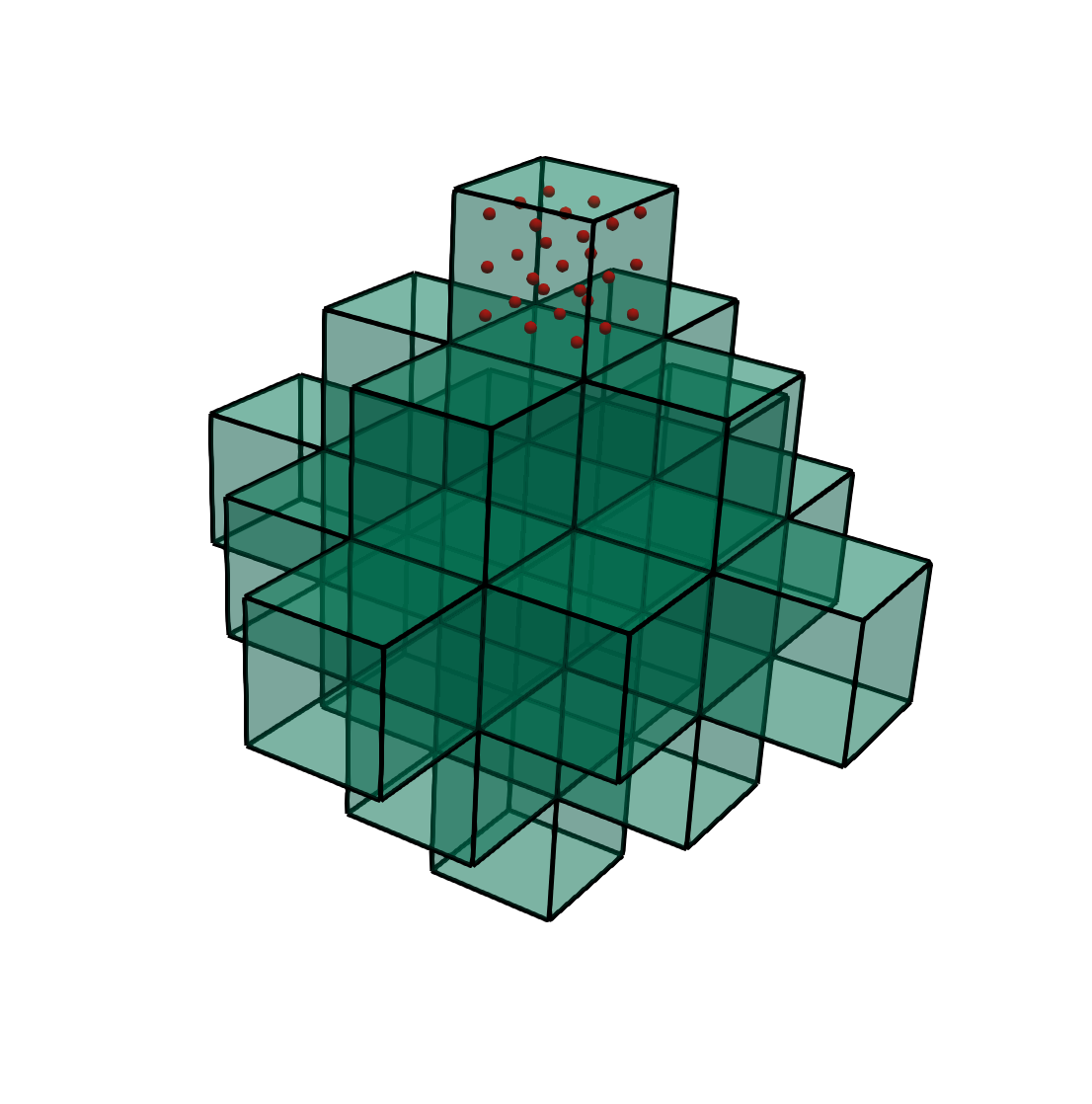}
     \caption{Viscous operator.}
     \label{fig:ComputationalStencil:b}
 \end{subfigure}
\caption{A schematic drawing of the computational stencil for the spectral difference scheme. Stencil for the inviscid (left) and viscous (right) operators.}
\label{fig:ComputationalStencil}
\end{figure}

It is important to note that the computational stencil differs between the viscous and inviscid operators~\cite{marchal_phd_thesis,MarchalEXTENSION}. For a purely inviscid operator, each spectral element depends only on elements located at most one cell away, resulting in a stencil of total $N_{\mathrm{dof}}=(\mathrm{p}+1)\times7$ degrees of freedom, where $\mathrm{p}$ is the polynomial order. For the viscous operator, instead, the dependency extends up to two cells away, resulting in a wider stencil involving $N_\mathrm{dof}=(\mathrm{p}+1)\times 25$ total degrees of freedom. A schematic drawing of the computational stencil for both viscous and inviscid operators is given in figure \ref{fig:ComputationalStencil}. 

The wider stencil arises from the coupling introduced by the gradient computation. In fact, as explained in Section \ref{sec:sd}, the computation of second-order terms requires an additional interpolation step and an extra correction at the interfaces between elements with respect to first-order derivatives. As a result, each DoF within a given element is coupled with all DoFs belonging to the elements included in the stencil.

Since the right-hand side operator includes both viscous and inviscid contributions, two layers of ghost cells per processor are required. This ensures that the vector–Jacobian product can be assembled locally.

In the case of the one-dimensional Burgers’ equation, the vector–Jacobian product is constructed by explicitly assembling the Jacobian matrix, as this step is not particularly computationally demanding in the one-dimensional case and allows for straightforward and efficient evaluation of the linearized operator. In contrast, for the three-dimensional Navier–Stokes equations, a matrix-free approach is adopted due to the significantly higher computational cost and memory requirements associated with assembling and storing the full Jacobian.

The vector–Jacobian product, which has to be resolved for the adjoint variables, is computed by directly using automatic differentiation in reverse (adjoint) mode. This is achieved using the tool \texttt{Tapenade}~\cite{Tapenade}. Such an approach has the double advantage of saving computational time while also avoiding the storage of the Jacobian matrix. 

However, the framework requires careful planning and implementation, since the adjoint mode automatic differentiation requires to reverse the entire code of the right hand side operator.

\begin{figure}[h!]
\centering
\includegraphics[width=.80\textwidth]{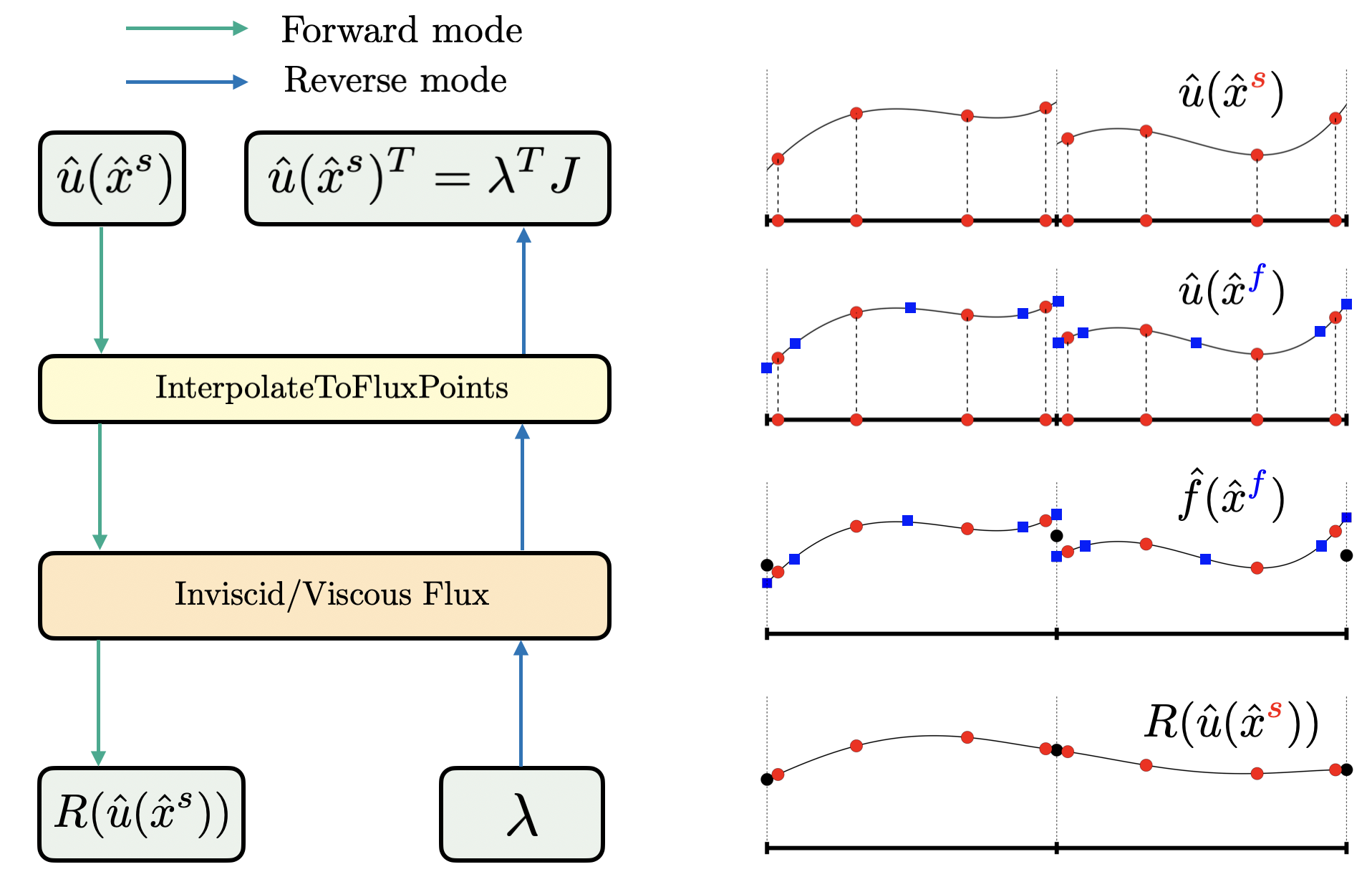}
\caption{Assembly of the right hand side operator and reverse-mode mode computation of the vector-jacobian product. For the computation of the right hand side, following the green arrows: the state vector is interpolated to flux points, where inviscid and viscous fluxes are evaluated and combined to form the residual. Each of these steps is schematically shown on the right side of the scheme. Following the blue arrows, instead, we show the reverse-mode mode computation of the vector-jacobian product: the adjoint variable $\lambda$ is propagated backward trough the same sequence of operations.} 
\label{fig:Opt-general-framework}
\end{figure}

For clarity, we report here a schematic view of the evaluation of the right-hand side operator for the Navier–Stokes equations in figure \ref{fig:Opt-general-framework}. Starting from the vector of the conservative variables stored at the solution points $\hat{u}(\hat{x}^{s})$, the variables are interpolated to the flux points through the ${\mathrm{\mathbf {InterpolateToFluxPoints}}}$ subroutine. The fluxes are then computed, distinguishing between inviscid and viscous contributions. While the inviscid flux depends directly on the interpolated state, the viscous flux additionally involves gradient computations, leading to a wider computational stencil. Finally, both contributions are combined to assemble the residual $R(\hat{u}(\hat{x}^{s}))$.
In contrast, the vector-jacobian product $\lambda^T J$ is computed by propagating the adjoint variable $\lambda$ backward through the same reversed sequence of transposed operations. Each single subroutine is differentiated by \texttt{Tapenade}~\cite{Tapenade} in backward mode.
After computing the local vector-jacobian product from the differentiated subroutines, a global gathering operation is performed for the assembly of the global vector-jacobian product. The ghost cells allow us to evaluate the vector-jacobian product in parallel in each processor. 
More specifically, DoFs of the ``neighbors'' processors are gathered into ghost cells. This operation ensures that all stencil-dependent contributions can be computed locally within each processor. The scattering and gathering are performed using the high performance computing library \texttt{PETSc}~\cite{Petsc}.

%
\section{Forced Burgers turbulence}\label{sec:forcedhit}
In order to assess the methodology of section \ref{sec:theadjoint} for the optimization of SGS models, the forced Burgers turbulence ~\cite{SOLANFUSTERO2021110246,MOURA2015695} test-case is considered here. We start by considering different Implicit Large-Eddy Simulations of the Forced Burgers test case for different resolutions in order to assess the impact of the numerical scheme on the overall solution. Details regarding the space discretization (resolution and polynomial order) are reported in table \ref{tab:fhit-iles}. 

\begin{figure}[h!]
 \centering  
 \begin{subfigure}{0.48\textwidth}
     \includegraphics[width=\textwidth]{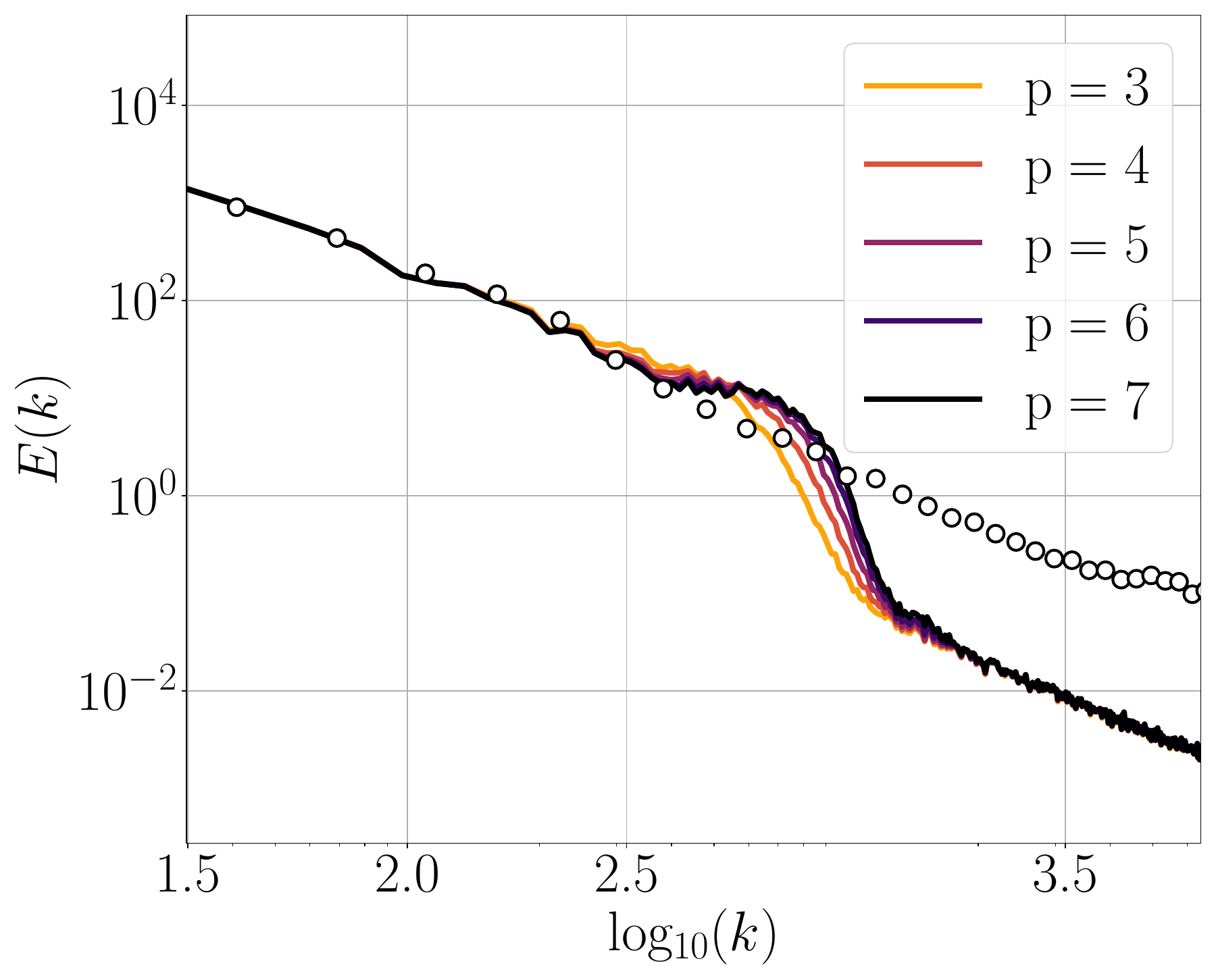}
     \caption{$\NDOF=1024$.}
     \label{fig:FHIT-ILES:a}
 \end{subfigure}
 \begin{subfigure}{0.48\textwidth}
     \includegraphics[width=\textwidth]{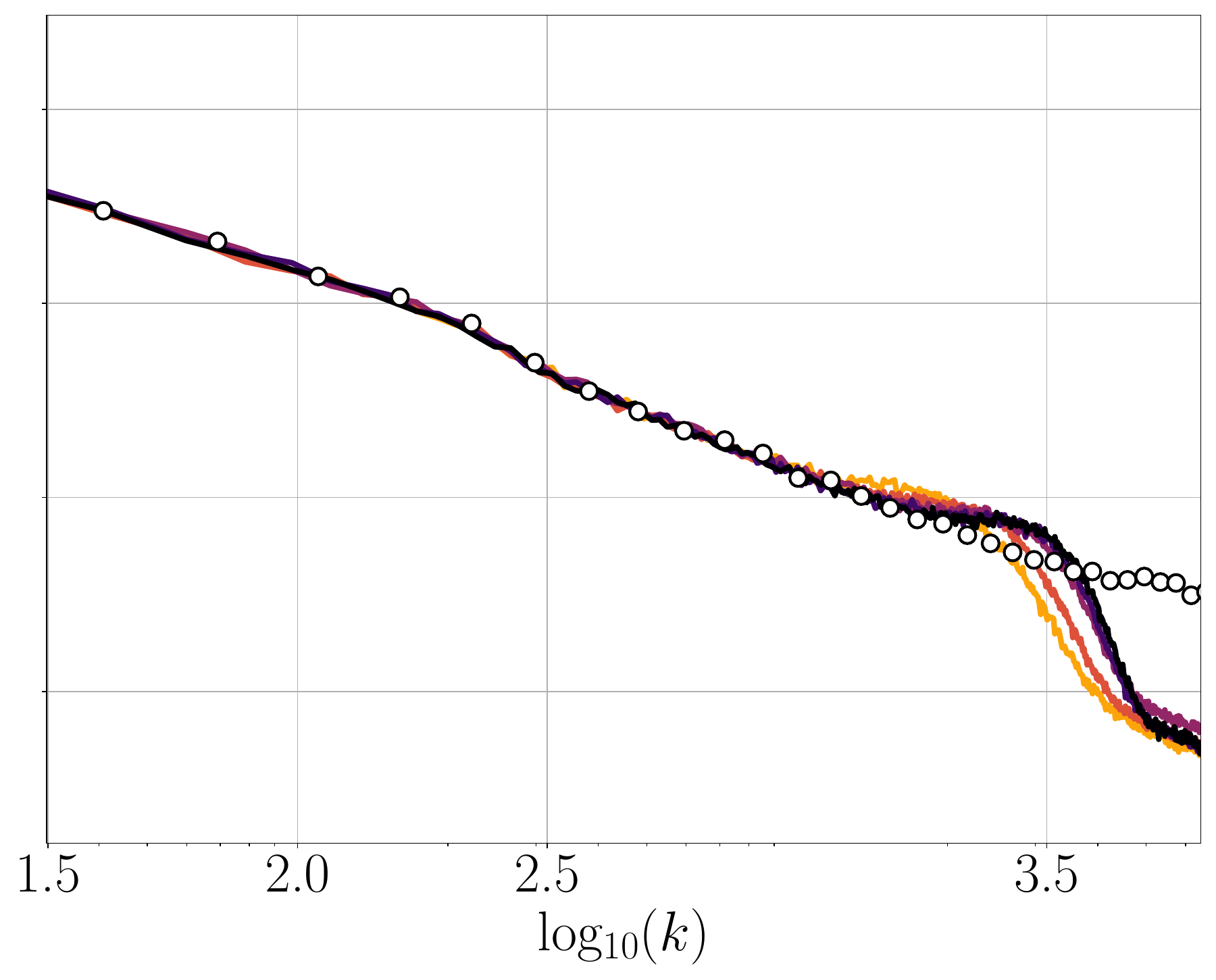}
     \caption{$\NDOF=4096$.}
     \label{fig:FHIT-ILES:c}
 \end{subfigure}
\caption{Kinetic energy spectra for various resolutions and for different polynomial orders considered in the study. Symbols indicate the reference.}
   \label{fig:FHIT-ILES}
\end{figure}

Given the total number of degrees of freedom and polynomial order of the SD scheme, the number of elements can be evaluated by $\NDOF/(\mathrm{p}+1)$. The domain is a one-dimensional segment $\Omega=[-1,1]$ with periodic boundary conditions. The initial solution is set to $u_0=1$. Time integration is carried out using the three-stages third order Runge-Kutta SSP scheme.
\begin{table}[!ht]
\begin{center}
\large
\begin{tabular}{|l | c c c c |} 
 \hline
  $\NDOF$ & $\mathrm{p}=4$ &$\mathrm{p}=5$ & $\mathrm{p}=6$ & $\mathrm{p}=7$ \\ 
 \hline
 1024 & a4 & a5 & a6  & a7\\ 
 \hline
 2048 & b4  & b5  & b6 & b7\\
 \hline
 4096 & c4 & c5  & c6 & c7\\
 \hline 
 8192 $(\mathrm{Ref})$ & & & & d4 \\
 \hline
\end{tabular}
\end{center}
\vspace{-0.4cm}
\caption{Test cases considered in this study.}
\label{tab:fhit-iles}
\end{table}
Figure \ref{fig:FHIT-ILES} shows the time-averaged kinetic energy spectra for the test cases considered. The spectra are obtained from successive applications of Taylor's hypothesis over the interval $300\leq t \leq 450$.
Each spectrum was evaluated by probing the solution at $x = 0$ over time windows of $\delta t  = L/u_0 = 2$. For all the cases considered here, the value $N_c = 80$ was employed in equation~\eqref{eq:burg_filt}. As a result, the kinetic energy spectra shown in figure~\ref{fig:FHIT-ILES} exhibit a $-5/3$ slope up to $\log_{10}(k_c) = \log_{10}(\pi N_c) \approx 2.4$, as expected from the form of the applied forcing. Beyond this range, the characteristic $-2$ slope of Burgers turbulence is recovered. All resolutions considered here have been selected so that the effect of numerical dissipation lies within this second range, while energy is injected into the system at large scales through the forcing term. By inspecting the ILES profiles in figure \ref{fig:FHIT-ILES}, similarly to what is commonly observed in the kinetic energy spectra of under-resolved three-dimensional turbulence, we can observe an accumulation of kinetic energy occurs near the cutoff. This effect suggests that, although simplified, Burgers turbulence can provide useful insights into kinetic energy cascade and, in particular, in the role played by the numerical scheme in its dissipation at small scales.

\begin{figure}[h!]
 \centering  
 \begin{subfigure}{0.48\textwidth}
     \includegraphics[width=\textwidth]{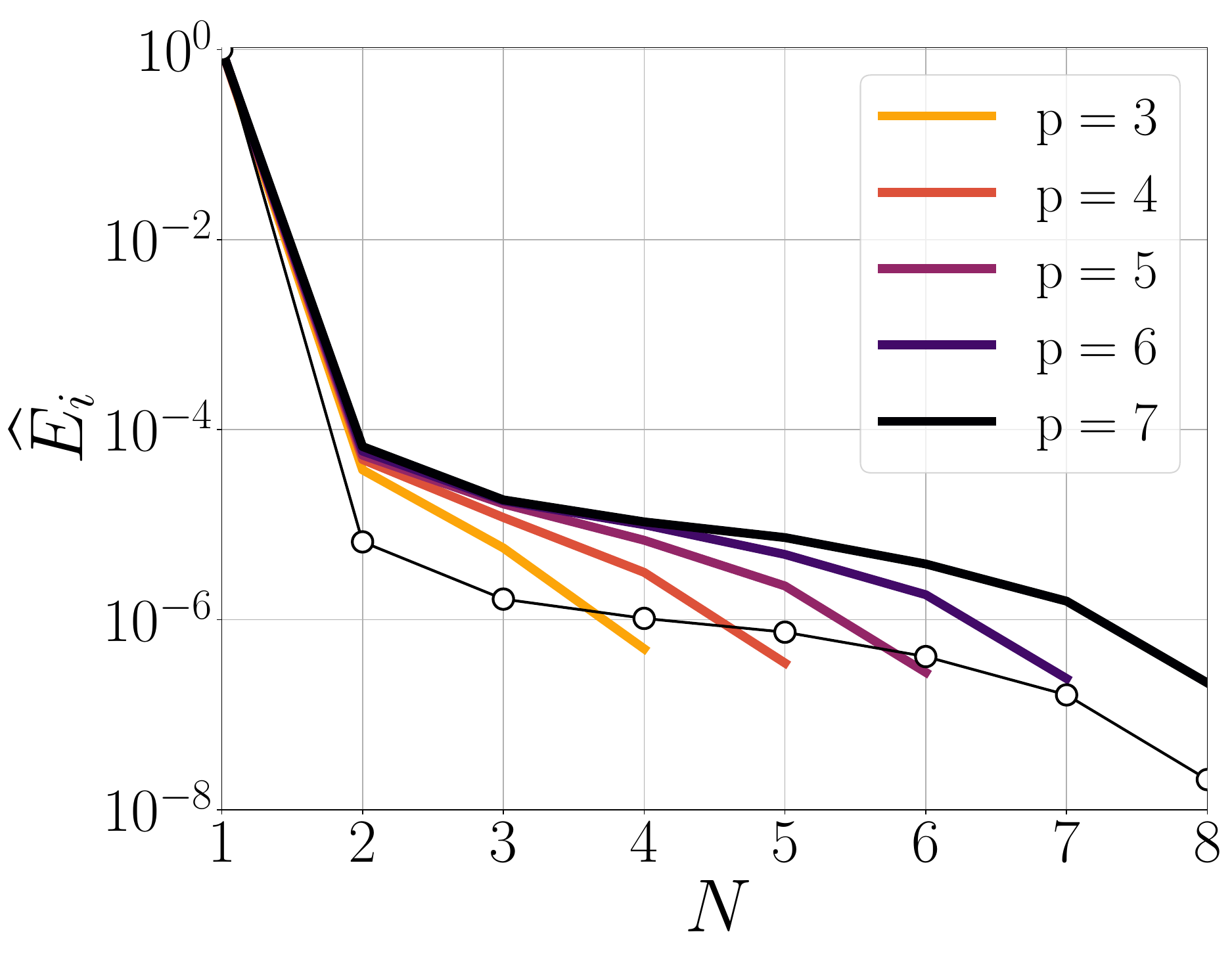}
     \caption{$\NDOF=1024$.}
     \label{fig:FHIT-ILES-1:a}
 \end{subfigure}
 \hspace{-0.2cm}\begin{subfigure}{0.48\textwidth}
     \includegraphics[width=\textwidth]{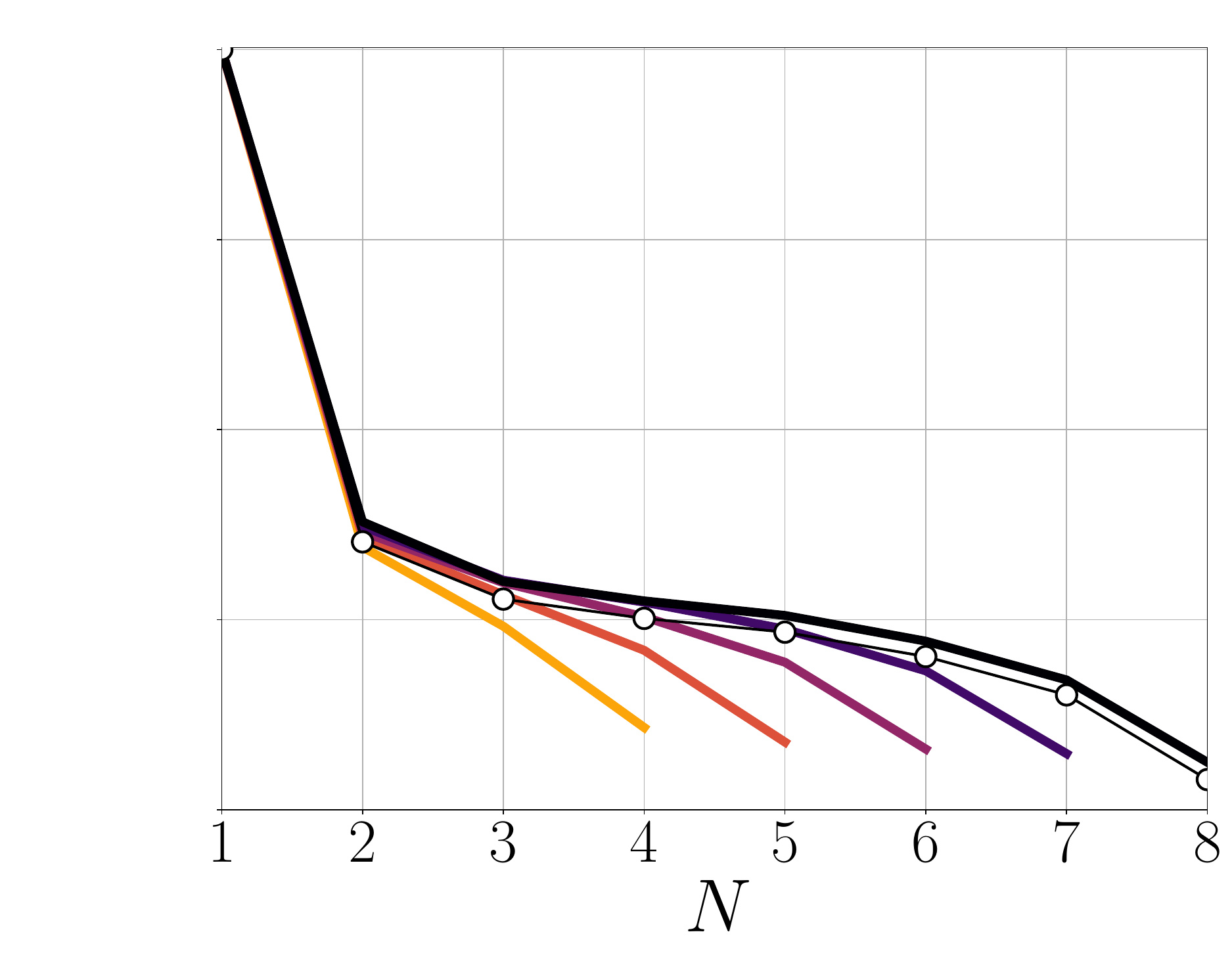}
     \caption{$\NDOF=4096$.}
     \label{fig:FHIT-ILES-1:c}
 \end{subfigure}
\caption{Time- and space-average of the modal energy associated by each mode for different resolutions and different polynomial orders. Line with symbols indicate the reference.}
   \label{fig:FHIT-ILES-1}
\end{figure}

Figure \ref{fig:FHIT-ILES-1} shows the modal energy \eqref{eq:AvgLegendre} for different polynomial orders and different resolutions in the Burgers case. 
In this framework, Legendre modes can be interpreted similarly to Fourier modes: low-order modes are associated with large-scale structures, whereas higher-order modes represent progressively finer scales. Note that lower resolutions present a less pronounced modal decay with respect to higher resolutions. This aspect was also found by Chapelier \& Lodato~\cite{CHAPELIER2016279} for the Navier-Stokes turbulence.

\begin{figure}[h!]
\centering
\includegraphics[width=.8\textwidth]{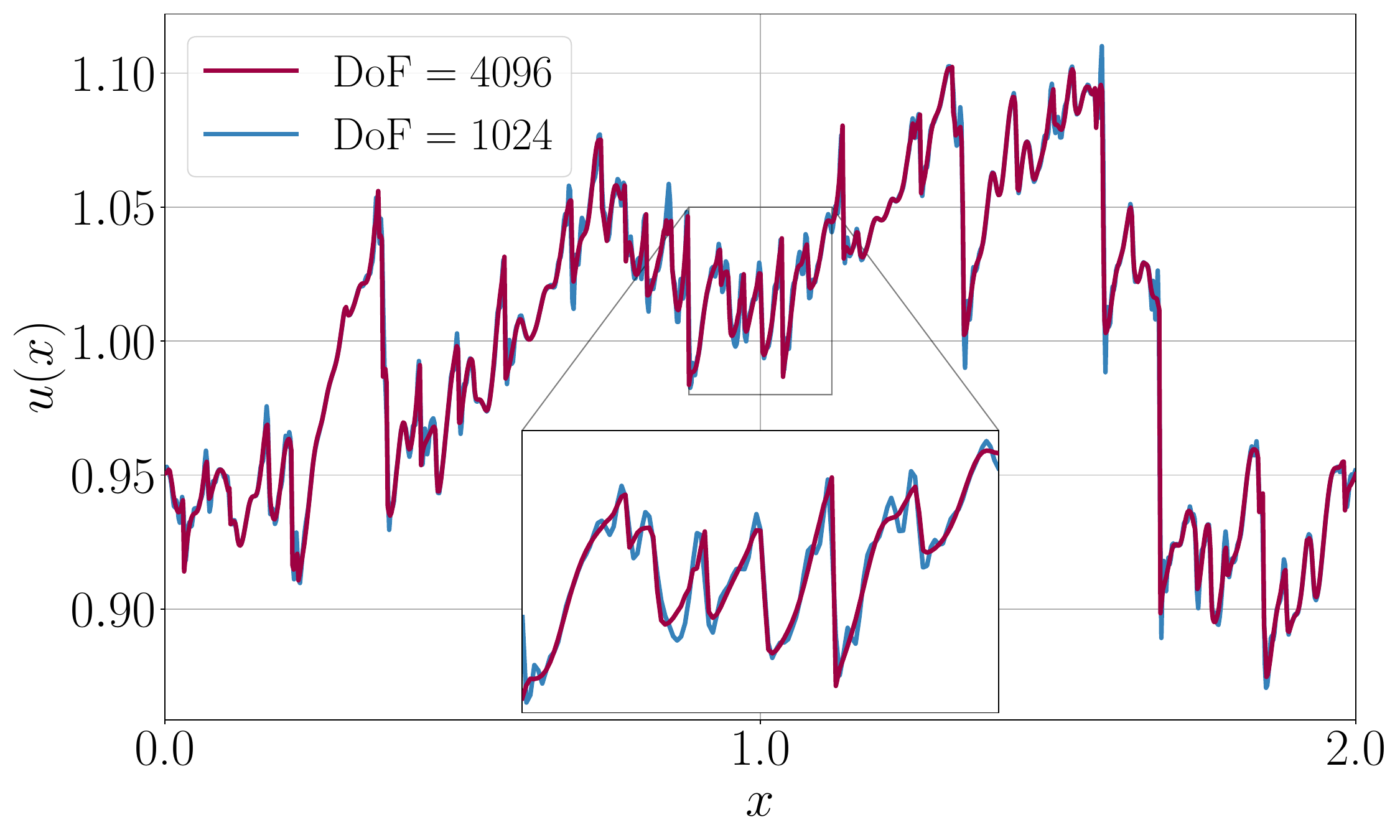}
\caption{Comparison between solutions at $t=400$s for $\mathrm{p}=7$ for the coarsest ($\NDOF=1024$) and the finest resolution ($\NDOF=4096$). The solution with $\NDOF=1024$ contains many wiggles as a consequence of under-resolution.}
\label{fig:SnapshotILESs}
\end{figure}

In order to assess how the number of DoFs influences the resolution capability of the scheme, figure \ref{fig:SnapshotILESs} compares the solution snapshots at $t=400$ for $\mathrm{p}=7$ using the coarsest and finest resolutions ($1024$ and $4096$ DoFs). From the zoomed view, we note that the low resolution simulation presents more spikes. This behavior is expected for the DSEM, which relies on high-order polynomial representations within each element. When the resolution is low, the polynomial basis cannot accurately capture all relevant scales. As a result, unresolved high-frequency content contaminates the numerical solution, manifesting itself as spurious oscillations or wiggles. Moreover, notice that although an upwind numerical flux is used for the inviscid flux computation, it is insufficient to suppress these oscillations when the polynomial order is significantly high.

In the previous paragraphs, we discussed the role played by numerics in this particular setting. As representative parameter associated to the discretization, we considered the polynomial order, noticing its significant impact on the overall accuracy of the solution. It is then clear that any closure model for such problem should vary with the polynomial order and consequently, in general, adapt for different hyper-parameters of the discretization. In developing such model it is then important both the model formulation but also the particular loss function we want to optimize the control parameter on.

\subsection{Optimization of the Smagorinsky constant}\label{subs:OptSmago}
In order to assess the framework, we start by optimizing the Smagorisnky constant for each test-case. Previous studies carried on the decaying Burgers test-case~\cite{MAULIK201812} revealed that the best value should be around $\CSMAGO\simeq0.5$.

In particular, referring to the nomenclature introduced in section \ref{sec:problemdef} we will consider the simple closure:
\begin{equation}
    \tau^{\mathrm{SGS}} = - \nu_{t} \frac{\partial \widetilde{u}}{\partial x} \quad \mathrm{with} \quad \nu_{t} = (C_{s} \Delta)^{2} \bigg \rvert \frac{\partial \widetilde{u}}{\partial x}\bigg \lvert
\end{equation}
Theoretically, the constant should be universal (\ie, equal at each resolution and for each polynomial). However, this assumption is based only on physical reasoning, without considering the impact of the numerical scheme. In the context of LES, the numerical scheme interacts with the SGS model. Consequently, the optimum value of the constant is expected to change for different hyper-parameters of the numerical scheme. Finally, optimizing the Smagorinsky constant can also be seen as optimizing the filter width $\Delta$ which can have multiple definitions in DSEMs.
\subsubsection{Training}\label{subs:OptSmago_training}
In order to have a sufficiently large number of snapshots for accumulating statistics, we set the time window to $\Delta T=4s$. Time integration is carried out using a three-stage Runge-Kutta time scheme with $\delta t=10^{-4}$. 
The reference modal energy was derived from high-fidelity data generated by a fine simulation with $N_e$=1024 elements and a polynomial order $\mathrm{p}$=7, corresponding to $8192$ DoFs.
It is important to highlight that we are considering a chaotic system. Consequently, it is expected that different time horizons, could, in principle, lead to different outcomes. Previous studies have examined the influence of the optimization horizon in the context of the incompressible Navier–Stokes equations~\cite{ListChenThuerey22}, where it was generally observed that longer trajectories tend to improve optimization outcomes. 

However, the approach presented in~\cite{ListChenThuerey22} is characterized by significant differences in terms of numerical scheme, optimization algorithm, and physical system. We consequently do not necessarily expect the same type of results. In this work, in order to ensure statistical reliability, we decided to gather data over a sufficiently long trajectory. Consequently, as we fix the time-step to $\delta t=10^{-4}$, we accumulate and store thousands of snapshots during the forward simulation. This is clearly feasible within the one-dimensional Burgers equation case whereas it requires several improvements in order to deal with three-dimensional Navier-Stokes turbulence. The challenges regarding such an extension will be discussed later on.

In this work, we set the initial value of the constant to $\CSMAGO=1$. Since we optimize only a single parameter, we employ a quasi-Newton–type optimizer. This method approximates the action of the inverse Hessian using a limited-memory quasi-Newton scheme~\cite{MartinsNing2021, NocedalWright2006}.

Table \ref{tab:smago-opt-results} shows the optimized Smagorinsky constants obtained at the end of the optimization for all the cases considered in this study. We can clearly notice that higher polynomial orders tend to have larger coefficients. This is in agreement with the results of the standard linear analysis, in which high order polynomials lead to better conservation of energy~\cite{moura2015linear,mengaldo2018spatial,mengaldo2018spatial2}.
\begin{table}[!ht]
\begin{center}
\large 
\begin{tabular}{|c | c c c c |} 
 \hline
 $\NDOF$ & $\mathrm{p}=7$ & $\mathrm{p}=6$ & $\mathrm{p}=5$ & $\mathrm{p}=4$  \\
 \hline
 1024 & 0.641  & 0.563 & 0.510 & 0.453  \\
 \hline
 2048 & 0.600 & 0.568 & 0.525 & 0.469  \\
 \hline
 4096 & 0.550  & 0.546 & 0.497 & 0.446 \\ 
 \hline
\end{tabular}
\end{center}
\vspace{-0.4cm}
\caption{Optimized Smagorinsky constants for different resolutions and different polynomial orders for $\Delta T=4$.}
\label{tab:smago-opt-results}
\end{table}
By considering $\mathrm{p=7}$, we observe that for a fixed time window, the optimal value of the constant decreases as the resolution increases. This behavior is expected: in fact, the under-resolution decreases as the the number of DoF increases. Consequently, a smaller coefficient is enough for the stability of the solution. This observation is also consistent with the qualitative analysis on the kinetic energy spectrum presented in Figure \ref{fig:FHIT-ILES}, where low-resolution simulations were shown to exhibit stronger spurious oscillations than well-resolved ones. Consequently, a larger stabilization constant is required at coarse resolutions to damp oscillations and ensure a stable solution.

A similar behavior is also found for lower polynomial orders. However, as the polynomial order decreases the optimization becomes more difficult. At low polynomial orders, the loss function includes only a limited number of modes, and the energy associated with the higher modes is significantly smaller compared to the dominant, lower-order ones. Consequently, achieving an exact match for the modal energy of the few higher-order modes becomes difficult.

\begin{figure}[h!]
\centering
\includegraphics[width=.99\textwidth]{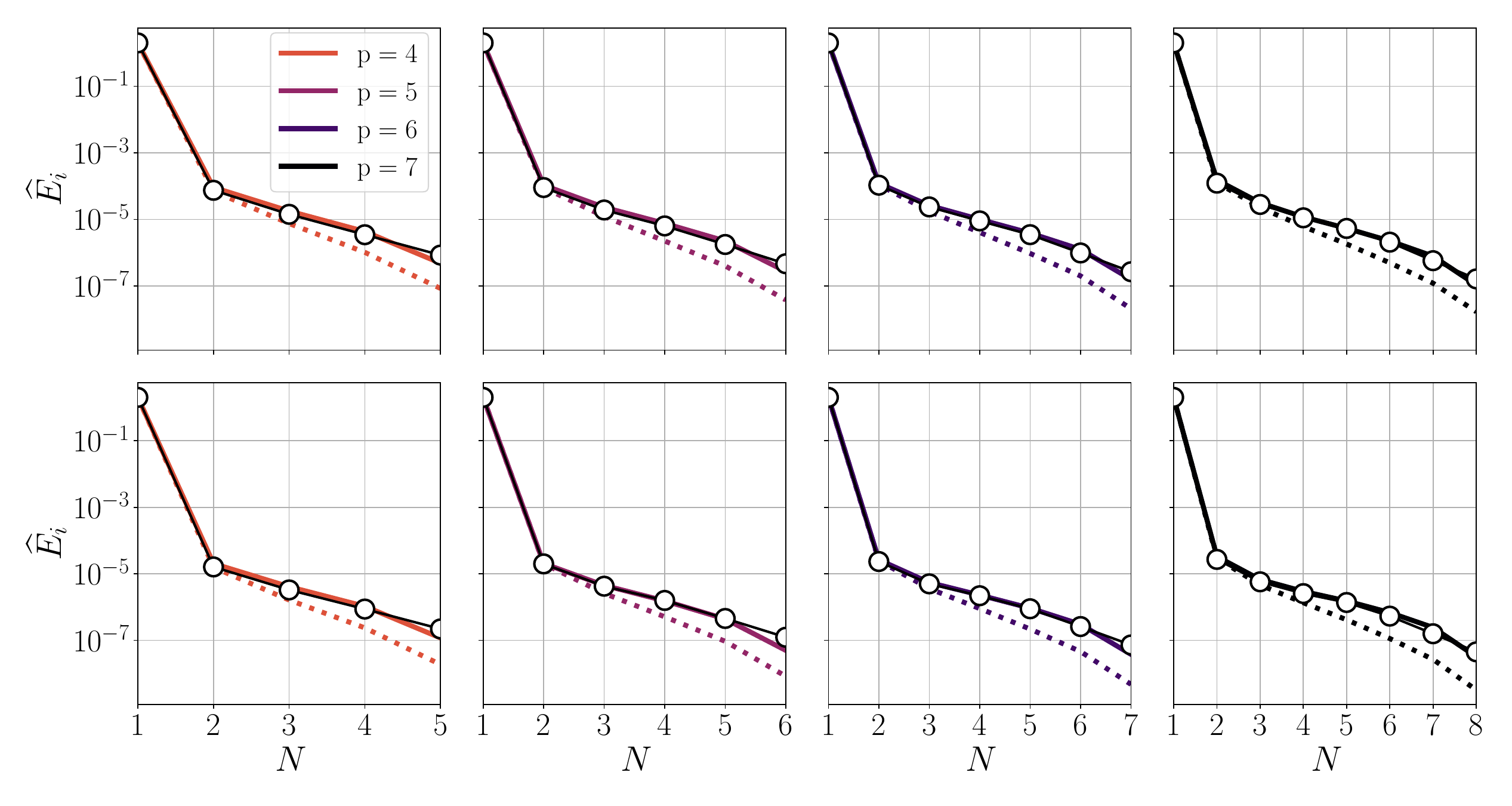}
\caption{Comparison of the modal decay obtained at the beginning of the optimization and the optimized one for the various polynomial orders. Top row, $N_{\mathrm{dof}}=1024$; Bottom row, $N_{\mathrm{dof}}=4096$.  Solid lines, optimized solution; dotted lines, baseline model ($C_s=1.0$). Lines with symbols indicate the reference.}
\label{fig:OptimizationModesSmago}
\end{figure}

In order to investigate the behavior of the Legendre modes for the different optimized models, figure \ref{fig:OptimizationModesSmago} shows the modal energy at the end of the optimization. These results are compared with the energy obtained using $C_s = 1$ and with the reference modal energy derived from the reference data (\ie, $\NDOF=8192$), considering all polynomial orders at both the highest and lowest resolutions (\ie, $\NDOF=1024$ and $\NDOF=4096$).
Since $C_s = 1$ is too high, its modal decay curve lies below the reference curve. Consequently, the energy associated with each mode is expected to be lower than in the reference case. Instead, the curves at the end of the optimization are perfectly in line with the reference. A slight difference is observed for the highest mode, where, in general, the modal energy of the optimized model is slightly lower with respect to the reference.

%
%

\subsubsection{Testing}\label{subs:OptSmago_testing}
In this subsection we will assess the optimized model in a-posteriori runs. The models will be tested a-posteriori outside the time window considered for the training.

\begin{figure}
    \centering
    \begin{subfigure}[b]{0.48\textwidth}
        \centering
        \includegraphics[width=\textwidth]{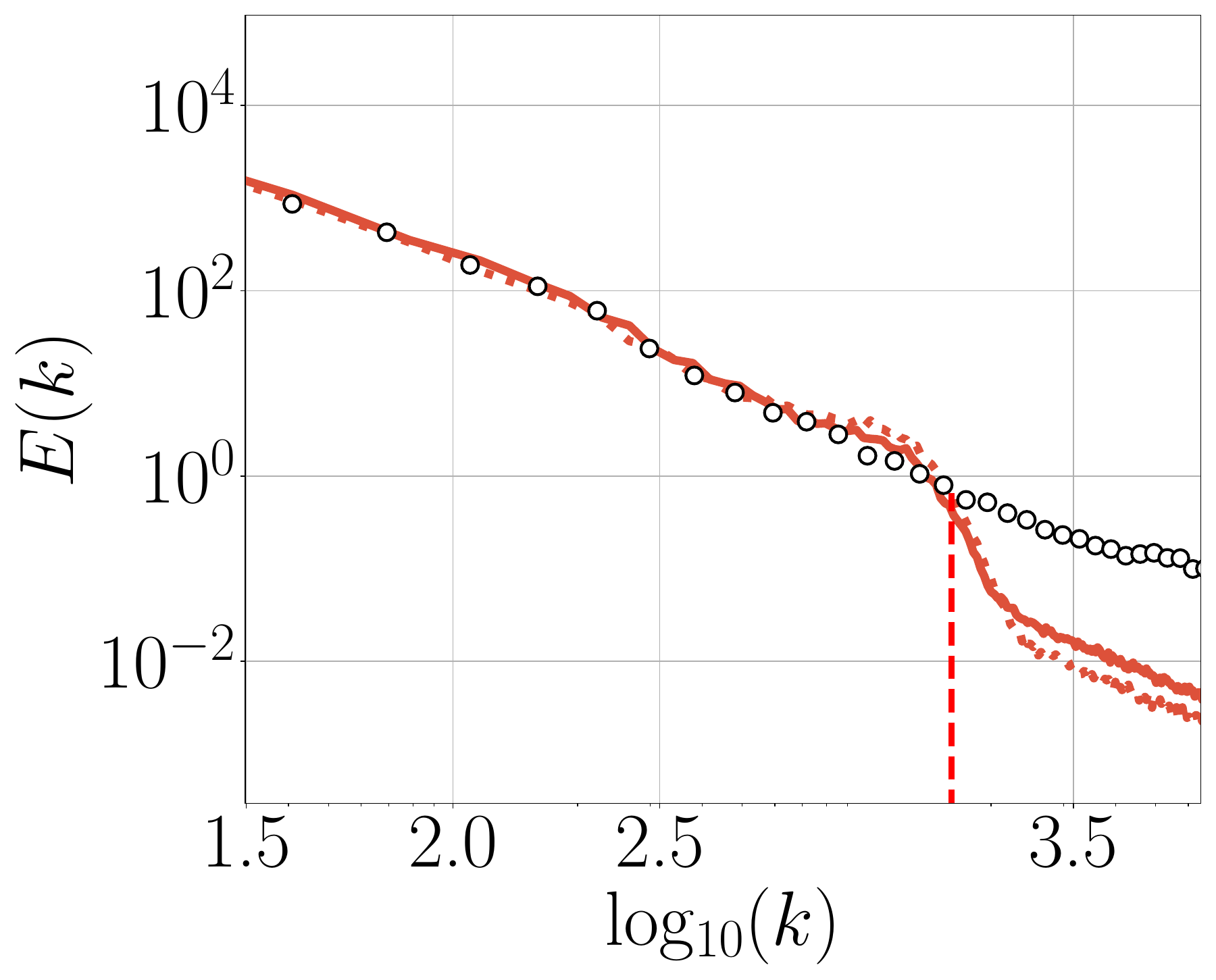}
        \caption{\normalsize $\mathrm{p}=4$}
        \label{fig:opt-smago-2048:a}
    \end{subfigure}
    \hfill
    \begin{subfigure}[b]{0.48\textwidth}  
        \centering 
        \includegraphics[width=\textwidth]{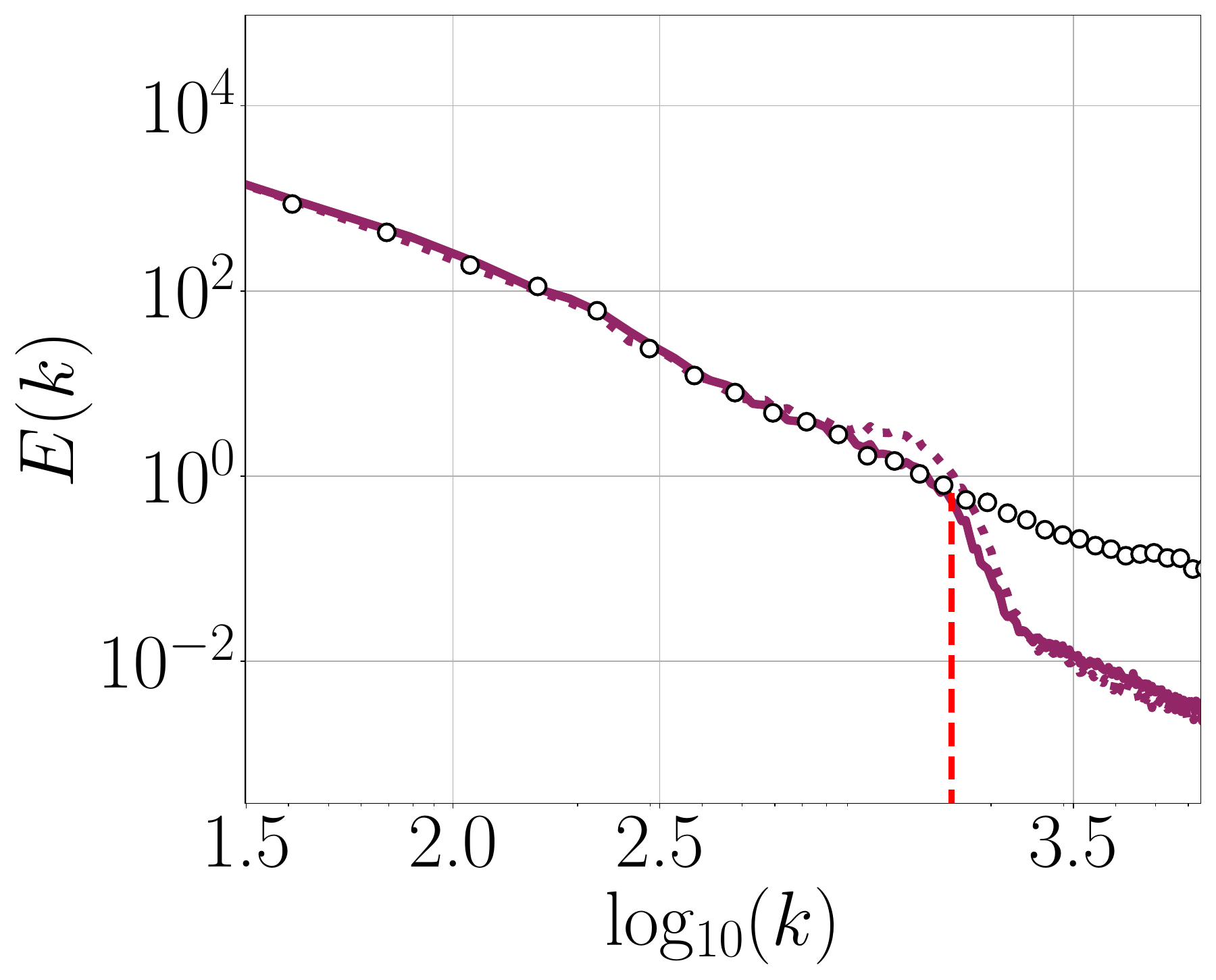}
        \caption{\normalsize $\mathrm{p}=5$}
        \label{fig:opt-smago-2048:b}
    \end{subfigure}
    \hfill
    \begin{subfigure}[b]{0.48\textwidth}   
        \centering 
        \includegraphics[width=\textwidth]{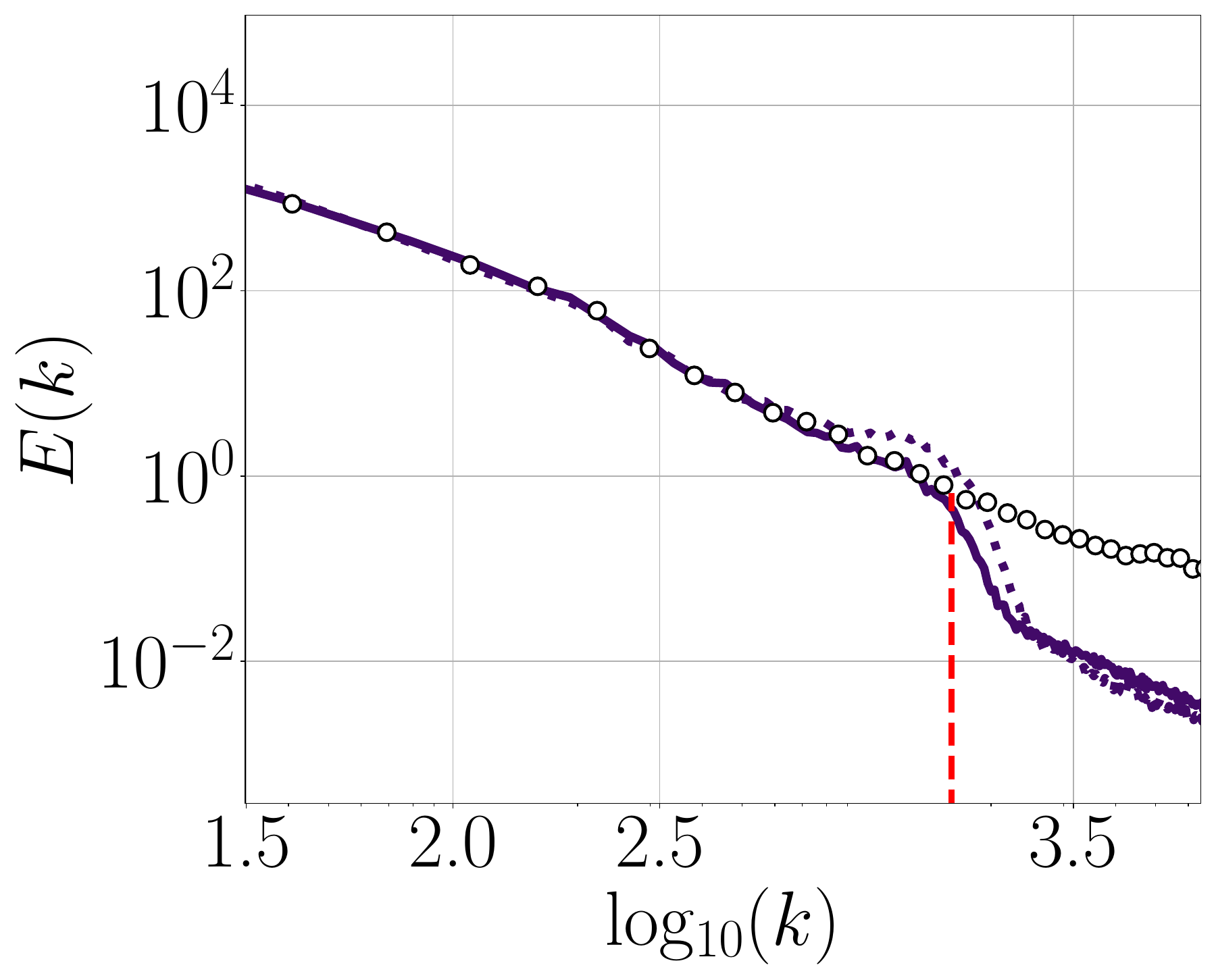}
        \caption{\normalsize $\mathrm{p}=6$}
        \label{fig:opt-smago-2048:c}
    \end{subfigure}
    \hfill
    \begin{subfigure}[b]{0.48\textwidth}   
        \centering 
        \includegraphics[width=\textwidth]{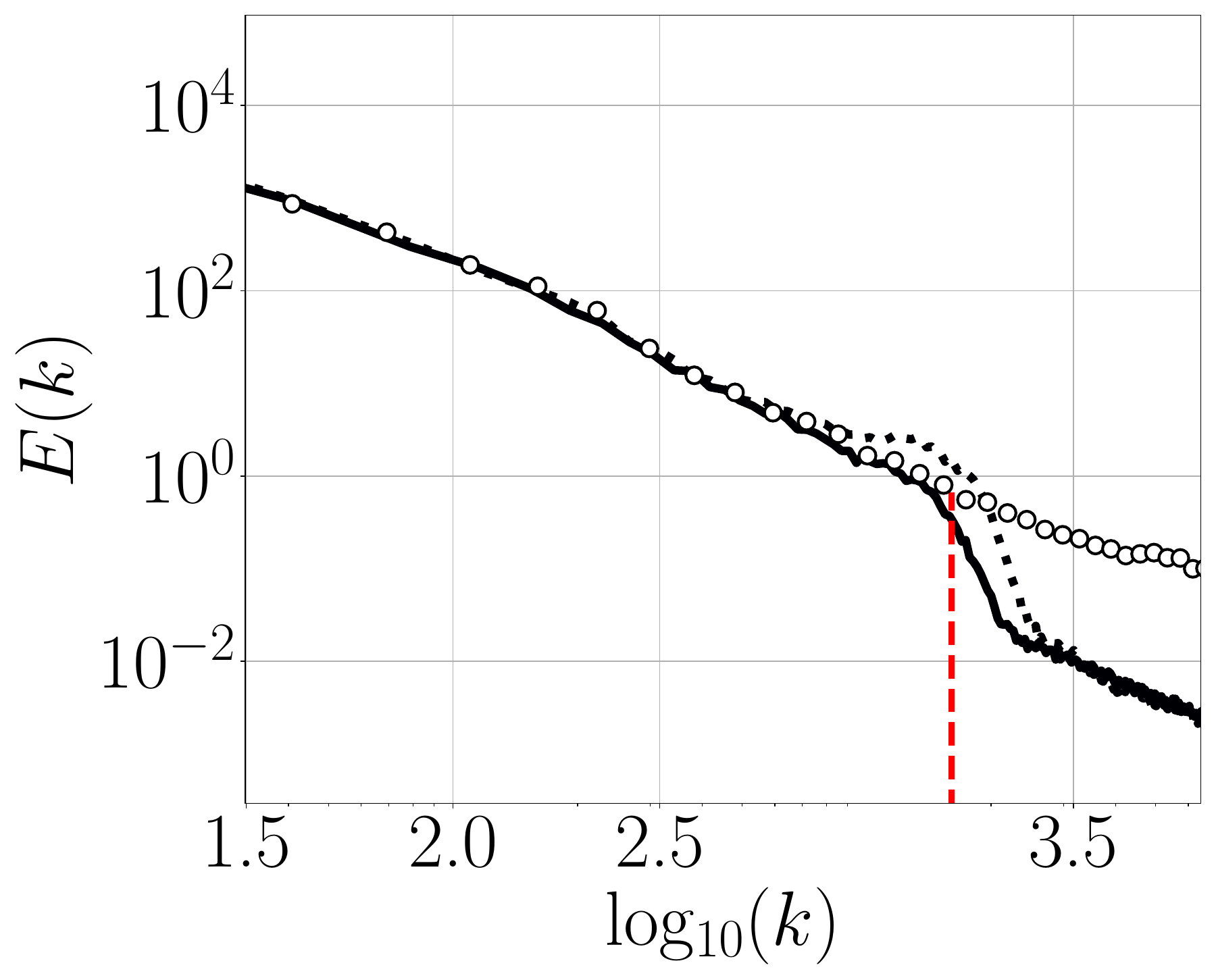}
        \caption{\normalsize $\mathrm{p}=7$}
        \label{fig:opt-smago-2048:d}
    \end{subfigure}
    \caption{Comparison of the kinetic energy spectrum between ILESs and optimized Smagorinsky for different polynomial orders for $\NDOF=2048$. Solid lines, optimized Smagorinsky; dotted lines, ILES. Symbols indicate the reference.}
    \label{fig:opt-smago-2048}
\end{figure}

Figure \ref{fig:opt-smago-2048} shows the kinetic energy spectrum obtained during the a-posteriori run by the optimized models across all the polynomial orders for the intermediate resolution $\NDOF=2048$. 
As shown by the figures, the optimized model successfully produces a clean scaling range in the inertial subrange in accordance with the reference and eliminates the spurious accumulation of energy at high wavenumbers. Notice, however, that $\mathrm{p}=4$ is slightly under-dissipative as the model does not fully eliminate the hump. This behavior is most probably related to the definition of the loss function, which is based on Legendre modes. By reducing the polynomial orders, fewer modes contribute to the loss, making the optimization inherently more challenging. Consequently, the optimization becomes more difficult: at low orders it is challenging to capture the influence of high order modes as their magnitude is several orders of magnitude smaller with respect to the first modes.

\begin{figure}
    \centering
    \begin{subfigure}[b]{0.48\textwidth}
        \centering
        \includegraphics[width=\textwidth]{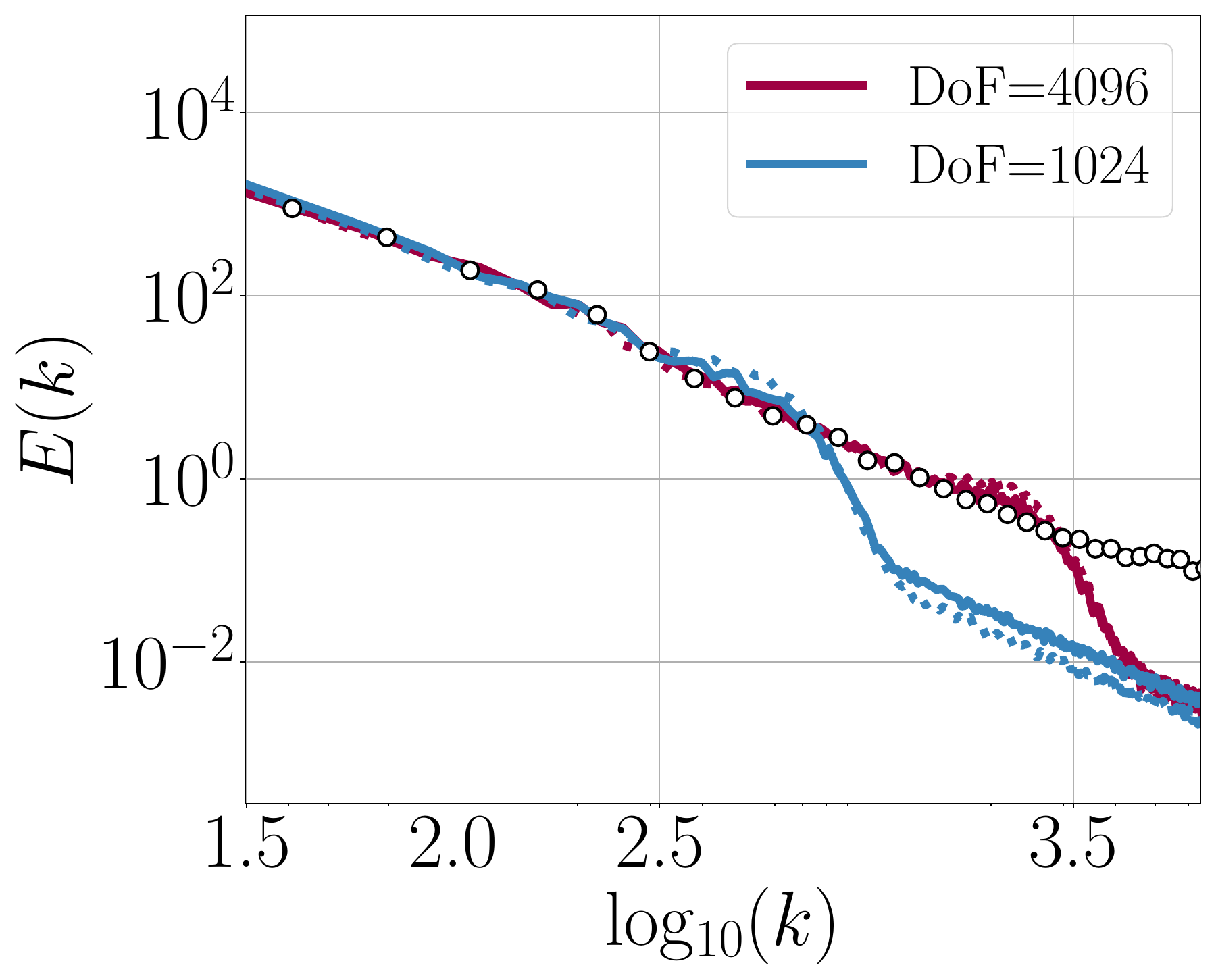}
        \caption{\normalsize $\mathrm{p}=4$}
        \label{fig:opt-smago-dofs:a}
    \end{subfigure}
    \hfill
    \begin{subfigure}[b]{0.48\textwidth}  
        \centering 
        \includegraphics[width=\textwidth]{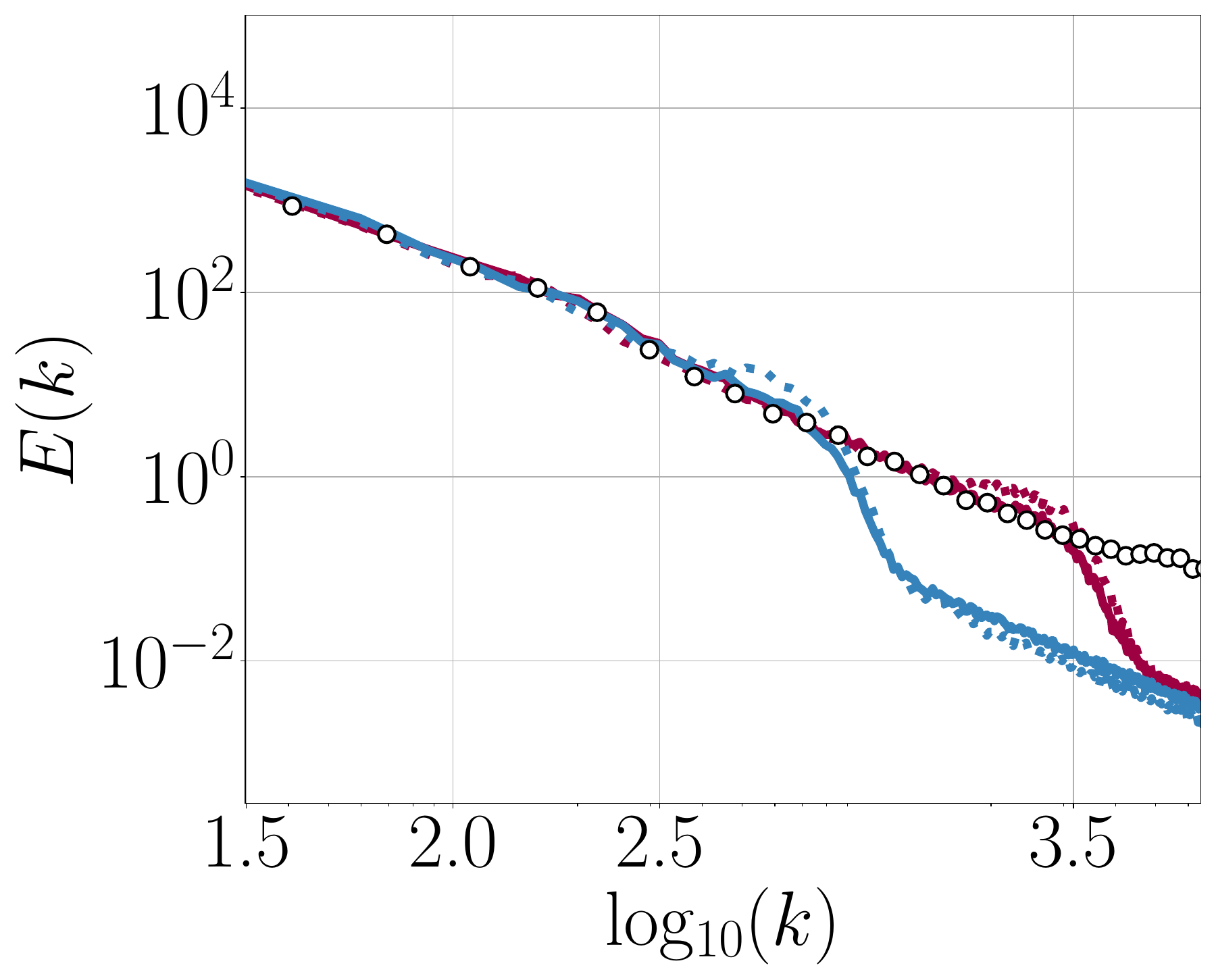}
        \caption{\normalsize $\mathrm{p}=5$}
        \label{fig:opt-smago-dofs:b}
    \end{subfigure}
    \hfill
    \begin{subfigure}[b]{0.48\textwidth}   
        \centering 
        \includegraphics[width=\textwidth]{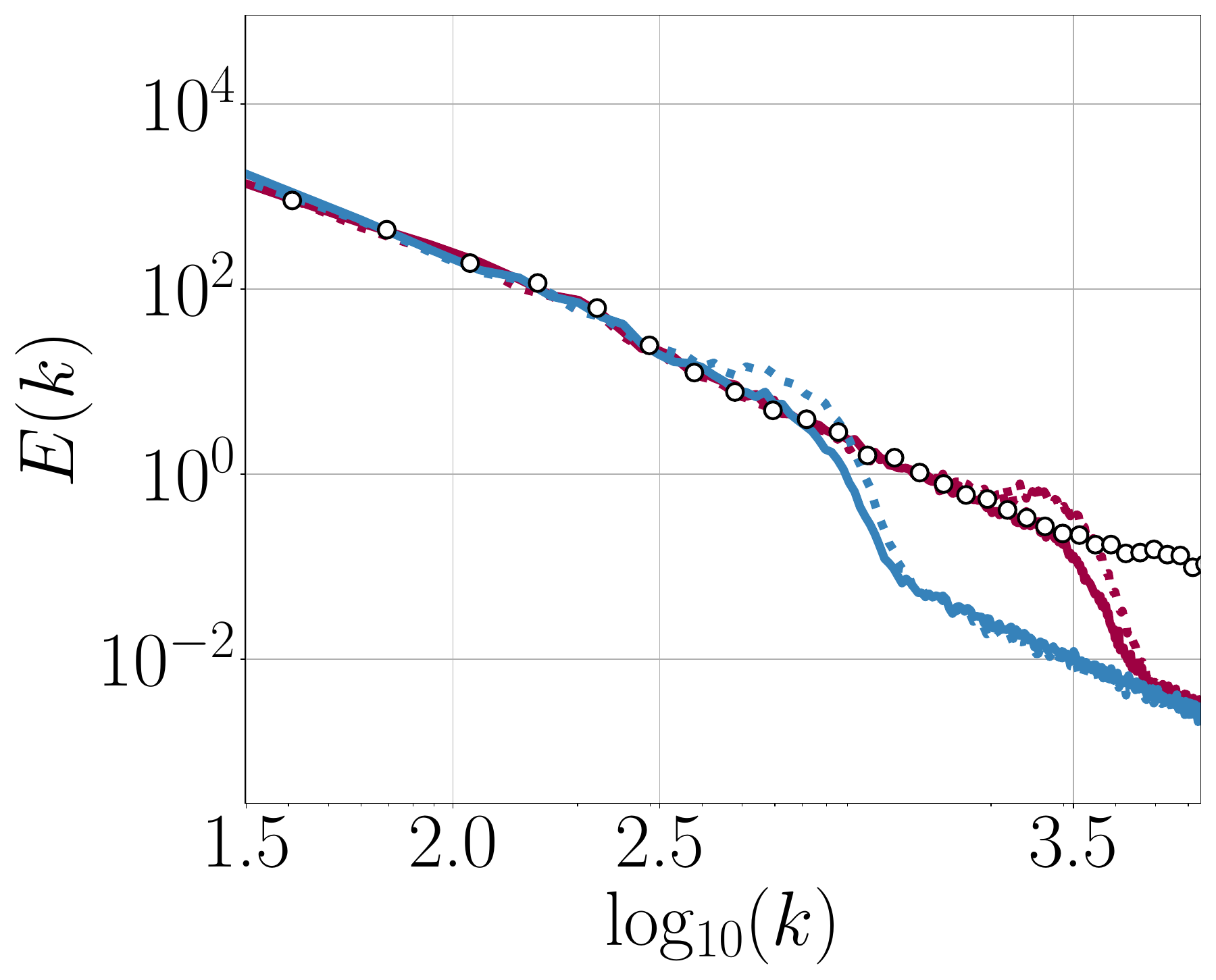}
        \caption{\normalsize $\mathrm{p}=6$}
        \label{ffig:opt-smago-dofs:c}
    \end{subfigure}
    \hfill
    \begin{subfigure}[b]{0.48\textwidth}   
        \centering 
        \includegraphics[width=\textwidth]{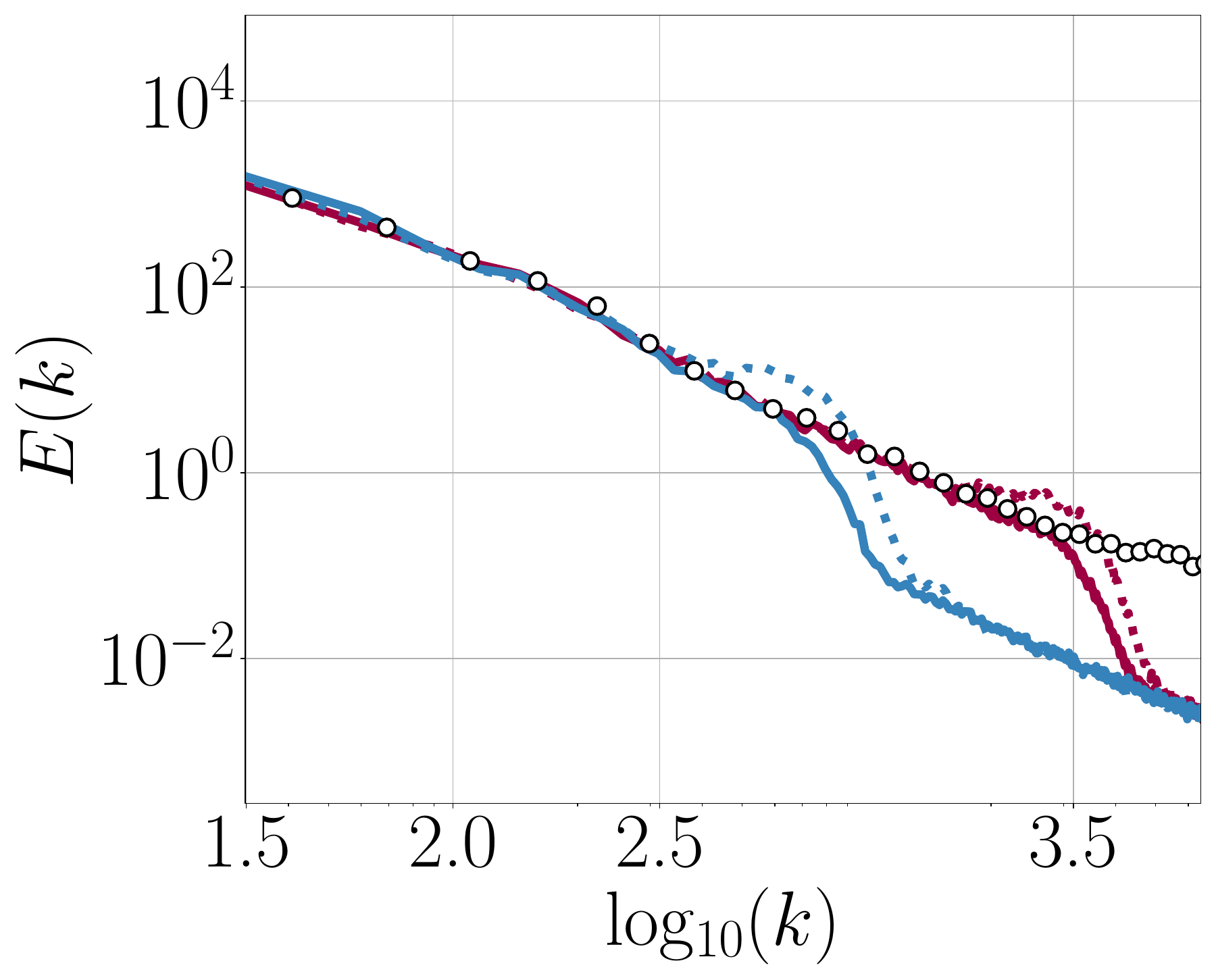}
        \caption{\normalsize $\mathrm{p}=7$}
        \label{fig:opt-smago-dofs:d}
    \end{subfigure}
    \caption{Comparison of the kinetic energy spectrum between ILESs and the optimized Smagorinsky for different polynomial orders for the coarsest ($\NDOF=1024$) and the finest resolution ($\NDOF=4096$). Solid lines, optimized Smagorinsky; dotted lines, ILES. Symbols indicate the reference.}
    \label{fig:opt-smago-dofs}
\end{figure}

In order to compare the a-posteriori results for other resolutions, figure \ref{fig:opt-smago-dofs} compares the kinetic energy spectrum obtained by the optimized model with the ILESs for different polynomial orders for the lowest and the highest resolutions considered in this study. 
Across all polynomial orders, the curves relative to the highest resolution ($\NDOF=4096$) consistently remain closer to the reference energy decay, while the curves relative to the coarsest resolution ($\NDOF=1024$) deviate earlier as the wavenumber increases. This behavior reflects the impact of spatial resolution: for the high resolution case small-scale energy is preserved for a longer range of modes, maintaining a smoother and more gradual decay.

Even for these resolutions, the curves obtained by the optimized model are almost overlapped to the reference curve up to the cutoff. Similarly to the results previously found (\ie, for $\NDOF=2048$), the curves obtained for $\mathrm{p}>4$ are in good agreement with reference data. The kinetic energy spectrum for $\mathrm{p}=4$, instead, is slightly under-dissipative with respect to the reference. This is expected as the optimization becomes more challenging as the polynomial order decreases.

All these analyses confirm that the specific choice of the loss function is effective in capturing the impact of the numerical scheme on the flow physics. Even though a direct and explicit relationship between the Legendre spectrum and the Fourier spectrum is not available in the DSEM framework~\cite{MOURA2015695,moura2019spatial,manzanero2018dispersion}, the optimization built entirely on Legendre modes has a beneficial effect on the global kinetic energy spectrum too. In other words, optimizing the model in modal space (Legendre) leads to improvements in physical space and wavenumber space (Fourier), demonstrating that the chosen loss function successfully drives the solution toward the desired spectral behavior.

\section{Three-dimensional compressible Navier-Stokes equations}\label{sec:ns3d}
In this section we extend our methodology to the much more challenging three-dimensional compressible Navier-Stokes equations.

From a computational point of view, the implementation of the framework becomes significantly more demanding. In particular, the discrete-adjoint approach requires the evaluation of vector–Jacobian products assembled at the fully discrete level, resulting in a substantial increase in implementation complexity and computational cost. Efficient memory storage and parallel computing are mandatory in order to make the whole setting feasible for practical applications.

Secondly, while the forced Burgers test case can be useful for highlighting certain features of the numerical footprint of the scheme~\cite{MOURA2015695,MAULIK201812}, it does not capture the richness of the dynamics encountered in fully three-dimensional turbulent flows. The latter involve more complex interactions across scales, making the assessment of model performance substantially more stringent. Although some resembling of kinetic energy cascade can be observed within the stochastically forced Burgers equations, it will only roughly approximate the complexity of three-dimensional turbulence.

Consequently, a model calibrated or formulated within the simplified setting of the Burgers test case may not retain its performance when applied to the fully compressible Navier–Stokes equations.
\subsection{Optimization of the Smagorinsky constant}
Like in the forced-Burgers test case, we start by optimizing the Smagorinsky constant within a statistically steady state setting (\ie, FHIT). In particular, referring to the nomenclature introduced in section \ref{sec:problemdef} we will consider the following closure:
\begin{equation}
    \tau_{ij}^{\mathrm{SGS}} = -2\nu_{t} \widetilde{S}_{ij} \quad \mathrm{with} \quad \widetilde{S}_{ij} = \frac{1}{2} \bigg( \frac{\partial \widetilde{u}_{i}}{\partial x_{j} } +\frac{\partial \widetilde{u}_{j}}{\partial x_{i} }\bigg) \quad \mathrm{and} \quad \nu_{t} = (C_{s} \Delta)^{2} \sqrt{2\widetilde{S}_{ij} \widetilde{S}_{ij}}.
\end{equation}
Although the framework herein proposed is fully compressible, we will always consider low-Mach test cases. Consequently, the closure for the SGS heat flux is simply set to zero for all models presented here. For the same reason, we will always use the fully compressible notation of $\widetilde{(\cdot)}$ even though the Favre-filtering will always be very close to the classical filtering operation $\overline{(\cdot)}$ for the flow regimes considered in this work.

\emph{A-priori} studies conducted by Meyers \& Sagaut~\cite{MeyersSagaut06} for decaying homogeneous isotropic turbulence revealed that the coefficient of the Smagorinsky model should range from $0.06$ to $0.10$, depending on the specific quantities of interest to be matched for a given grid resolution. Based on these findings, we expect the optimal Smagorinsky constant obtained in the present study lie in this range. At the same time, we also expect some differences arising from the a-posteriori nature of the present work. In particular, the numerical discretization, the training configuration, and the definition of the objective functional are all expected to influence the estimated coefficient.

\begin{figure}[h!]
\centering
\includegraphics[width=.85\textwidth]{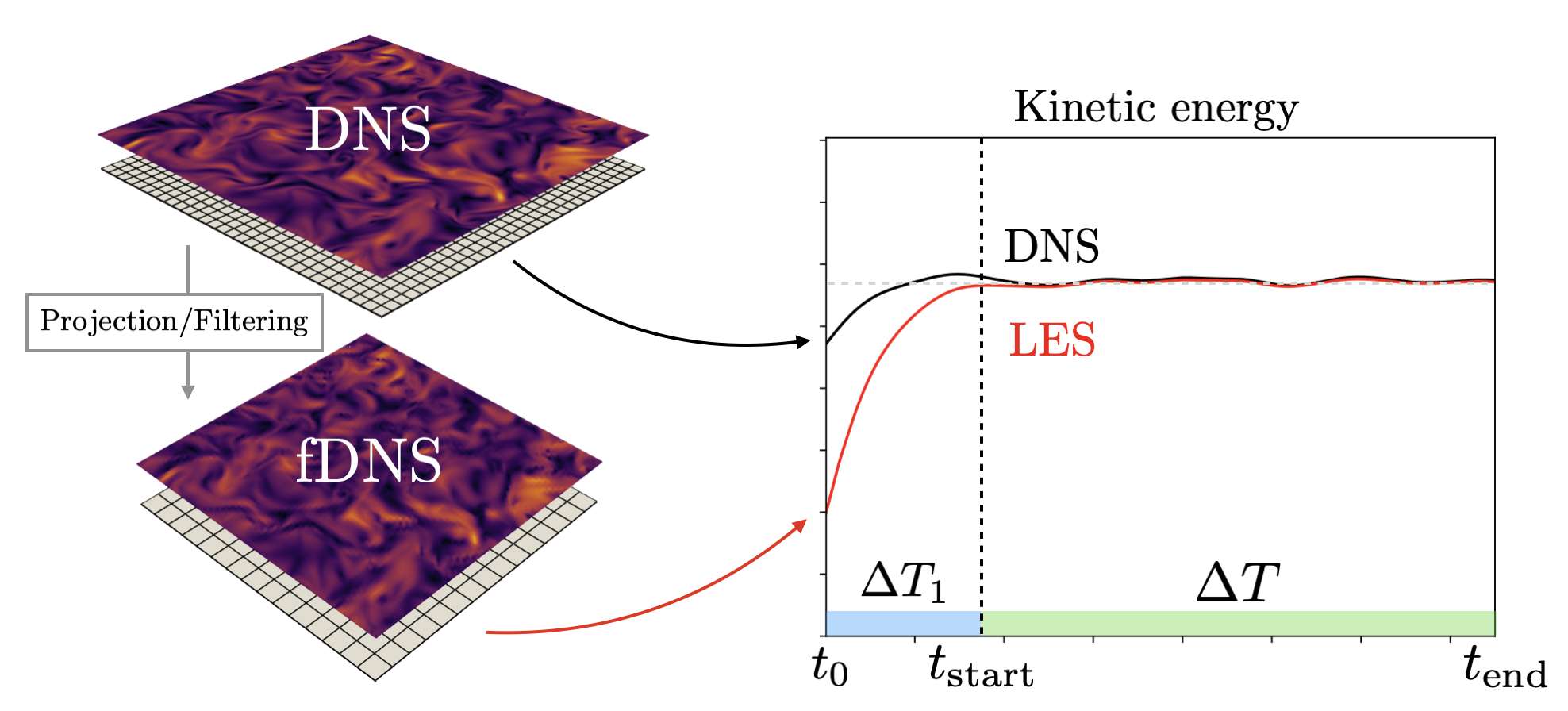}
\caption{Schematic of the 3D FHIT methodology: a forward run over $\Delta T_1$ is first performed, after which sensitivities are computed using the adjoint loop over the interval $\Delta T$.}
\label{fig:TwindowNS}
\end{figure}

The initial velocity field for the LESs are generated by filtering the DNS fields followed by a projection onto the LES grids. Moreover, the optimization is carried out after performing a number of time steps sufficient to suppress the spurious effects caused by the projection. For clarity, a schematic diagram of the methodology is shown in figure \ref{fig:TwindowNS}. After projection, the LES solution goes through a short transient before recovering the expected kinetic energy enforced via the forcing term. In the first interval $\Delta T_1$ a forward run of the solver is performed. Only after this interval, sensitivities are computed using the adjoint loop in the second time window $\Delta T$.
%
\subsubsection{Training}\label{subs:OptSmagoNS_training}
For training, we consider a statistically steady state turbulent flow scenario, namely the forced homogeneous isotropic Turbulence test case. The computational domain consists of a three-dimensional box of of size $[0,2\pi]^3$ with periodic boundary conditions.
An initial turbulent kinetic energy spectrum is imposed~\cite{Chasnov95} and the Rogallo procedure is employed for the initialization~\cite{rogallo1981numerical}. To maintain a statistically steady state, an isotropic linear forcing term is then applied in the momentum balance~\cite{LinearForcingDeLaage}. 

The Reynolds number based on the Taylor microscale is fixed to $\mathrm{Re}_\lambda\simeq160$. As DNS we consider a simulation based on a mesh made of $N_{el}=43$ element with a polynomial degree $\mathrm{p}=5$ resulting in a total number of DoF $N_{\mathrm{dof}}=258^3$. In order to gather significant statistics, the DNS is advanced up to $t=15$, which corresponds to around $20$ eddy turn-over times.
\begin{figure}
    \centering
    \begin{subfigure}[b]{0.45\textwidth}
        \centering
        \includegraphics[width=\textwidth]{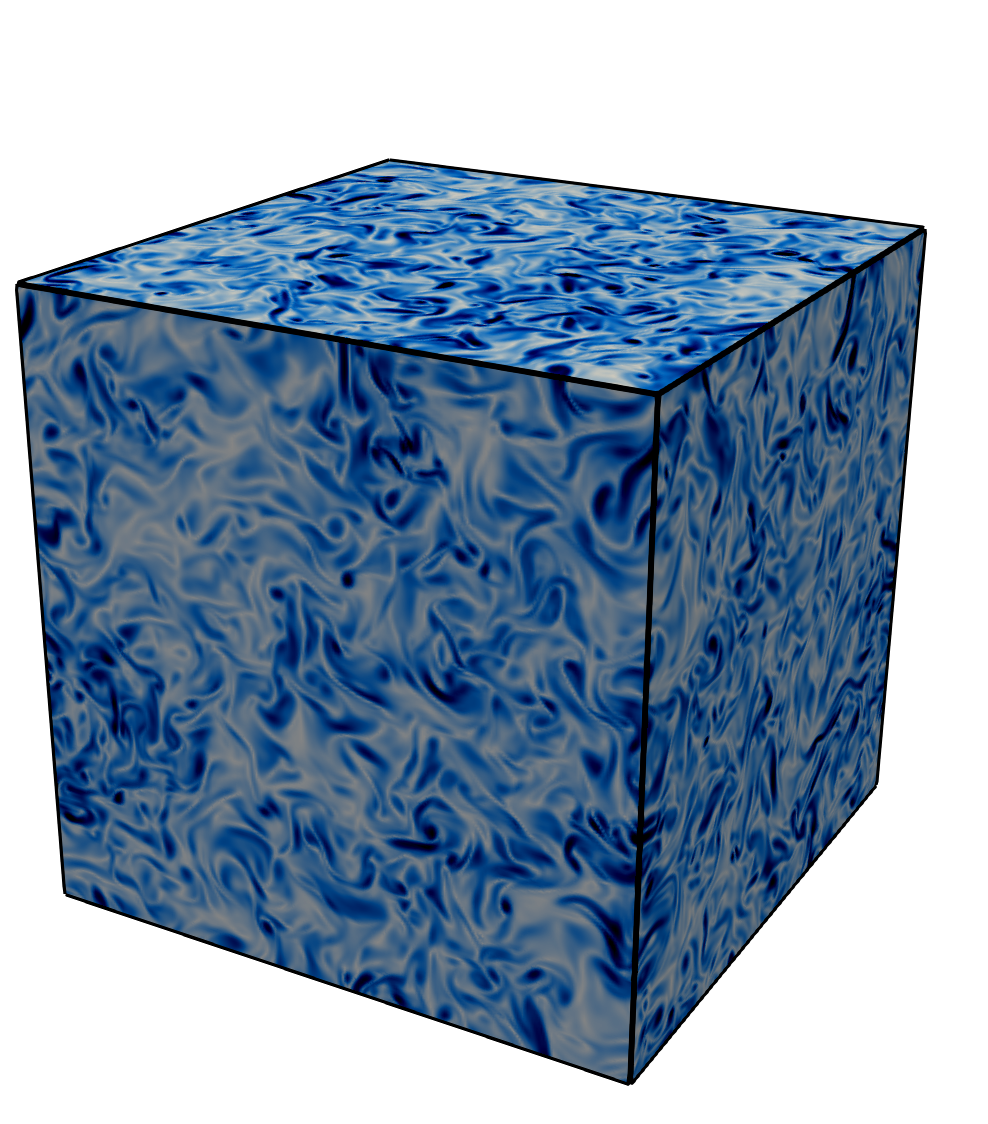}
        \end{subfigure}
        \begin{subfigure}[b]{0.44\textwidth}  
            \centering 
            \includegraphics[width=\textwidth]{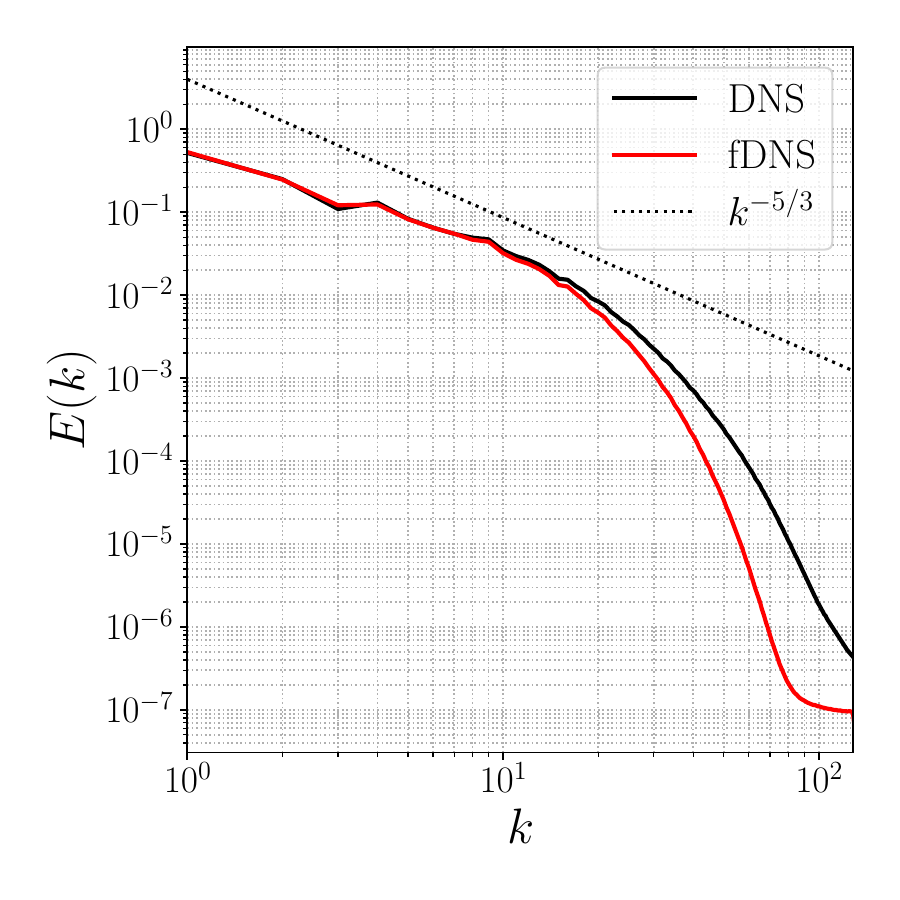}
        \end{subfigure}
        \hfill
        \caption{Example of instantaneous vorticity field (left) and spatio-temporally averaged kinetic energy spectrum (right) for the FHIT case. The filtered DNS, shown in the right figure in red, refers to the cutoff $\Delta_{\mathrm{LES}}=4\Delta_{\mathrm{DNS}}$ which corresponds to a total of number of degrees of freedom of $N_{\mathrm{dof}}=64^3$.}
        \label{}
\end{figure}
Given the high chaotic nature of the Navier-Stokes equations, the time window used for the optimization should be sufficiently small to avoid the adjoint variable to blow-up \cite{ashley2019towards,blonigan2013new} and sufficiently large to yield non-negligible sensitivities. In this work, we choose to optimize our parameters on a time window which contains about $1200$ Runge-Kutta steps. From numerical experiments presented later on, this choice of time window was found to be large enough to provide good convergence of the modal energy during the forward run.
We choose a third-order, three stages Runge-Kutta time scheme with a fixed time step of $\delta t=5.0e-5$. This time step is small enough to make the backward loop for the adjoint stable. As a consequence, the time window used for the optimization is $\Delta T=0.02$ which corresponds to approximately $0.027$ eddy turn-over times. Although such an interval can seem insufficiently small to provide fully converged statistics, from the numerical experiments contained in the following section, it was found to be large enough to capture the main characteristics of coarse-grained simulations and to carry the necessary information needed to match the filtered DNS data.

Within the optimization framework presented in section \ref{sec:theadjoint}, the control parameter to optimize coincides with the Smagorinsky coefficient, which was initially set to $\theta^{(0)}=C_s=0.187$.

The optimization is performed on two different coarse resolutions with respect to DNS, namely we consider coarse-grain simulations involving $N_{\mathrm{dof}}=64^3$ and $N_{\mathrm{dof}}=80^3$ total solution points. Within this framework, we consider different polynomial orders varying from $\mathrm{p}=4$ to $\mathrm{p}=6$.

\begin{figure}
    \centering
    \begin{subfigure}[b]{0.475\textwidth}
        \centering
        \includegraphics[width=\textwidth]{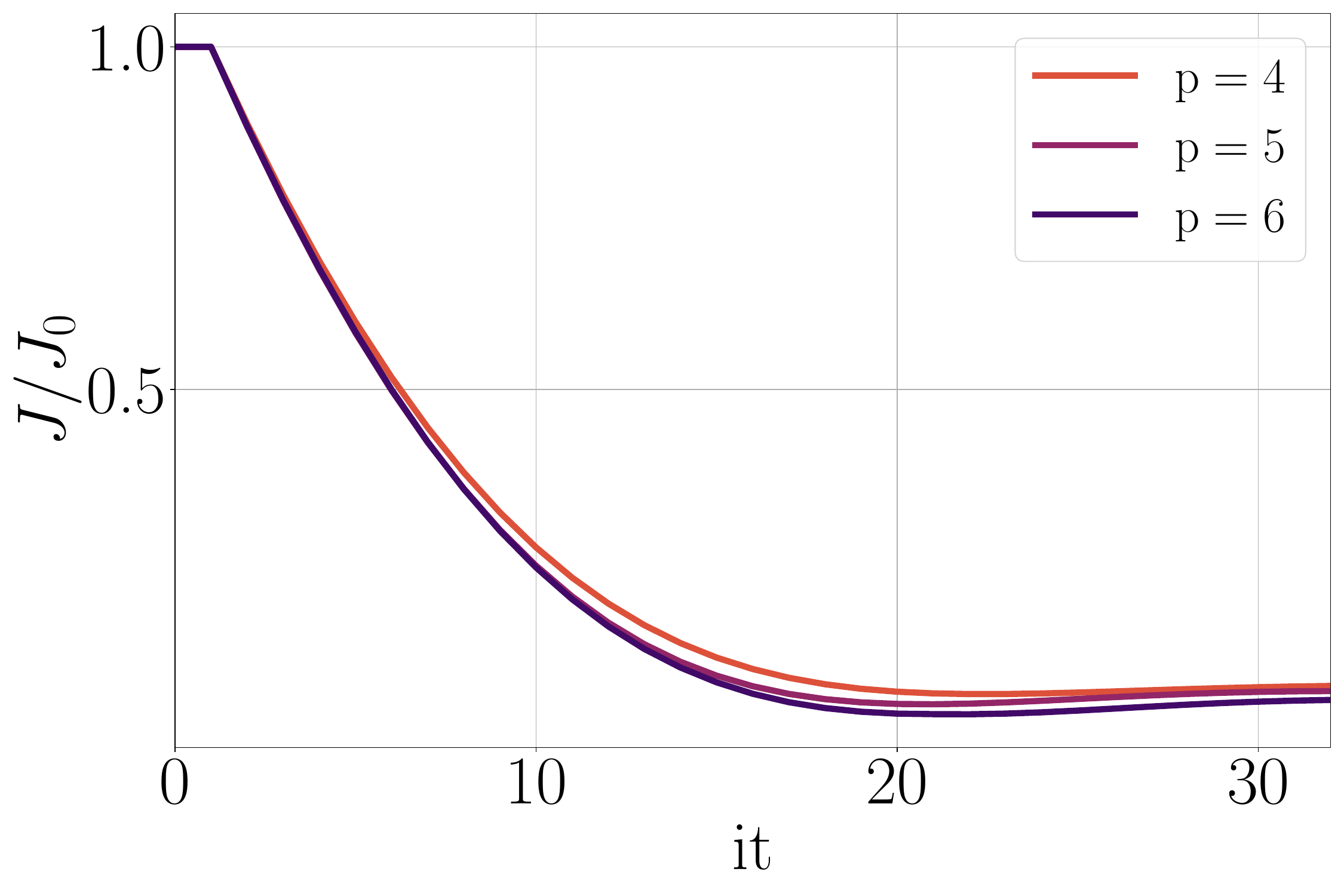}
        \caption{$N_{\mathrm{dof}}=64$}
        \label{fig:ns_decay_opt_smago_64:a}
        \end{subfigure}
        \begin{subfigure}[b]{0.486\textwidth}  
            \centering 
            \includegraphics[width=\textwidth]{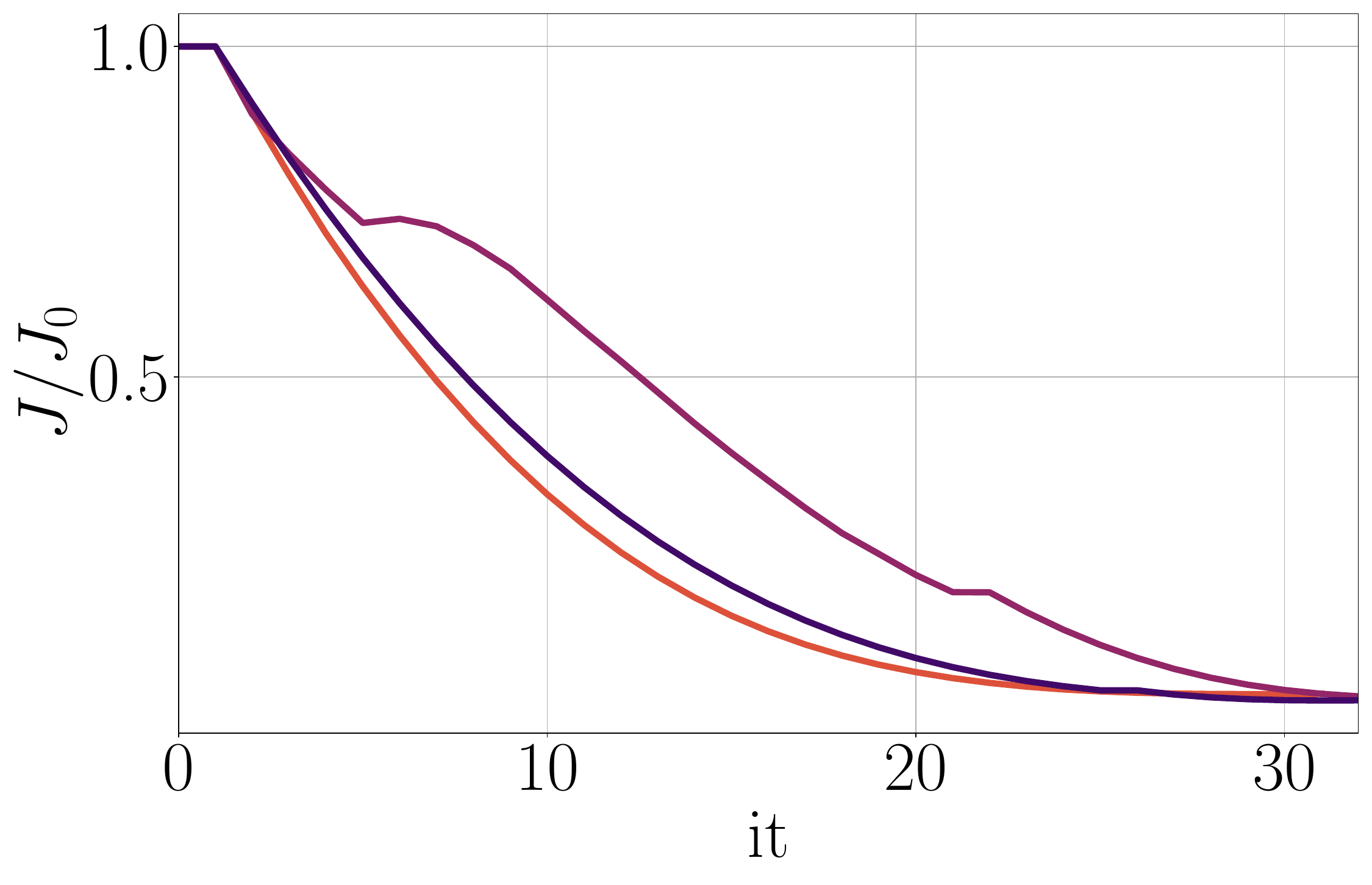}
             \caption{$N_{\mathrm{dof}}=80$}
             \label{fig:ns_decay_opt_smago_64:b}
        \end{subfigure}
        \hfill
        \caption{Normalized loss functions for different polynomial orders  for $N_{\mathrm{dof}}=64$ (left) and for $N_{\mathrm{dof}}=80$ (right).}
        \label{fig:ns_loss_opt_smago}
\end{figure}

Figure \ref{fig:ns_loss_opt_smago} shows the evolution of the normalized loss functions for the two resolutions and for all the polynomial orders considered. The objective function becomes one-tenth of the initial value after about $30$ iterations for both resolutions. 
The loss function decreases smoothly for all polynomial orders indicating that the overall procedure is stable and, with limited number of iterations, gives significant improvements in terms of objective function minimization. A slightly different trend is observed at the highest resolution ($N_{\mathrm{dof}} = 80$) for $\mathrm{p} = 5$. Nevertheless, after $30$ iterations, the loss function converges to values very similar to those obtained for the other polynomial orders.

\begin{figure}[h!]
\centering
\includegraphics[width=.99\textwidth]{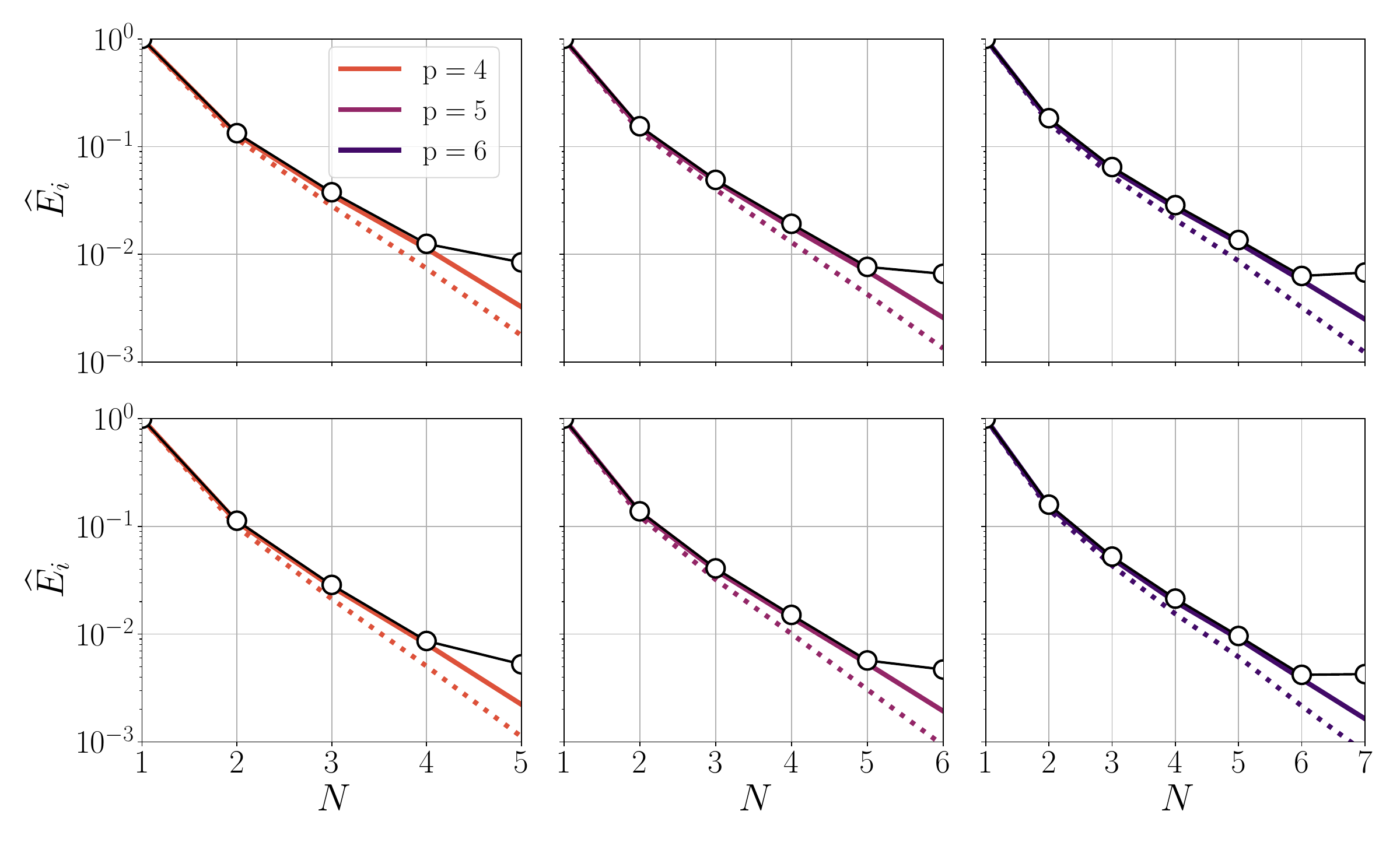}
\caption{Comparison of the modal decay obtained at the beginning of the optimization and the optimized one for the various polynomial orders. Top row, $N_{\mathrm{dof}}=64$; Bottom row, $N_{\mathrm{dof}}=80$.  Solid lines, optimized Smagorinsky model; dotted lines, baseline model ($\mathrm{C_s}=0.187$). Lines with symbols indicate the filtered DNS data.}\label{fig:smago_training_modes}
\end{figure}

In analogy with the Burgers test case, figures \ref{fig:smago_training_modes} reports the space- and time-averaged modal energy at the end of the optimization procedure for the two resolutions considered in this study. These results are compared against the modal energy obtained using the baseline model with $C_s=0.187$ (dotted lines) and the reference decay obtained by filtering and then projecting the DNS onto the LESs grids (symbols).

In agreement with the observations from the forced Burgers test case, the baseline model exhibits excessive dissipation, resulting in a modal decay curve that lies below the reference. Instead, the optimized model shows good agreement with the reference decay. However, a difference is observed for the largest mode, where, in general, the modal energy of the optimized model is lower with respect to the reference. This difference may be due to projection effects, which can artificially increase the energy content of the highest mode.

Table \ref{tab:smago-opt-results-ns} shows the optimized Smagorinsky constants obtained at the end of the optimization for the two resolutions considered in the study. The optimized constants range from $0.080$ to $0.105$ overall. This is broadly consistent with the a-priori analysis of Meyers \& Sagaut~\cite{MeyersSagaut06}, where optimal constants were found to vary between $0.06$ and $0.10$ depending on the resolution. 

The values obtained in the present study are in good agreement with those reported by Meyers \& Sagaut~\cite{MeyersSagaut06}, despite the significant differences between the two frameworks. In particular, the numerical solver, the optimization objective, the test case under consideration, and the type of analysis performed (a-priori versus a-posteriori) differ substantially. Consequently, although the optimal coefficients are found to be of similar magnitude, an exact agreement for a fixed resolution is not necessarily expected, given the methodological and numerical differences between the two studies.

This observation further highlights that the coefficients of any turbulence model are inherently dependent on the specific numerical discretization. It is therefore expected that different numerical schemes, characterized by distinct intrinsic properties, will lead to different optimal parameters. This, in turn, underscores the importance of optimizing these coefficients using the numerical solver itself within the optimization loop, as the discretization directly impacts the overall prediction.

Even within the same numerical strategy, by varying the polynomial order, we notice significant differences in the optimal coefficients. In fact, as observed in the forced Burgers test case, the optimal value decreases when the resolution increases. Moreover, the constant decreases as the polynomial order decreases. This is again in agreement with the results of the standard linear analysis, in which high order polynomials lead to
better conservation of energy.
\begin{table}[!ht]
\begin{center}
\large
\begin{tabular}{|c|c c c |} 
 \hline
 $\NDOF$ & $\mathrm{p}=6$ & $\mathrm{p}=5$ & $\mathrm{p}=4$  \\
 \hline
 64 & 0.1032  & 0.104 & 0.0995 \\ 
 \hline
 80 & 0.0915 & 0.0845 & 0.0805  \\
 \hline
\end{tabular}
\end{center}
\vspace{-0.4cm}
\caption{Optimized Smagorinsky constants for the different resolutions and different polynomial orders considered in this study.}
\label{tab:smago-opt-results-ns}
\end{table}
 %
\subsubsection{Testing}

Hereafter, the optimized model is evaluated on three unseen flow configurations: forced homogeneous turbulence over an extended time horizon (using a time window longer than that employed during training), decaying homogeneous isotropic turbulence, and the Taylor–Green vortex breakdown at $\mathrm{Re}=5000$. These test cases are designed to assess the ability of the optimized model to generalize beyond the flow configuration considered during training.

\subsubsection{Forced Homogeneous turbulence} \label{smago:fhit}
Starting from the FHIT test case, we run our simulations a-posteriori in a time window $\Delta T=1$, corresponding to approximately $1.2$ large-eddy turnover times. This duration is significantly longer than the time window used during the optimization phase. This choice allows us to assess the model’s ability to generalize beyond the temporal conditions it was exposed to during training.
The initial field is the same field used for the optimization, namely the field obtained by filtering and projecting the initial DNS data. The models are evaluated with the same resolution used for the training $N_{\mathrm{dof}}=64$. We consider the a-posteriori space- and time-averaged modal decay as evaluation metric in order to assess our model generalization.

\begin{figure}[h!]
\centering
\includegraphics[width=.99\textwidth]{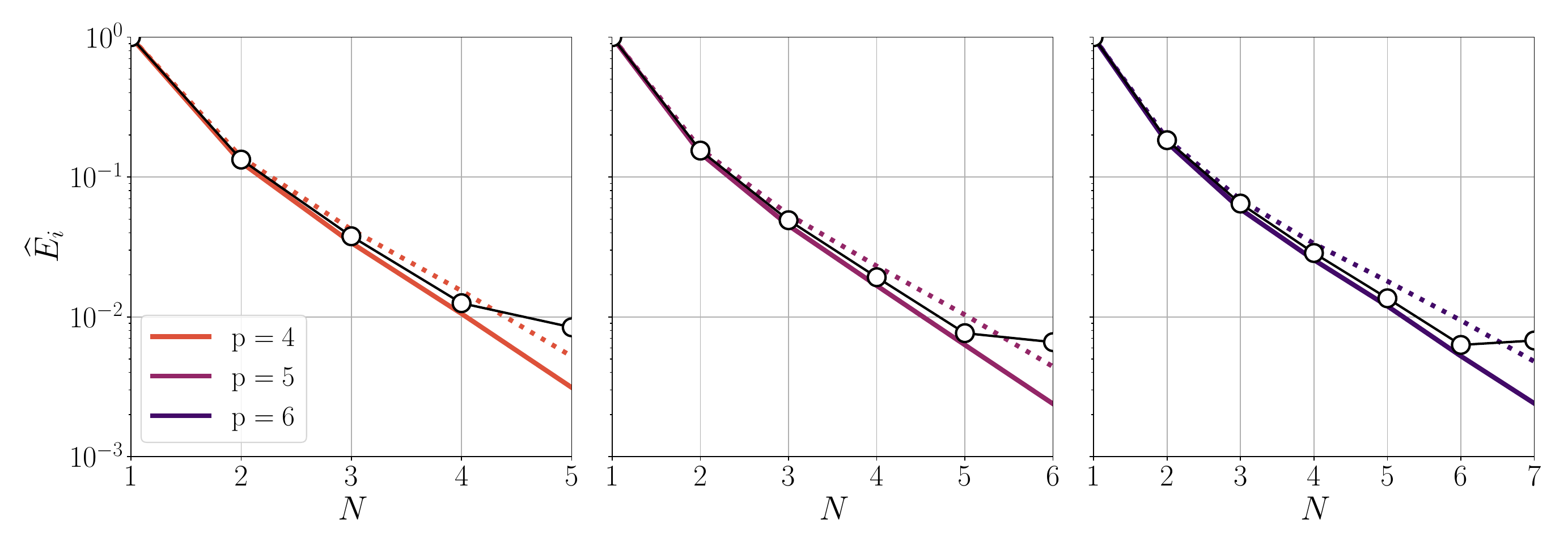}
\caption{Time and space-averaged modal energy for different polynomial orders. Solid lines, optimized Smagorinsky; dotted lines ILES. Line with symbols indicate the filtered DNS.}
\label{fig:modes_fhit_validation_64}
\end{figure}

Figure \ref{fig:modes_fhit_validation_64} presents the space- and time-averaged modal energy for all polynomial orders considered in the study. 
For all polynomial orders, the optimized model is able to achieve a good matching with the modal energy given by filtered DNS. In contrast, the ILESs exhibit significantly higher modal energy compared to the filtered DNS. This behavior indicates insufficient numerical dissipation, leading to an accumulation of energy that it is not properly dissipated.

\begin{figure}[h!]
\centering
\includegraphics[width=.99\textwidth]{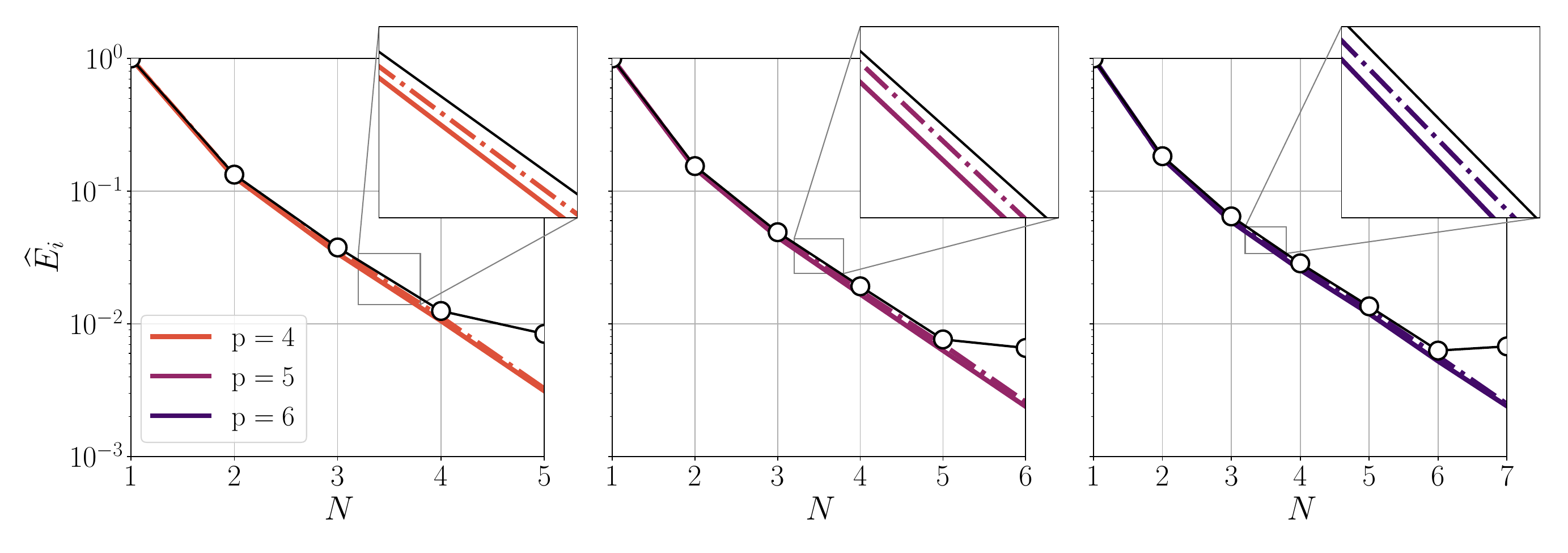}
\caption{Time and space-averaged modal energy for different polynomial orders. Solid lines, $\Delta T=1$; dash-dotted lines $\Delta T=0.02$. Line with symbols indicate the filtered DNS.}
\label{fig:modes_dt_validation}
\end{figure}

We can therefore conclude that, even though the time window considered during training is relatively short, the modal energy is already close to statistical convergence. Indeed, extending the a-posteriori time interval to $\Delta T = 1$ results in negligible changes in the modal energy (see figure \ref{fig:modes_dt_validation}). This behavior is likely due to the combined spatial and temporal averaging procedure. As a result, despite the limited temporal extent, the spatial distribution of information is sufficient to ensure good statistical convergence.
Notice that the highest mode of the reference solution slightly differs from the optimized modal. This discrepancy is not unexpected, as the highest mode of the reference is likely affected from errors arising from the projection of the filtered DNS onto the LES grid. 
Consequently, all the analyses on the modal energy should be considered valid for all the modes except the last one, which is the one most likely affected by the projection error.
\subsubsection{Decaying Homogeneous Turbulence} \label{smago:dhit}
In this section, we will assess the optimized models in a DHIT test case \cite{BIFERALE03DECAY,THORNBER2007IMPLICIT}. In the present work, the DHIT test case is employed to further validate the proposed modeling strategy under conditions that differ from the forced turbulence setup, thereby providing a complementary assessment of its predictive capabilities.
This test case is widely used to assess the performance of numerical methods and subgrid-scale models, in particular for data-driven SGS models \cite{beck2023toward,agdestein2025discretize,bezgin2025ml,huang2026consistency}. In this context, an initial velocity field characterized by an imposed kinetic energy spectrum is allowed to evolve freely under no external forcing.
The incompressible velocity field is constructed as proposed
by Rogallo~\cite{rogallo1981numerical}, with an initial spectrum of the kinetic energy given by Chasnov in~\cite{Chasnov95}.
We use a total of $258^3$ DoF for the DNS~\cite{DECROUYCHANEL2024106287,huang2026consistency}.

The decay of kinetic energy is then dictated by the viscous dissipation and nonlinear interactions developed during the time evolution.

\begin{figure}
    \centering
    \begin{subfigure}[b]{0.485\textwidth}
        \centering
        \includegraphics[width=\textwidth]{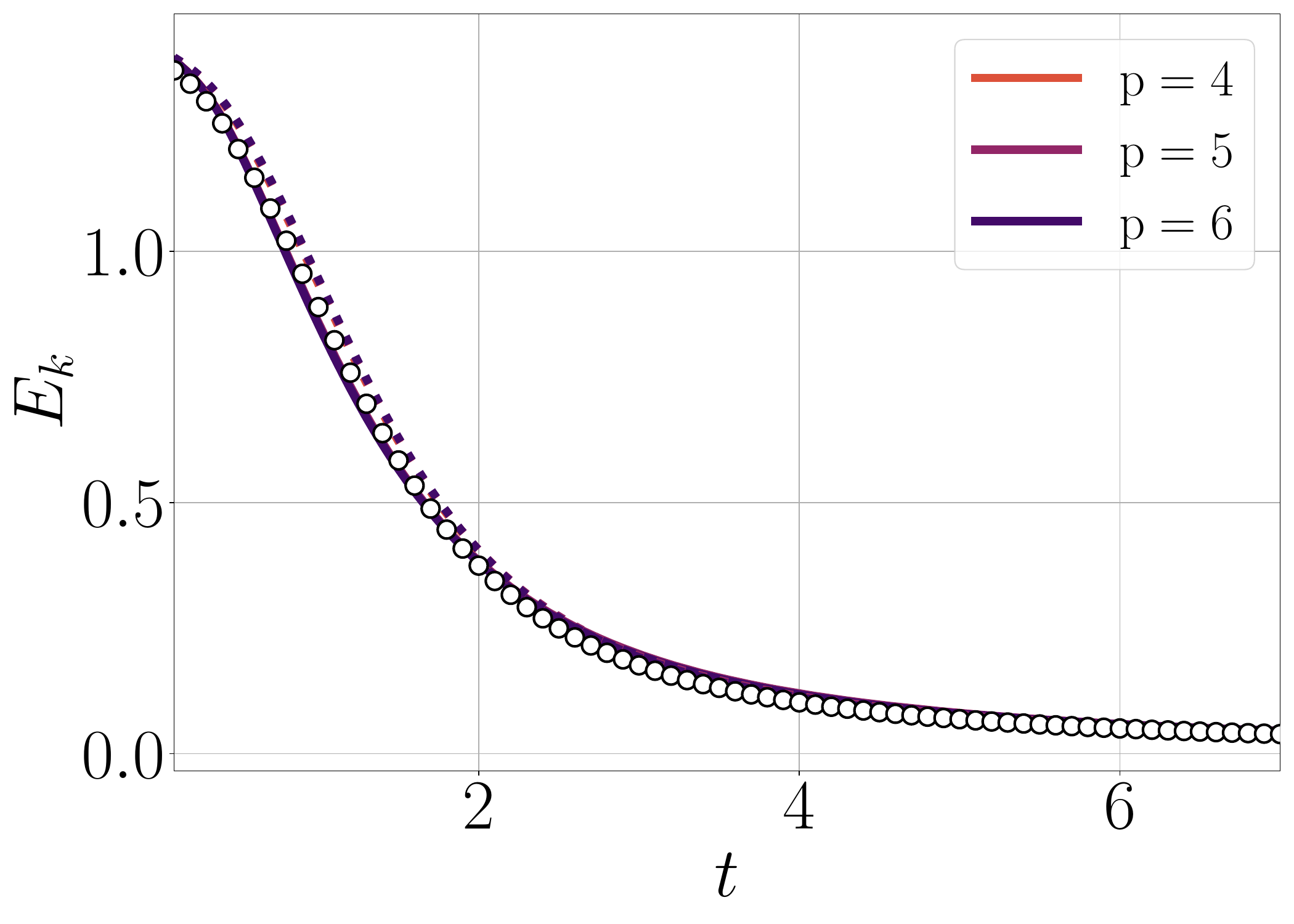}
        \caption{Kinetic Energy.}
        \label{fig:dhit_opt_smago_64:a}
        \end{subfigure}
        \begin{subfigure}[b]{0.485\textwidth}  
            \centering 
            \includegraphics[width=\textwidth]{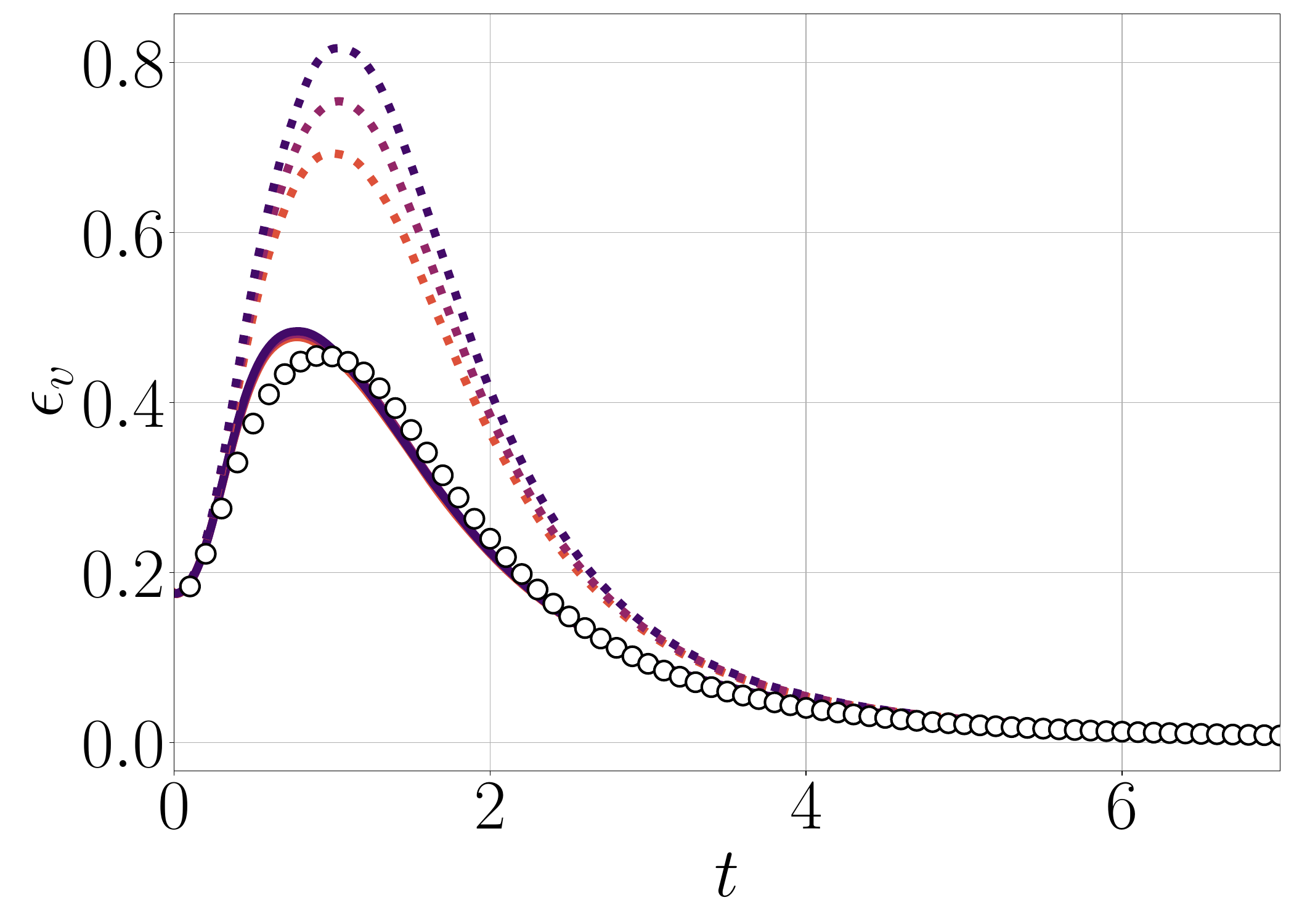}
             \caption{Viscous Dissipation.}
             \label{fig:dhit_opt_smago_64:b}
        \end{subfigure}
        \hfill
        \caption{Time evolution of the kinetic energy (left) and viscous dissipation (right) for $N_\mathrm{DoF}=64$. Solid lines, optimized Smagorinsky model; dotted lines, ILES. Symbols indicate the filtered DNS data at the given resolution.}
        \label{fig:dhit_opt_smago_64}
\end{figure}

Starting from $N_{\mathrm{dof}} = 64$, figure \ref{fig:dhit_opt_smago_64} illustrates the temporal evolution of the kinetic energy and viscous dissipation for all polynomial orders.

The ILES approaches exhibit a clear under-dissipative behavior, as both the kinetic energy and viscous dissipation levels remain significantly higher than those of the reference solution. This is again related to the inability of the numerical scheme to properly dissipate small scale features which ultimately lead to an accumulation of kinetic energy at those scales.

In contrast, the optimized model shows good agreement with the reference data for both quantities of interest. Small differences can still be identified, particularly in the viscous dissipation curves shown in figure \ref{fig:dhit_opt_smago_64:b}. For these models, the peak of viscous dissipation occurs at approximately $t \simeq 0.90$, which is reached slightly earlier compared to the filtered DNS.

\begin{figure}
    \centering
    \begin{subfigure}[b]{0.485\textwidth}
        \centering
        \includegraphics[width=\textwidth]{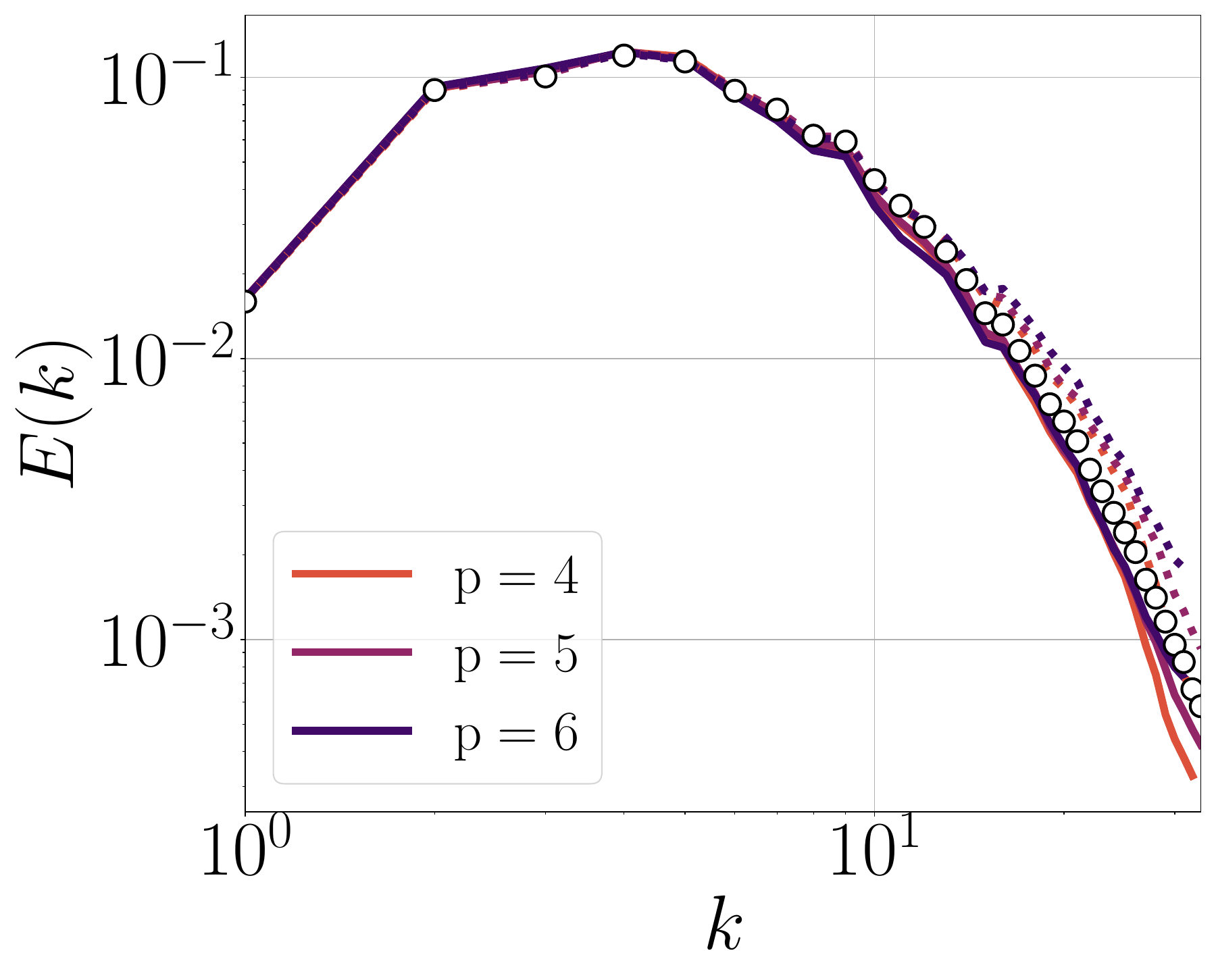}
        \caption{Kinetic Energy spectrum.}
        \label{fig:dhit_spectrum:a}
        \end{subfigure}
        \begin{subfigure}[b]{0.485\textwidth}  
            \centering 
           \includegraphics[width=\textwidth]{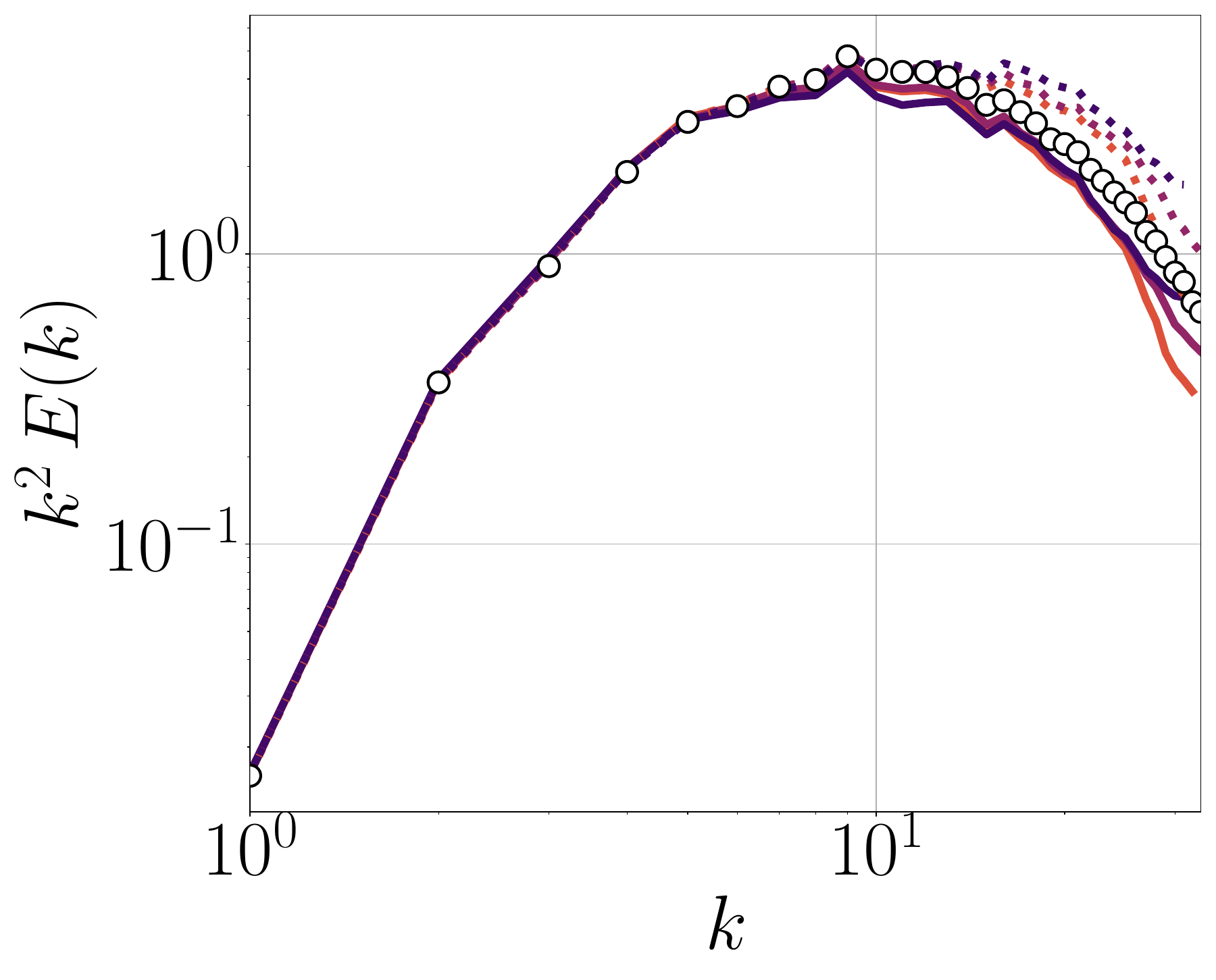}
             \caption{Spectrum of the viscous dissipation.}
             \label{fig:dhit_spectrum:b}
        \end{subfigure}
        \hfill
        \caption{Kinetic energy (left) and viscous dissipation (right) spectra for $N_{\mathrm{dof}}=64$ computed at $t=1$. Solid lines, optimized Smagorinsky model; dotted lines, ILES. Symbols indicate the filtered DNS data at the given resolution.}
        \label{fig:dhit_spectrum}
\end{figure}

To better understand how the optimized models behave, we compute the spectrum for the kinetic energy and the viscous dissipation near the enstrophy peak, specifically, at $t=1$ in figure \ref{fig:dhit_spectrum}.

By inspecting the kinetic energy spectrum on the left plot, we can notice that ILESs are under-dissipative, leading to an excess of energy at the smallest resolved scales compared to the reference solution. 
In contrast, the optimized Smagorinsky model provides an appropriate level of dissipation, yielding a kinetic energy spectrum that closely matches the reference, although minor discrepancies remain near the cutoff region, where some differences between the various polynomial orders are visible. 
Notice, however, that although the kinetic energy spectrum provides information regarding the amount of energy across the scales, it does not explain the differences in viscous dissipation shown in figure \ref{fig:dhit_opt_smago_64:b}. To further investigate this aspect, we show in figure \ref{fig:dhit_spectrum:b} the corresponding spectrum of viscous dissipation. 
The spectrum of viscous dissipation amplifies the discrepancies at high wave numbers. The excess of energy for the ILESs directly translates into an over prediction of viscous dissipation, explaining the differences observed in its temporal evolution (see figure \ref{fig:dhit_opt_smago_64}). In contrast, the optimized model remains in closer agreement with the reference, indicating a more accurate balance between energy transfer and dissipation at the smallest resolved scales.
%
 
\subsubsection{Taylor-Green Vortex}
Presenting the numerical results, we started by evaluating the performance of the optimized model for a-posteriori runs of the same test case it was trained on (\ie, FHIT). Subsequently, we passed to a case of decaying HIT, where a different type of dynamics takes place with satisfactory results. In this section, we want to test how the optimized model performs even further from the training condition. We then consider in this section the Taylor-Green Vortex problem. Although it shares some similarities with homogeneous isotropic turbulence (in particular at late stages of the dynamics) it also includes a transition to turbulence from the laminar initial condition. Furthermore, within this setting, we will also investigate how the model performs varying both resolution and Reynolds number.

\begin{figure}[h!]
\centering
\includegraphics[width=.99\textwidth]{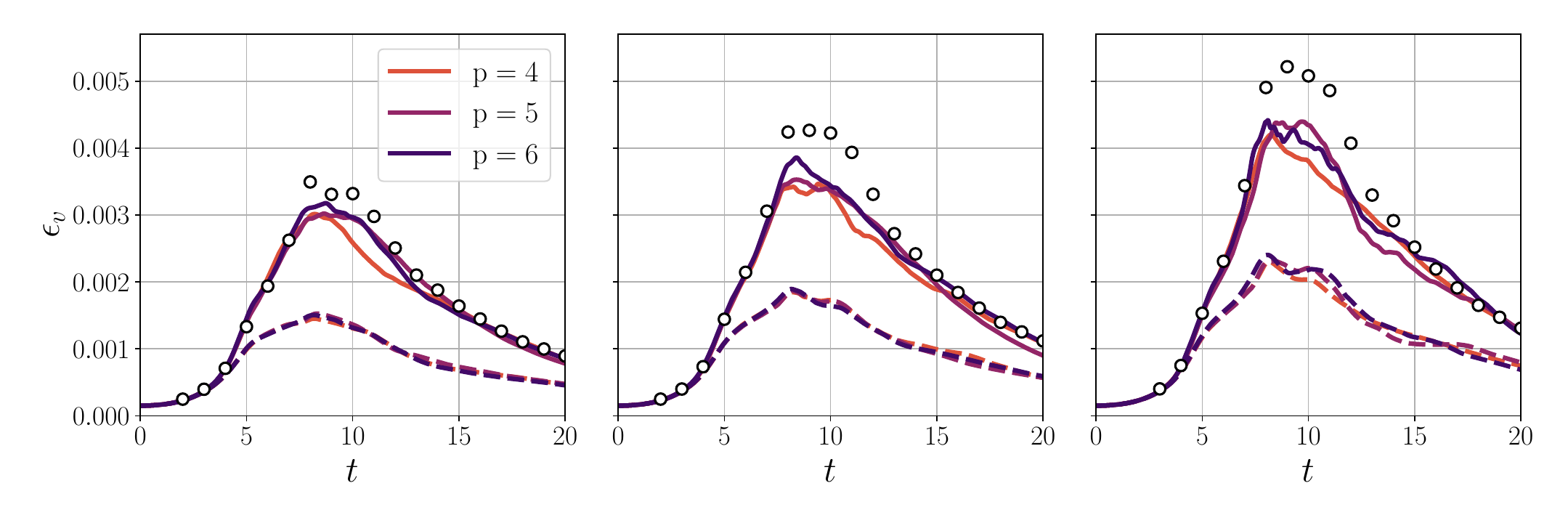}
\caption{Averaged viscous dissipation for the TGV case at $\mathrm{Re}=5000$ using the optimized Smagorinsky model (solid lines) and baseline Smagorinsky (dashed lines) for different resolutions and different polynomial orders. From left to right: $64^{3}$, $80^{3}$ and $96^{3}$. Symbols indicate the filtered DNS data at the given resolution.}
\label{fig:TGV_res_p}
\end{figure}

As observed in table \ref{tab:smago-opt-results-ns} the coefficient for the Smagorinsky constant varies with respect to the polynomial order. In fact, each discretization has its own numerical characteristics such as numerical dispersion and diffusion that significantly influence the overall result for coarse simulations. In figure \ref{fig:TGV_res_p} we show the viscous dissipation for the TGV case at $\mathrm{Re}=5000$ at different resolutions and different polynomial orders, comparing the baseline Smagorinsky model and its optimized version.
The DNS is obtained from a high-order SD simulation using $N_{el}=60$ elements and a polynomial degree $\mathrm{p}=7$ resulting in a total of $480^3$ DoF~\cite{CHAPELIER2016279,chapelier2018coherent}.
We can observe good agreement with the filtered reference data for all the resolutions, with significant improvements with respect to the baseline model. We can notice that the optimized coefficients are so that the overall influence of the numerical discretization is contained: all the curves are quite close to one another. The variations of the Smagorinsky constant observed in table \ref{tab:smago-opt-results-ns} compensate for the different numerical characteristics of the specific polynomial order. This is in clear agreement with the results presented in the previous sections regarding forced and decaying HIT. We can also notice that the best agreement between LES and filtered DNS is found for the coarsest resolution (\ie, $64^{3}$ DoF). This is somewhat expected since this is the resolution used for training. We can also notice that the optimized coefficient extrapolates quite nicely also to other resolutions.
\begin{figure}[h!]
\centering
\includegraphics[width=.99\textwidth]{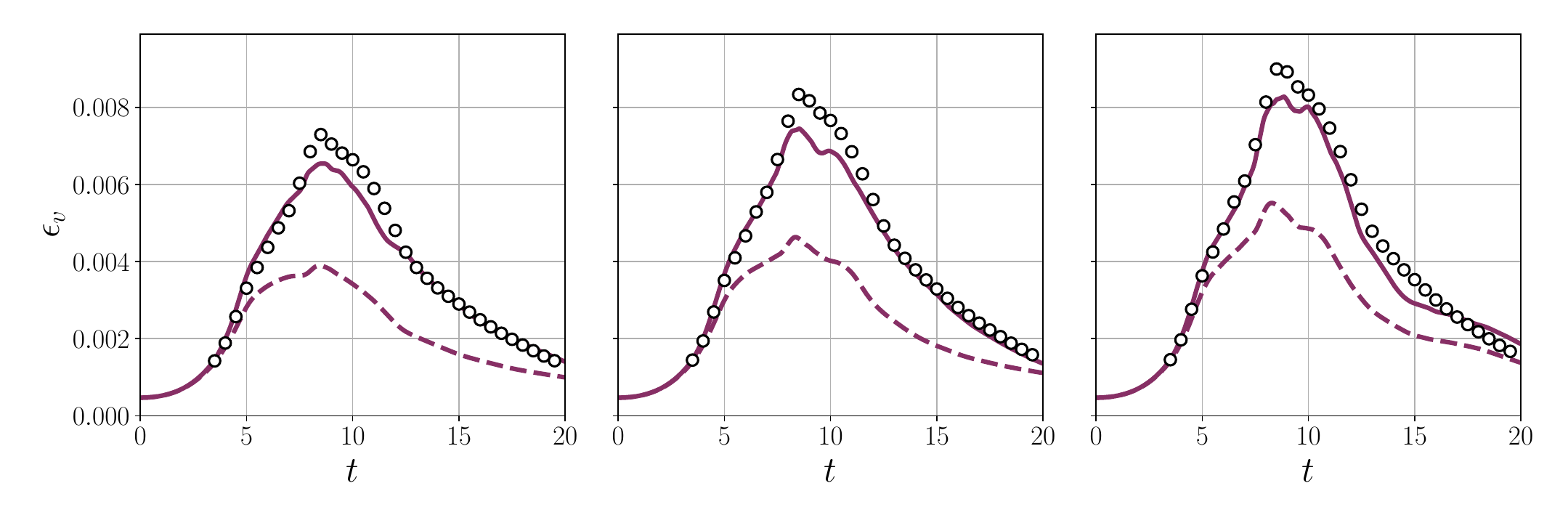}
\caption{Averaged viscous dissipation for the TGV case at $\mathrm{Re}=1600$ for different resolutions. From left to right: $64^{3}$, $80^{3}$ and $96^{3}$. Purple solid lines, optimized Smagorinsky model; purple dashed lines, baseline Smagorinsky model. Symbols indicate the filtered DNS data at the given resolution.}
\label{fig:TGV_res3}
\end{figure}
To further assess the robustness of the proposed approach we also consider the TGV case at a different Reynolds number. In particular, the TGV case is often considered at $\mathrm{Re}=1600$ too~\cite{reissmann2021application}. In this case, we restrict our analyses to $\mathrm{p}=5$ only. 

\begin{figure}[h!]
\centering
\includegraphics[width=.99\textwidth]{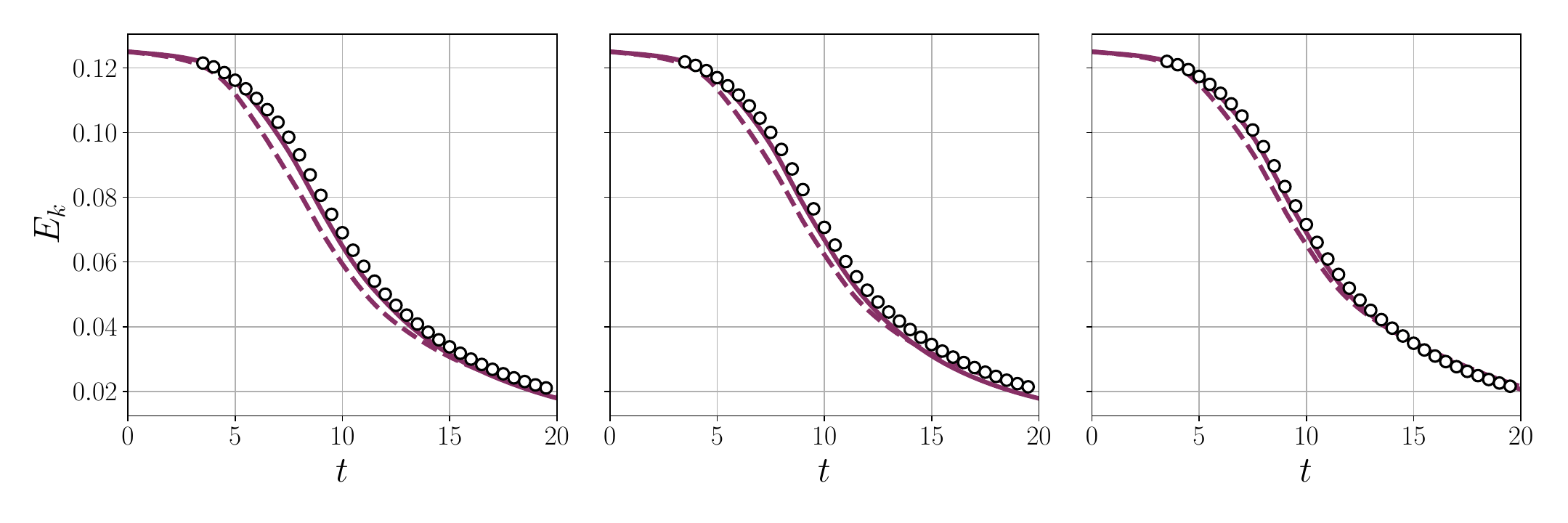}
\caption{Averaged kinetic energy for the TGV case at $\mathrm{Re}=1600$ for different resolutions. From left to right: $64^{3}$, $80^{3}$ and $96^{3}$. Purple solid lines, optimized Smagorinsky model; purple dashed lines, baseline Smagorinsky model. Symbols indicate the filtered DNS data at the given resolution.}
\label{fig:TGV_res4}
\end{figure}

\begin{figure}[h!]
\centering
\includegraphics[width=.99\textwidth]{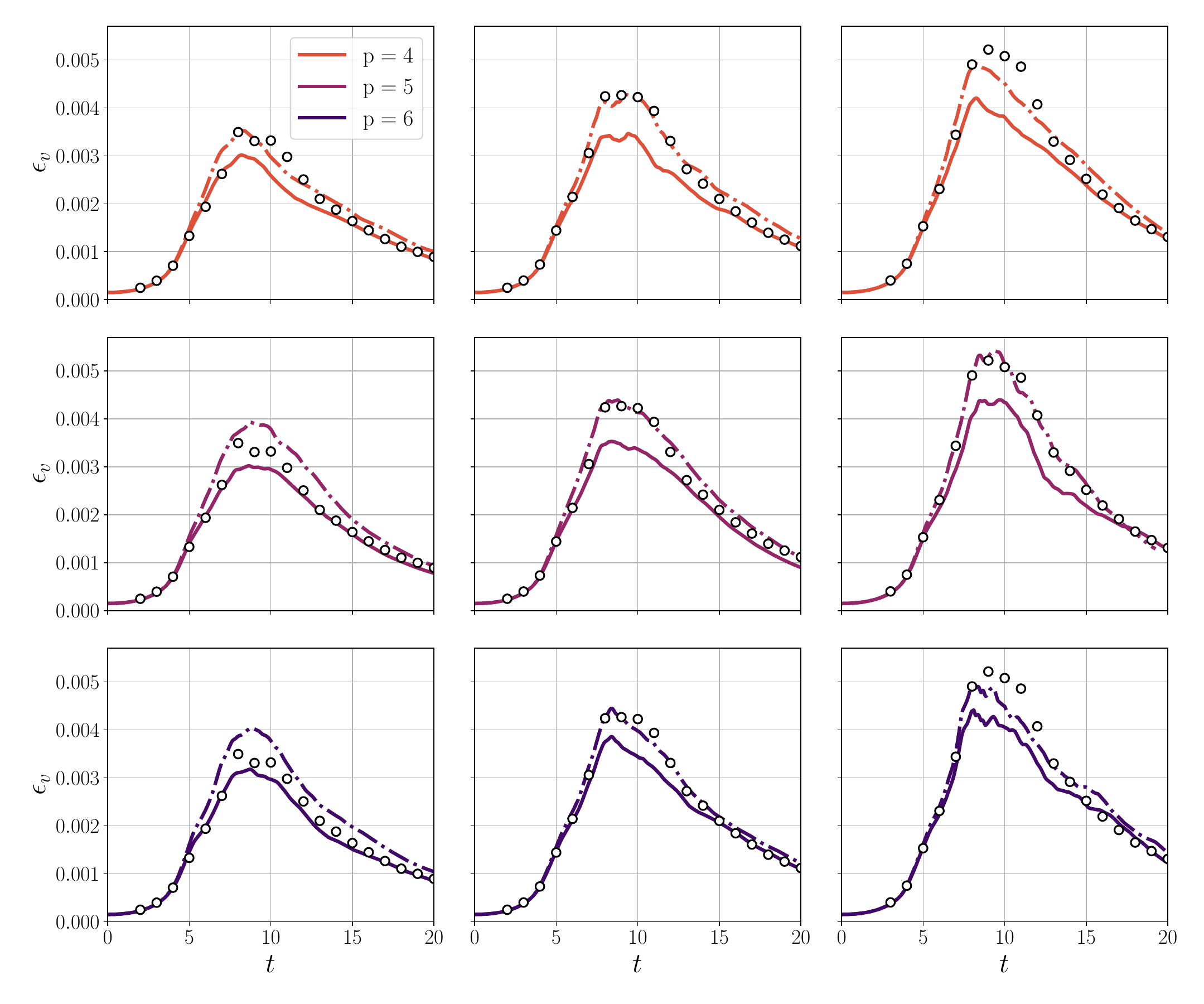}
\caption{Averaged viscous dissipation for the TGV case at $\mathrm{Re}=5000$ for different resolutions and polynomial orders. From left to right: $64^{3}$, $80^{3}$ and $96^{3}$. From top to bottom: $\mathrm{p}=4$, $\mathrm{p}=5$ and $\mathrm{p}=6$. Solid lines, optimized Smagorinsky model at resolution $64^{3}$; dash-dotted lines, optimized Smagorinsky model at resolution $80^{3}$. Symbols indicate the filtered DNS data at the given resolution.}
\label{fig:TGV_res7}
\end{figure}

The averaged viscous dissipation and kinetic energy are respectively shown in figures \ref{fig:TGV_res3} and \ref{fig:TGV_res4}. We can also observe that for $\mathrm{Re}=1600$ the optimized Smagorinsky coefficient provides significant improvements with respect to the baseline model.
Finally, up to this point we only showed comparisons using the  optimized Smagorinsky model trained on $64^{3}$ FHIT. As specified in the training 
section we also trained the same constant at $80^{3}$. In figures \ref{fig:TGV_res7} we compare the optimized Smagorinsky models trained on different resolutions ($64^{3}$, $80^{3}$, $96^{3}$) and different polynomial orders ($\mathrm{p}=4,5,6$) for the TVG case at $\mathrm{Re}=5000$.
As expected from inspection of the first column of figure~\ref{fig:TGV_res7}, the solid lines are significantly closer to the reference than the dashed ones. This indicates that, for all polynomial orders considered, the model optimized at a given resolution yields the most accurate prediction of the filtered DNS data at that same resolution. In particular, the model trained on the $64^{3}$ dataset performs best on the corresponding filtered DNS data (first column), while the Smagorinsky model optimized on the $80^{3}$ forced HIT configuration shows the best agreement with the filtered DNS data at the $80^{3}$ resolution (second column).

In the third column, both optimized models are assessed in an extrapolation setting at $96^{3}$. Here, the model trained on a $80^{3}$ resolution gives overall improved predictions compared to the one trained on the coarser resolution. This advantage is particularly evident in the peak enstrophy, while differences are less marked in the decay phase.

It is important to emphasize that all models were trained on forced HIT, without any knowledge of transition to turbulence. As such, there is no \emph{a priori} expectation regarding their relative performance in this scenario. Nonetheless, the results indicate that both models generalize well to the TGV case, across different resolutions and polynomial orders.
\subsection{The Tensor basis approach}
In the previous section, we focused on optimizing the coefficient of the Smagorinsky model and assessing its performance across different test cases. In this section, we instead aim to optimize more complex SGS models involving a larger number of control parameters. The motivation is twofold. First, this increases the model expressivity by enabling more sophisticated formulations, potentially leading to improved performance. Second, it allows us to assess the robustness of the present framework in the presence of multiple control parameters.

In this study, we consider an additional model that depends on the strain-rate tensor $\widetilde{S}_{ij}$ and the rotation-rate tensor $\widetilde{\Omega}_{ij}$:
\begin{equation}
   \tau^{\mathrm{SGS}}_{ij} = \tau^{\mathrm{SGS}}_{ij}({\widetilde{S}}_{ij},\widetilde{\Omega}_{ij}), \label{eq:TB_eq0}
\end{equation}
with
\begin{equation}
    \widetilde{S}_{ij} = \frac{1}{2}\left( \pd{\widetilde{u}_i}{x_j} + \pd{\widetilde{u}_j}{x_i}\right) \;\;\;\;\; \widetilde{\Omega}_{ij}=\frac{1}{2} \left(\pd{\widetilde{u}_i}{x_j}-\pd{\widetilde{u}_j}{x_i} \right). \label{eq:TB_eq1}
\end{equation}
We consider a model form based on a combination of the two second-order tensors $\widetilde{S}_{ij}$ and $\widetilde{\Omega}_{ij}$. The number of independent second-order tensor functions that can be formed from $\widetilde{S}_{ij}$ and $\widetilde{\Omega}_{ij}$ is limited to ten due the Caley-Hamilton theorem. Consequently, $\tau^{\mathrm{SGS}}_{ij}$ can be written as a linear combination of a finite set of basis functions ${T}^{k}_{ij}$:
\begin{equation}
\tau_{ij}^{\mathrm{SGS}} = -2 \left(\Delta \sqrt{2 \widetilde{S}_{ij}\widetilde{S}_{ij}} \right)^{2} \left[ \sum_{k=1}^{n}\alpha_k T_{ij}^{k}\right]. \label{eq:TBexpansion}
\end{equation}
In order to reduce the computational complexity of the model, we restrict our model to the first three invariants:
\begin{align}
    T^{1}_{ij} &=  s_{ij},  \\
    T^{2}_{ij} & = (s_{ik}\omega_{kj}-\omega_{ik}s_{kj}),\\
    T^{3}_{ij} & = (s_{ik}s_{kj}-s_{mk}s_{km} \delta_{ij}/3),
\end{align}
where $s_{ij}$ and $\omega_{ij}$ are the normalized strain-rate tensor and rotation-rate tensor
\begin{equation}
    s_{ij} = \frac{\widetilde{S}_{ij}}{\sqrt{2\widetilde{S}_{ij}\widetilde{S}_{ij}}} \;\;\;\;\; \omega_{ij}=\frac{\widetilde{\Omega}_{ij}}{\sqrt{2\widetilde{S}_{ij}\widetilde{S}_{ij}}}. \label{eq:TB_eq1}
\end{equation}
Note that this model can be seen as a simple correction of an eddy-viscosity model with a structural part given by second order terms $T^2$ and $T^3$. 
It is well known that purely eddy-viscosity models are inherently dissipative and therefore unable to represent backscatter \cite{PIOMELLI1999335,BARDINASSM91,MCMILLANFERZIGER79}. At the same time, purely structural models typically show good a priori agreement with filtered DNS data, although they often exhibit poor a posteriori performance. In this framework, the proposed model can be viewed as a mixed formulation, in which the second-order terms introduce a structural component capable of capturing backscatter effects, thereby compensating for the deficiencies of purely dissipative closures while retaining their intrinsic stability properties. Moreover, we aim to develop a simple model that is easily interpretable from a physical standpoint, while still being able to adapt to the underlying flow physics. For a given set of optimal coefficients $\alpha_{k}$, we will refer to the correspondent model as $\mathrm{TB}^3$.

In the following sections we will compare the present model with the similar models developed in~\cite{reissmann2021application}. Their formulation is basically equivalent to the one presented here except for a coefficient of $\sqrt{2}$ in the first term of the expansion. This is simply due to the adimensionalisation using $\sqrt{2\widetilde{S}_{ij}\widetilde{S}_{ij}}$ instead of $\sqrt{\widetilde{S}_{ij}\widetilde{S}_{ij}}$. In our formulation, in addition, we only consider the first three terms of the expansion instead of four. As we will see in the results, we do not expect particularly different results including additional terms of the expansion. For completeness, the formulation and models proposed by~\cite{reissmann2021application} are reported in \ref{:GEPappendix}.
\subsubsection{Training}\label{subs:OptTB_training}
Following the same approach adopted for the Smagorinsky model, we train the proposed model using the FHIT test case introduced in the previous section.
To ensure consistency, we employ the same numerical setup used for the Smagorinsky optimization. Accordingly, the training is performed at the two resolutions considered above (\ie, $N_{\mathrm{dof}} = 64$ and $N_{\mathrm{dof}} = 80$) and for polynomial orders $\mathrm{p}=4,5,6$. 
As the optimal values of the scalar coefficients are not known a-priori, we set the initial values to $\boldsymbol{\alpha}=(0.035,0.0,0.0)$. This is equivalent to start from a Smagorinsky-like model with a constant $C_s=\sqrt{\alpha_1}\simeq 0.187$.

\begin{figure}
    \centering
    \begin{subfigure}[b]{0.475\textwidth}
        \centering
        \includegraphics[width=\textwidth]{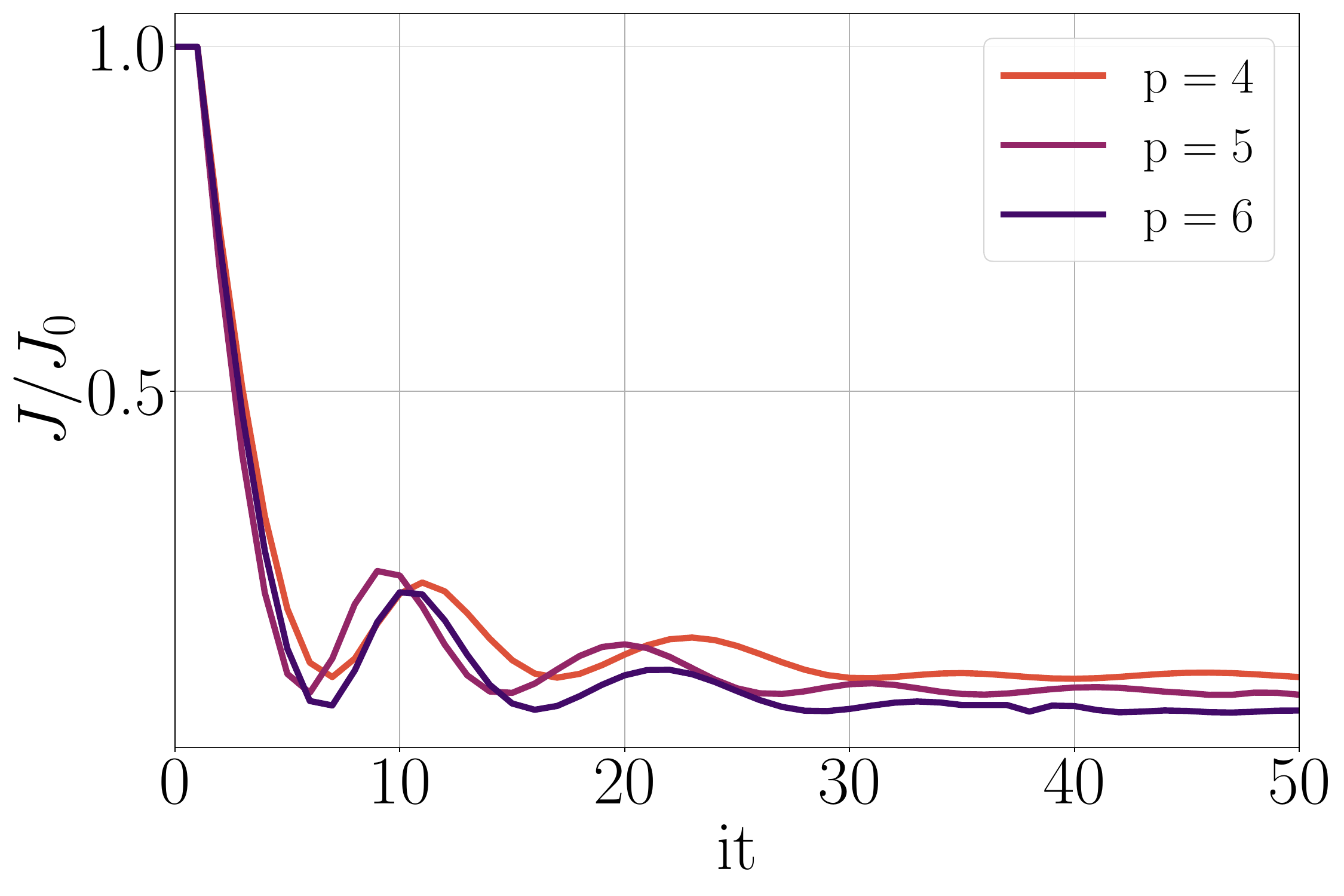}
        \caption{$N_{\mathrm{dof}}=64$}
        \label{fig:ns_loss_tb3:a}
        \end{subfigure}
        \begin{subfigure}[b]{0.485\textwidth}  
            \centering 
            \includegraphics[width=\textwidth]{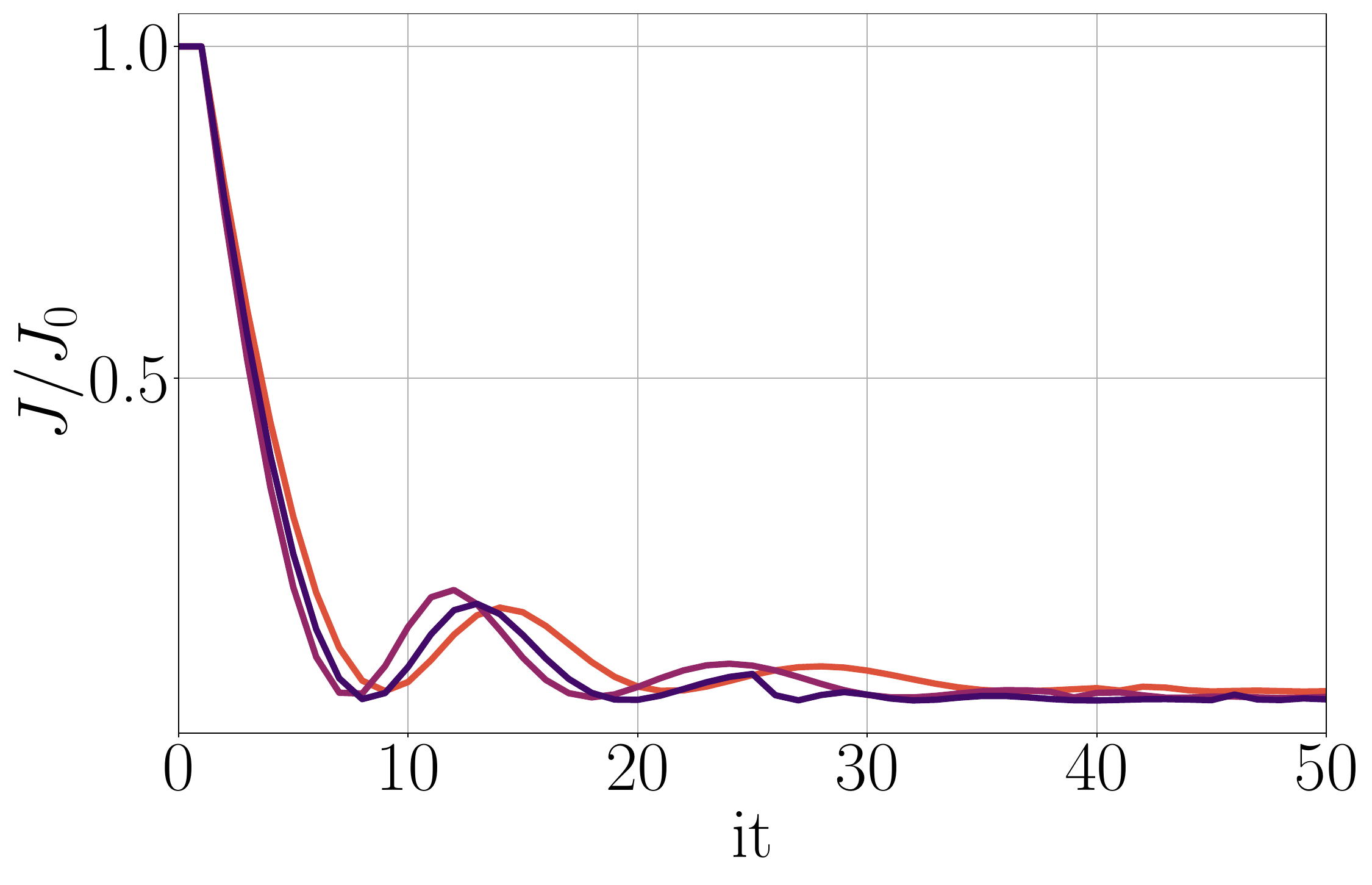}
             \caption{$N_{\mathrm{dof}}=80$}
             \label{fig:ns_loss_tb3:b}
        \end{subfigure}
        \hfill
        \caption{$\mathrm{TB^3}$ model. Normalized-loss functions for different polynomial orders for $N_{\mathrm{dof}}=64$ (left) and for $N_{\mathrm{dof}}=80$ (right).}
        \label{fig:ns_loss_tb3}
\end{figure}

Figure \ref{fig:ns_loss_tb3} shows the normalized loss functions for both the resolutions and all polynomial orders. In both cases, the loss function decreases to approximately one tenth of its initial value after about $40$ iterations. Since in this case the model incorporates multiple parameters, the optimization is more challenging with respect to the Smagorisnky model. Consequently, at the early stage of the optimization some oscillations can be observed. These oscillations are dampened as the iterations increase, until the loss function reaches a plateau. Consequently, we stop the number of iterations when $\mathrm{it}=50$.

Regarding the modal energy, we do not report here the modal energy behavior found during the optimization as we obtain almost the same results found in the optimized Smagorisnky model. Consequently, the same conclusions of previous sections apply: the baseline approach (which is basically a standard Smagorinky model) is dissipative, leading to excessive damping of the modes. In contrast, the optimized models show good agreement with the reference decay.

The left half of table \ref{tab:tb-opt-results-ns} reports the optimized coefficients for the case $N_{\mathrm{dof}} = 64$. First, it can be observed that the coefficient $\alpha_1$ is lower for $\mathrm{p} = 4$ than for the higher polynomial orders, namely $\mathrm{p} = 5$ and $\mathrm{p} = 6$. This behavior may be attributed to the nature of the first term in the expansion, which is purely dissipative and therefore is expected to increase with polynomial order.

Regarding the coefficients relative to the second order terms, we can notice that the coefficient $\alpha_2$ is small in magnitude compared to $\alpha_3$. The latter is consistently negative across all the polynomials, with its largest magnitude observed for $\mathrm{p}=6$.

Correspondingly, the right half of table \ref{tab:tb-opt-results-ns} shows the optimized coefficients for $N_{\mathrm{dof}}=80$. It is worth noting that the same trends observed at the lower resolution are also recovered here. Furthermore, comparing the same polynomial order for the two resolutions, it can be noted that the first coefficient decreases as the resolution increases. This behavior is expected, as the leading-order coefficient plays the dominant role in the expansion. Consequently, with increasing resolution, the required amount of dissipation is reduced, leading to smaller values of this coefficient.

It is worth mentioning that the Clark model~\cite{ClarkModel79} can be written in terms of the second order expansion terms,
\begin{equation}
    \tau^{Clark}_{ij} = -2\left(\Delta \sqrt{2 \widetilde{S}_{ij}\widetilde{S}_{ij}} \right)^{2} \frac{1}{24} \bigg(T^{2}_{ij}-T^{3}_{ij}+T^{4}_{ij} \bigg).\label{eqn:TBclark}
\end{equation}
This observation establishes a link between the proposed tensor-basis model and the Clark model, showing that the latter can be interpreted as a specific instance of the expansion with fixed coefficients.
In this context, it is interesting to note that the second coefficient is greater, in magnitude, with respect to the third. According to Horiuti~\cite{Hariouti03} this term is responsible for generating vorticity although it is neutral for the generation of subgrid-scale dissipation.
\begin{table}
\centering
\begin{tabular}{|c|ccc|ccc|ccc|}
\hline

& \multicolumn{3}{c|}{$\mathrm{p}=4$} 
& \multicolumn{3}{c|}{$\mathrm{p}=5$} 
& \multicolumn{3}{c|}{$\mathrm{p}=6$} \\
\hline

$N_{\mathrm{dof}}$ 
& $\alpha_1$ & $\alpha_2$ & $\alpha_3$
& $\alpha_1$ & $\alpha_2$ & $\alpha_3$
& $\alpha_1$ & $\alpha_2$ & $\alpha_3$ \\
\hline

64 
& 0.0149 & -0.0282 & 0.0061
& 0.0159 & -0.0267 & 0.0062
& 0.0154 & -0.0405 & -0.0002 \\
\hline

80 
& 0.0115 & -0.0289 & 0.0135
& 0.0139 & -0.0339 & 0.0192
& 0.0138 & -0.0368 & 0.0155 \\
\hline

\end{tabular}
\caption{Optimized coefficients of the model expansion for two resolutions and different polynomial degrees.}
\label{tab:tb-opt-results-ns}
\end{table}

%
\subsubsection{Testing}
\begin{figure}[h!]
\centering
\includegraphics[width=.99\textwidth]{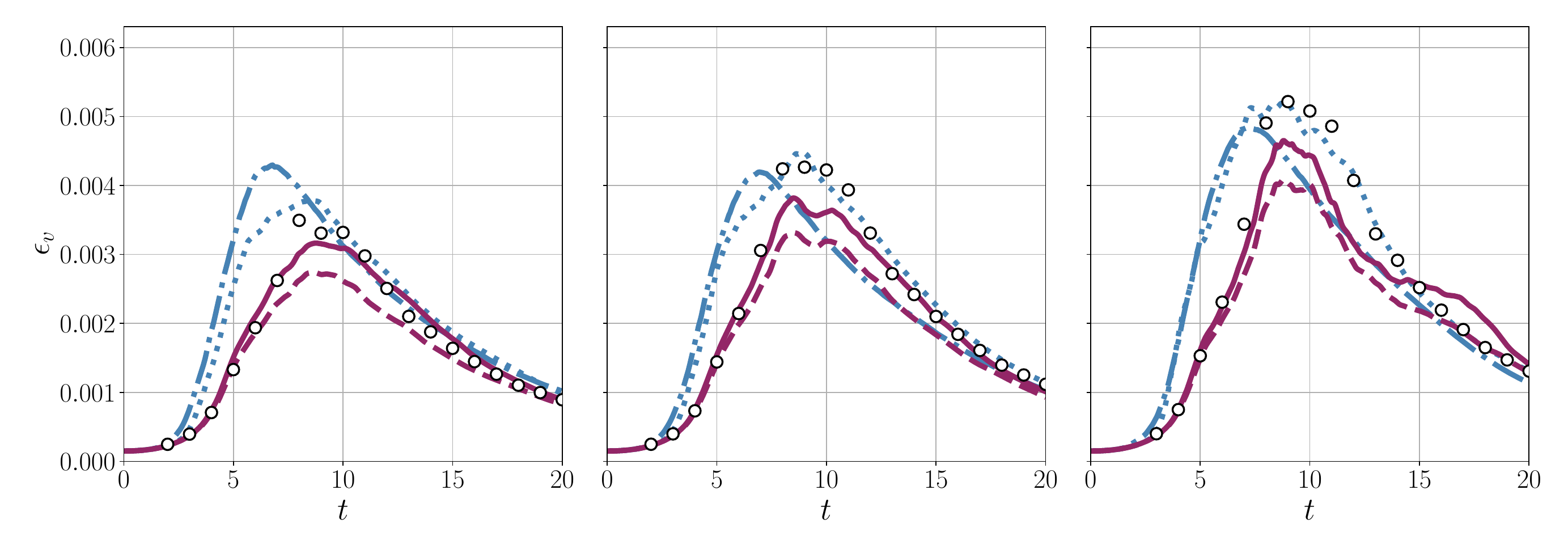}
\caption{Averaged viscous dissipation for the TGV case at $\mathrm{Re}=5000$ for different resolutions and polynomial orders. From left to right: $64^3$, $80^3$, $96^3$. Solid lines, TB3 model at resolution $80^3$; dashed lines, TB3 model at resolution $64^3$; blue dash-dotted lines, GEP2 from~\cite{reissmann2021application}; blue dotted lines, GEP1 from~\cite{reissmann2021application}. Symbols indicate the filtered DNS data at the given resolution.}
\label{fig:TGV5k_tb3_validation_epsV}
\end{figure}

In figure \ref{fig:TGV5k_tb3_validation_epsV} we show the viscous dissipation for the TGV case at $\mathrm{Re} = 5000$ at different resolutions and different polynomial orders, comparing the $\mathrm{TB}^3$ model optimized on both resolutions $N_\mathrm{dof}=64^{3}$ and $N_\mathrm{dof}=80^{3}$ with the GEP1 and GEP2 models from~\cite{reissmann2021application}. We can observe that the model trained on $N_\mathrm{dof}=80^{3}$ shows a good agreement with the reference filtered data for all the resolutions.
For the coarsest resolution examined in the study (i.e., $64^3$), the model reproduces the peak viscous dissipation of the reference filtered data with high accuracy. Furthermore, at the $80^3$ resolution, the model agrees well with the reference data during both the laminar and decay phases, although the peak viscous dissipation is slightly under predicted.

However, in contrast, the model trained on the lowest resolution, namely, $N_{\mathrm{dof}}=64^{3}$, seems to be a bit more dissipative compared to the model optimized on $N_\mathrm{dof}=80^{3}$ as the peak of the viscous dissipation is lower in this case. We can also notice that the model trained on the coarsest resolution performs slightly better in the decaying phase (\ie, for $t>12$) when tested on $N_\mathrm{dof}=64^{3}$, while the other is a bit under dissipative, leading to an excessive viscous dissipation.

\begin{figure}[h!]
\centering
\includegraphics[width=.99\textwidth]{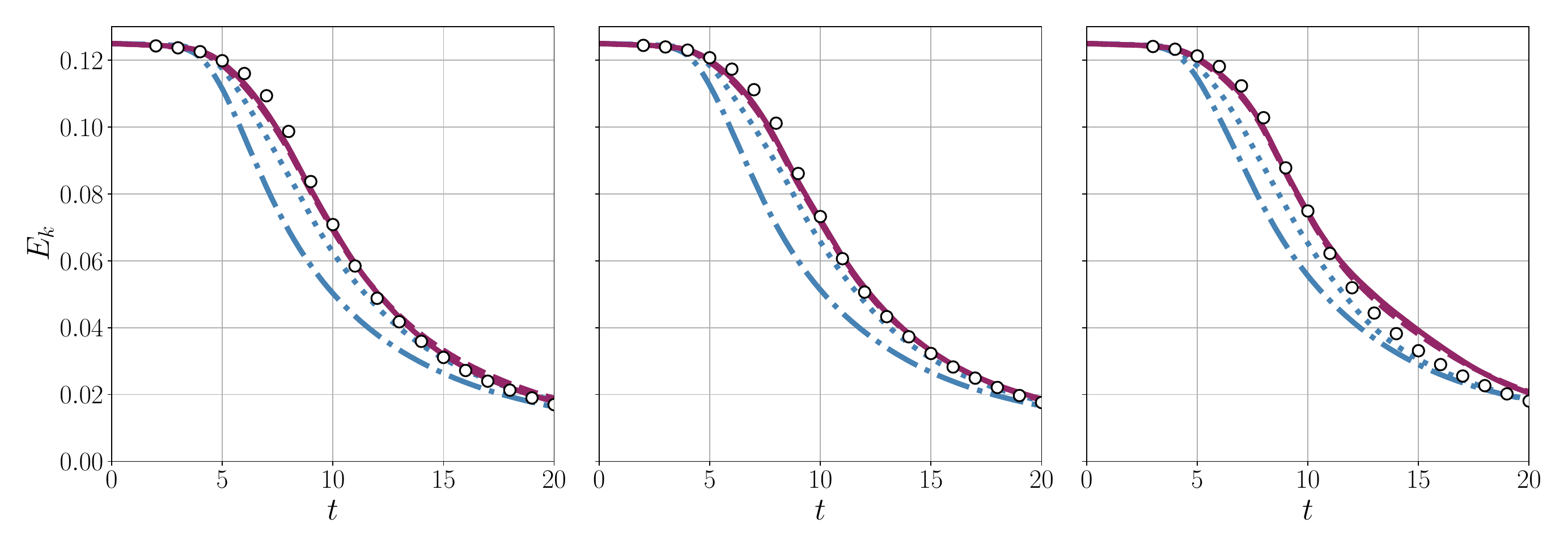}
\caption{Averaged kinetic energy for the TGV case at $\mathrm{Re}=5000$ for different resolutions and polynomial orders. From left to right: $64^3$, $80^3$, $96^3$. Solid lines, TB3 model at resolution $80^3$; dashed lines, TB3 model at resolution $64^3$; blue dash-dotted lines, GEP2 from~\cite{reissmann2021application}; blue dotted lines, GEP1 from~\cite{reissmann2021application}. Symbols indicate the filtered DNS data at the given resolution.}
\label{fig:TGV5k_tb3_validation_ke}
\end{figure}

Next, we consider the time evolution of the averaged kinetic energy in figure \ref{fig:TGV5k_tb3_validation_ke}. Also for this quantity of interest, the model shows a good agreement with the reference filtered data for all the resolutions. However, some discrepancies are found when extrapolating at $N_{\mathrm{dof}}=96$. In this scenario, the kinetic energy in the decaying phase is slightly over-predicted for both the models trained on $80$ and $96$. 

It is interesting to note that although the models trained on different resolutions show different behavior in the viscous dissipation, the curves of the kinetic energy are almost overlapped. This may due to the fact that the models are optimized on the modal energy, which is a quantity related to the kinetic energy rather than the viscous dissipation. 
Since the resolutions used for the training are not too far away each other, the budget of the energy contained in each mode it is not different between the two resolutions. 
Consequently, the models are able to provide comparable results in terms of kinetic energy.

%

%
\section{Conclusions} \label{sec:conclusions}
In this work, we have presented a discrete-adjoint-based framework for the optimization of SGS model parameters for LESs. The proposed approach enables \emph{in-the-loop} training by leveraging automatic differentiation of a high-order spectral difference solver, thereby accounting for both modeling and discretization errors during the optimization process. The methodology has been applied to the calibration of classical SGS closures, including the Smagorinsky and tensor-basis models, applied to one-dimensional Burgers equation and to fully three-dimensional turbulence.

The training procedure has been carried out on coarse-grained simulations of FHIT, using filtered DNS data as reference. The optimization has been performed across different polynomial orders of the spatial discretization, highlighting the strong coupling between SGS model parameters and the underlying numerical scheme. The resulting optimized coefficients exhibit a clear dependence on the discretization properties, confirming the importance of consistent model calibration.

The generalization capabilities of the trained models have been assessed on a range of out-of-sample test cases, including decaying homogeneous isotropic turbulence and the Taylor-Green vortex. Variations in polynomial order, grid resolution, and Reynolds number have been considered. In all cases, the optimized models demonstrate significant improvements over standard baseline closures, providing more accurate predictions of turbulence statistics.

\section*{Aknowledgements}
This study was funded by the European Union - NextGenerationEU, in the framework of the iNEST - Interconnected Nord-Est Innovation Ecosystem (iNEST ECS00000043 – CUP G93C22000610007). The views and opinions expressed are solely those of the authors and do not necessarily reflect those of the European Union, nor can the European Union be held responsible for them. The authors NC, NT and GR acknowledge the support by INdAM-GNCS: Istituto Nazionale di Alta Matematica –– Gruppo Nazionale di Calcolo Scientifico. 
Paola Cinnella acknowledges support by the French National Research Agency (ANR) under the France 2030 program, within the framework
of the IA Cluster initiative (grant ANR-23-IACL-0007).
The use of the SD solver originally developed by Antony Jameson’s group at Stanford University is gratefully acknowledged. We are also grateful to Guido Lodato for the invaluable suggestions and guidance, which helped in improving the present work. The authors would like to thank Jean-Baptiste Chapelier for providing the DNS data used as reference in this work.

\appendix

\section{Gene Expression Programming models} \label{:GEPappendix}
In~\cite{reissmann2021application} Gene Expression Programming (GEP) was used in order to develop new SGS models within the framework of incompressible Large-Eddy Simulations. In particular, the Caley-Hamilton theorem was used to formulate the unknown SGS stress tensor as a finite series of tensor basis functions $T^{\alpha}$ and coefficients $G_{\alpha}$ which depend on the invariants of the velocity gradient:
\begin{equation}
\tau_{ij}^{\mathrm{mod}} = -2 \Delta^{2} \sum_{\alpha=1}^{n}T^{\alpha}G_{\alpha} \quad \mathrm{with} \quad G_{\alpha}=G_{\alpha}(I_{1},I_{2},...,I_{m})
\end{equation}
In particular, in~\cite{reissmann2021application}, they set $n=4$ and define:
\begin{align}
    T^{1}_{ij} &=  s_{ij} \cdot |\overline{S}|, & I_{1} &= s_{mn}s_{nm}, \\
    T^{2}_{ij} & = (s_{ik}\omega_{kj}-\omega_{ik}s_{kj}) \cdot |\overline{S}|^{2}, & I_{2} &= \omega_{mn}\omega_{nm}, \\
    T^{3}_{ij} & = (s_{ik}s_{kj}-s_{mk}s_{km} \delta_{ij}/3) \cdot |\overline{S}|^{2}, & I_{3} &= s_{km}s_{mn}s_{nk}, \\  
    T^{4}_{ij} & = (\omega_{ik}\omega_{kj}-\omega_{mk}\omega_{km} \delta_{ij}/3) \cdot |\overline{S}|^{2}, & I_{4} & = \omega_{km}\omega_{mn}s_{nk},
\end{align}
with $s_{ij}=\overline{S}_{ij}/|\overline{S}|$, $\omega_{ij}=\overline{\Omega}_{ij}/|\overline{S}|$,
\begin{equation}
\overline{S}_{ij} = \frac{1}{2} \bigg( \frac{\partial \overline{u}_{i}}{\partial x_{j}} + \frac{\partial \overline{u}_{j}}{\partial x_{i}} \bigg), \quad \overline{\Omega}_{ij} = \frac{1}{2} \bigg( \frac{\partial \overline{u}_{i}}{\partial x_{j}} - \frac{\partial \overline{u}_{j}}{\partial x_{i}} \bigg) \quad \mathrm{and} \quad |\overline{S}|=\sqrt{\overline{S}_{ij} \overline{S}_{ij}}.
\end{equation}
In~\cite{reissmann2021application} they obtained these two optimal models
\begin{align}
    \tau_{ij}^{\mathrm{GEP^{1}}} &= 2 \Delta^{2} (I_{3}+0.04)T_{ij}^{2} \\
    \tau_{ij}^{\mathrm{GEP^{2}}} &= -2 \Delta^{2} (C_{1} |\overline{S}| T_{ij}^{1}-C_{2}T_{ij}^{2} +C_{3}T_{ij}^{3} -C_{4} T_{ij}^{4})
\end{align}
with $C_{1}=0.01$, $C_{2}=0.146$, $C_{3}=0.01$ and $C_{4}=0.11$.

\bibliographystyle{abbrv}
\bibliography{bibexport}

\end{document}